\documentclass[]{aastex631}

\usepackage{makecell}
\usepackage{subfigure}
\usepackage{placeins}

\begin{document}

\title{LHAASO observation of Mrk~421 during 2021 March - 2024 March: a comprehensive VHE catalog of multi-timescale outbursts and its time average behavior}

\author{Zhen Cao}
\affiliation{State Key Laboratory of Particle Astrophysics \& Experimental Physics Division \& Computing Center, Institute of High Energy Physics, Chinese Academy of Sciences, 100049 Beijing, China}
\affiliation{University of Chinese Academy of Sciences, 100049 Beijing, China}
\affiliation{TIANFU Cosmic Ray Research Center, 610000 Chengdu, Sichuan,  China}
 
\author{F. Aharonian}
\affiliation{TIANFU Cosmic Ray Research Center, 610000 Chengdu, Sichuan,  China}
\affiliation{University of Science and Technology of China, 230026 Hefei, Anhui, China}
\affiliation{Yerevan State University, 1 Alek Manukyan Street, Yerevan 0025, Armenia}
\affiliation{Max-Planck-Institut for Nuclear Physics, P.O. Box 103980, 69029  Heidelberg, Germany}
 
\author{Y.X. Bai}
\affiliation{State Key Laboratory of Particle Astrophysics \& Experimental Physics Division \& Computing Center, Institute of High Energy Physics, Chinese Academy of Sciences, 100049 Beijing, China}
\affiliation{TIANFU Cosmic Ray Research Center, 610000 Chengdu, Sichuan,  China}
 
\author{Y.W. Bao}
\affiliation{Tsung-Dao Lee Institute \& School of Physics and Astronomy, Shanghai Jiao Tong University, 200240 Shanghai, China}
 
\author{D. Bastieri}
\affiliation{Center for Astrophysics, Guangzhou University, 510006 Guangzhou, Guangdong, China}
 
\author{X.J. Bi}
\affiliation{State Key Laboratory of Particle Astrophysics \& Experimental Physics Division \& Computing Center, Institute of High Energy Physics, Chinese Academy of Sciences, 100049 Beijing, China}
\affiliation{University of Chinese Academy of Sciences, 100049 Beijing, China}
\affiliation{TIANFU Cosmic Ray Research Center, 610000 Chengdu, Sichuan,  China}
 
\author{Y.J. Bi}
\affiliation{State Key Laboratory of Particle Astrophysics \& Experimental Physics Division \& Computing Center, Institute of High Energy Physics, Chinese Academy of Sciences, 100049 Beijing, China}
\affiliation{TIANFU Cosmic Ray Research Center, 610000 Chengdu, Sichuan,  China}
 
\author{W. Bian}
\affiliation{Tsung-Dao Lee Institute \& School of Physics and Astronomy, Shanghai Jiao Tong University, 200240 Shanghai, China}
 
\author{J. Blunier}
\affiliation{APC, Universit\'e Paris Cit\'e, CNRS/IN2P3, CEA/IRFU, Observatoire de Paris, 119 75205 Paris, France}
 
\author{A.V. Bukevich}
\affiliation{Institute for Nuclear Research of Russian Academy of Sciences, 117312 Moscow, Russia}
 
% ...
\collaboration{327}{The LHAASO Collaboration}


\begin{abstract}

The Large High Altitude Air Shower Observatory (LHAASO) monitors sources within its field of view for up to 7 hours daily, achieving a duty cycle exceeding 98\% and an annual point-source sensitivity of 1.5\% Crab Units (CU) in the very high energy (VHE) band. This unbiased sky-survey mode facilitates systematic monitoring and investigation of outburst phenomena.
In this paper, we present results from an unprecedented three-year monitoring campaign (March 2021–March 2024) of Mrk 421 using LHAASO, spanning energies from 0.4\,TeV to 20\,TeV. We find that the blazar stayed in a quiescent state in 2021 and became active starting in 2022 with a total of 23 VHE outburst events identified, where the highest observed daily significance reaches $20 \sigma$ with a flux equivalent to approximately 3.3 CU. LHAASO's continuous monitoring suggests the flaring occupancy of Mrk~421 to be around 14\%.
During long-term monitoring, multiwavelength (MWL) variability and correlation analyses are conducted using complementary data from Fermi-LAT, MAXI-GSC, SWIFT-XRT, and ZTF. A significant correlation ($>3~\sigma$) is observed between X-ray and VHE bands with no detectable time lag, while the correlation between GeV and TeV bands is weaker. The flux distribution of the TeV emission during the quiescent state is different from that in the active state, implying the existence of two modes of energy dissipation in the blazar jet. 
Using simultaneous MWL data, we also analyzed both the long-term and outburst-period SEDs, and discussed the possible origin of the outburst events. 

\end{abstract}

\keywords{Active galactic nuclei, BL Lacertae objects, Mrk 421, LHAASO}

\section{Introduction}

Blazars belong to a subclass of active galactic nuclei (AGNs) with their jets closely aligned with the line of sight of observers. Hence, the radiation of a blazar is dominated by the approaching jet\citep{1984RvMP...56..255B}. The radiation spectrum spans from the radio band to VHE gamma-ray band, exhibiting pronounced variability on timescales from years down to minutes \citep{1995ARA&A..33..163W}. The SED of blazars can be generally characterized by two distinct humps. The low-energy hump, peaking between the optical and X-ray bands, is commonly attributed to synchrotron radiation of ultra-relativistic electrons accelerated within the the jet. The frequency of the synchrotron peak in the blazar spectrum defines the classification for low, intermediate, and high-frequency-peaked objects (LBL, IBL, and HBL). On the other hand, the physical mechanism responsible for the high-energy hump is still under debate. The most common interpretation of the high-energy emission is the inverse Compton (IC) scattering of electrons. The seed photons for the IC process can be either the synchrotron photons produced by the same population of electrons\citep{1985ApJ...298..114M} known as, synchrotron self-Compton (SSC), or from external sources such as the broad line region (BLR), the accretion disk, and the cosmic microwave background (external Compton, EC).

Mrk~421 (Markarian 421), an HBL at a redshift of $z = 0.031$, was first identified as a VHE gamma-ray emitter in 1992 by the Whipple Observatory \citep{1992Natur.358..477P}. It is one of the brightest known VHE gamma-ray blazars exhibiting highly variable emission; its flux increases by more than an order of magnitude over timescales ranging from years down to minutes. 
And it is also one of the most extensively observed sources in the X-ray band.
Mrk~421 often exhibits rapid variability in its X-ray emissions, with timescales ranging from minutes to days, indicating highly dynamic processes near the supermassive black hole at its core. This variability is often accompanied by spectral changes in the X-ray band, with a clear ‘harder-when-brighter’ trend commonly observed \citep{2020ApJS..247...27K, 2020ApJS..248...29A}.

From the perspective of VHE band observations, numerous detailed studies have been conducted using imaging atmospheric Cherenkov telescopes (IACTs). For example, during an extreme outburst event in April 2013, the MAGIC and VERITAS telescopes detected the most intense VHE flare ever observed from Mrk~421, with the $>$300 GeV ﬂux reaching $>$11 CU \citep{2020ApJS..248...29A}. 
However, atmospheric conditions and the need to accommodate other observational tasks often limit the time available for long-term studies of individual objects. Consequently, blazar observations are frequently biased towards follow-up of alerts, which hinders unbiased coverage of low flux states or flares without multi-wavelength correlations. 
Two long-term monitoring campaigns of Mrk~421 were conducted in 2011 and 2016 using the 
wide-field-of-view instrument ARGO-YBJ \citep{2016ApJS..222....6B}. Due to limited sensitivity, however, these studies only provided light curves integrated over week-to-month-long timescales. Similarly, in 2017, the HAWC Observatory reported daily light curves of Mrk~421 \citep{2017ApJ...841..100A}, but its sensitivity was also insufficient for detailed ﬂare-by-ﬂare analysis.

LHAASO is an extensive air shower (EAS) array composed of three sub-arrays featuring four types of detectors, designed to measure cosmic ray and gamma ray induced showers from sub-TeV to beyond 1 PeV. For gamma-ray induced showers, LHAASO offers a 2/3 sky field of view and can observe Mrk~421 for up to 7 hours daily, achieving a duty cycle of over 98\% and a year-averaged sensitivity of 1.5\% CU in the VHE band. 
These features make it highly suitable for studying transients and performing full sky surveys. Several notable alerts, such as flares from 1ES 1959 in 2023 and IC 310 in 2024 have been issued via Astronomer's Telegrams \citep{2024ATel16437....1X,2024ATel16513....1X}, prompting the multi-wavelength and multi-messenger follow-up studies \citep{2024ATel16449....1K, 2024ATel16456....1B, 2024ATel16462....1T}. This unbiased monitoring is essential for determining the true probability distribution of flux states, studying the correlation between TeV emission and other wavelengths, and investigating potential spectral changes with flux.

In this paper, we present a detailed study of the light curves and spectra of Mrk~421 based on a 3-year observation period from March 8, 2021 to March 8, 2024. The period included two different states: a long-term averaged period and outburst episodes. Using continuous LHAASO monitoring as a trigger, we identiﬁed 23 outburst events and examined their contemporaneous Fermi-LAT, the Swift X-ray Telescope (XRT), the Monitor of All-sky X-ray Image (MAXI) and the Zwicky Transient Facility (ZTF) light curves and spectra. We analyse each of the 23 outbursts for its observational features, spectral evolution, and physical properties, expanding the unbiased sample and probing acceleration mechanisms. In addition, we also discussed the correlation studies between different wavelength bands.

This paper is structured as follows. Section 2 briefly introduces LHAASO and details of the data analysis. Section 3 describes the other instruments participating in the 2021–2024 campaign and their corresponding data analysis. In Section 4, we present a detailed study of the light curves and spectra of Mrk~421 on both long-term and short-term timescales. Based on LHAASO's continuous monitoring, we present a detailed analysis of the 23 outbursts detected over three years, describing their observational and inferred properties for each event. We then outline the theoretical models used to interpret the observed spectra. Section 5 summarizes our conclusions and future prospects.

\section{LHAASO Observation and Data Analysis} \label{sec:intro}

LHAASO is a multi-purpose EAS array, located at latitude +29$^\circ$21'31'' N, longitude +100$^\circ$8'15'' E with the altitude of 4410\,m above sea level in Daocheng, Sichuan Province, China. It consists of three sub-arrays: the Water Cherenkov Detector Array (WCDA), the Kilometer Square Array (KM2A) and Wide Field-of-view air Cherenkov/Fluorescence Telescope Array (WFCTA) \citep{2019arXiv190502773C}. LHAASO is designed to investigate cosmic-ray and gamma-ray in a wide energy range from hundreds of GeV to EeV.
In this work, the data of Mrk~421 are from WCDA and KM2A observation (Sections 2.1 and 2.2).

\subsection{TeV observations from WCDA}

WCDA, a sub-array of LHAASO, is a survey instrument sensitive to gamma rays with energies above 200~\rm GeV. The data used in this study include all available observations from 8th March 2021 to 8th March 2024. In order to process the data in units that do not contain more than one full transit for the source, all reconstructed events are sorted into sidereal days, starting from when the source enters the field of view of WCDA. For this study, we restricted our analysis to events with zenith angles less than 50 degrees to ensure high-quality reconstruction. The selected events are then categorized into eight groups based on the effective number of triggered PMT units, denoted as $N_{\rm hit}$, with ranges in [30, 60), [60, 100), [100, 200), [200, 300), [300, 500), [500, 700), [700, 1000), and [1000, 2000)
Additionally, gamma/proton separation parameter thresholds are set to 1.12, 1.02, 0.90, 0.88, 0.84, 0.84, 0.84 and 0.84, which helps to achieve better selection of gamma-like events through the energy-dependent threshold optimization techniques. For instance, in the $N_{\rm hit}$ range [100, 200), approximately $1.1\times10^{-2}$ of cosmic-ray events are retained. The rejection factor improves with increasing energy, reaching $6.6\times10^{-4}$ for $N_{\rm hit}$ in [700, 1000).
Ultimately, the effective live time for observations of Mrk~421 was recorded as 1069 transits, with a total of $1.33\times 10^{5}$ gamma-like events detected. The details about WCDA detector performance and reconstruction details can be by \cite{2021ChPhC..45h5002A, 2024ApJS..271...25C}

To calculate the excess signal from the source, event and background maps for corresponding to $N_{\rm hit}$ are generated using WCDA data within a $10^{\circ}\times 10^{\circ}$ area with a grid of $0.1^{\circ}\times 0.1^{\circ}$, centered at the position of Mrk~421. The event maps are created as histograms of the arrival direction of the reconstructed events.
The background maps are computed using a direct integration method \citep{2004ApJ...603..355F}. For this analysis, a sliding time window of 10 hours was used to estimate the detector acceptance, with an integration time set to the same value to estimate the relative background. Events within the region of the Galactic plane ($|b| < 10^{\circ}$ ) and LHAASO 1st catalog gamma-ray sources \citep{2024ApJS..271...25C} (with a spatial angle less than 5 degrees) were excluded from the background estimation. The excess map is then obtained by subtracting the background map from the event data map. 

To understand the statistical properties of the VHE light curve obtained from WCDA, we analyzed the daily signal events. This light curve is obtained by using events with $N_{\rm hit} > 100$, corresponding to a median gamma-ray energy of 1.8~\rm TeV, assuming a Crab-like spectrum. The parameter $N_{\rm s}$ represents the number of daily signal events extracted from the excess map. To accommodate variations in observation time and detection efficiency, a weighting factor is introduced to adjust $N_{\rm s}$, ensuring the daily light curves accurately reflect the actual excess emission from the source. Specifically, the scaled $N_{\rm s} = N_{\rm bk, mean} \times ({N_{\rm on}-N_{\rm bk}})/{N_{\rm bk}}$, where $N_{\rm on}$ is the total number of events within the point spread function ($r_{68}$) region, $N_{\rm bk}$ is the background events in the same region for a single transit, and $N_{\rm bk, mean}$ is the three-year average.

The spectrum energy distribution (SED) is evaluated based on the forward-folding method \citep{2021ChPhC..45h5002A}, with different functions of the spectrum being tested.
Additionally, considering that Mrk~421 is an extragalactic source, we corrected the $\gamma\gamma$ absorption based on the extragalactic background light (EBL) model provided by \citet{2021MNRAS.507.5144S} to the observed spectrum. The detailed spectral analysis can be found in Sections 4.2.4 and 4.3.3. 

\subsection{TeV gamma-ray data from KM2A}

KM2A is another sub-array besides WCDA, with 5216 electromagnetic detectors (ED) and 1180 muon detectors (MD), detecting gamma rays in the energy range from 25~\rm TeV to 1~\rm PeV. In this study, we extend the energy down to 4~\rm TeV. Detailed information and performance of the detector have been given by \citet{2021ChPhC..45b5002A}. The full array of KM2A started operation in 2021, the effective live time for Mrk~421 was recorded as 1000 transits during the whole observation period. 
 
For the analysis presented in the paper, only events with a zenith angle less than 50 degrees are used. Most of the events recorded by KM2A are cosmic ray-induced showers, which constitute the main background for gamma-ray observations. Gamma-ray-like events are identified using the parameter R=$\log_{10} \frac{N_{\mu}+0.0001}{N_{\rm e}}$.
Here, $N_{\mu}$ represents the total number of muons detected by MDs, while $N_{\rm e}$ denotes the total number of particles detected by EDs. The data sets are divided into five groups according to the reconstructed energy (in units of TeV), i.e., [$10^{0.6}$,$10^{0.8}$), [$10^{0.8}$,$10^{1.0}$), [$10^{1.0}$,$10^{1.2}$),[$10^{1.2}$,$10^{1.4}$).
The corresponding cut values for the parameter R are -5.1, -5.1, -5.1 -5.24, respectively. 
For each data set, the sky map is divided into a grid of $0.1^{\circ} \times 0.1^{\circ}$ pixels, with each pixel populated by the number of detected events based on their reconstructed arrival directions, forming an event map. To obtain the excess of gamma-induced showers in each pixel, the direct integral method is also adopted to estimate the number of cosmic-ray background events in the bin. 
In this work, we integrated 24 hours of data to estimate the detector acceptance for different directions. The integral acceptance combined with the event rate is used to estimate the number of background events in each pixel (background map). Then the excess map, which is used to extract gamma-ray signals from specific sources, is generated by subtracting the background map from the event map. 

The light curve (ranging from 6 to 40~\rm TeV) analysis method aligns with the methodology outlined in Section 2.1. As to the spectral analysis, the flux or upper limits within the energy range 6 – 40 TeV were derived using the same index obtained from WCDA, and the same EBL model is applied for corrections of absorption.

\section{Multi-wavelength Data Analysis}

In this section, we analyze data from multiple instruments during the same observation period, which cover various energy bands. 
We use Fermi-LAT data for high-energy (HE, 100 MeV \textless E \textless 100 GeV) gamma-ray (Section 3.1), and X-ray data from Swift-XRT and MAXI-GSC (Sections 3.2, 3.3). Additionally, we include optical band data from ZTF. Here, we provide an overview of these data sets, briefly summarizing the operational range of each detector and the main data processing steps. For further details, please refer to the cited references.  

\subsection{HE observations from Fermi-LAT}

Fermi-LAT is designed to cover the energy band from $20~\rm MeV$ to greater than $100~\rm GeV$ \citep{2009ApJ...697.1071A}. The data surrounding Mrk~421 were obtained from Fermi Science Support, accounting for all sources within a 15-degree radius and standard cuts were applied to select the good time intervals (Z\_max \textless $90^\circ$, DATA\_QUAL \textgreater 0 and LAT\_config == 1). To investigate the gamma-ray flux variations of Mrk~421, we obtain the light curve variation in the $100~\rm MeV$ - $100~\rm GeV$ energy range from the Fermi-LAT Light Curve Repository (LCR) \footnote{\url{https://fermi.gsfc.nasa.gov/ssc/data/access/lat/LightCurveRepository/}}. The appropriate instrument response function for this data set is P8R3\_SOURCE\_V3. For the background model, we include the diffuse Galactic interstellar emission (IEM, gll\_iem\_v07.fits), isotropic diffuse emission (iso\_P8R3\_SOURCE\_V3\_v1.txt), and all sources listed in the 4FGL-DR2. 
The spectral is modeled with a power law function. 
Spectral parameters for sources within a 5-degree radius are allowed to vary freely, whereas those beyond this range are fixed to the values reported in the 4FGL-DR2 catalog \citep{2020ApJS..247...33A}. For energy bins where the likelihood fit yields a test statistic $\mathrm{TS} < 4$, 95\% confidence level flux upper limits are provided. The light curve is binned over 3 days.

\subsection{X-ray observation from MAXI-GSC}

MAXI onboard the International Space Station (ISS) has been monitoring the entire sky continuously since August 2009 \citep{2009PASJ...61..999M}. Light curves for specific sources are publicly accessible online\footnote{\url{http://maxi.riken.jp/}}. In this study, we analyzed data from the MAXI-GSC, which operates in the $2 - 20~\rm keV$ energy range and covers 97\% of the sky daily. The XSPEC tool is used to construct the SED. A circular region with a radius of 1.6 degrees centered on the source was selected as the sample region. An annular background region with inner and outer radii of 1.6 and 3.0 degrees, respectively,
was defined for background subtraction. 
The spectral fitting employs a power-law model combined with a Galactic absorption model, where the neutral hydrogen column density was fixed at $n_{\rm H} = 1.34 \times 10^{20} \mathrm{cm}^{-2}$. The advanced features of the software suite enabled precise response calibration and effective background subtraction, successfully extracting the spectrum of Mrk~421 in the $2 - 20~\rm keV$ energy range.

\begin{figure*}[!ht]
\centering
\includegraphics[width=0.97\textwidth]{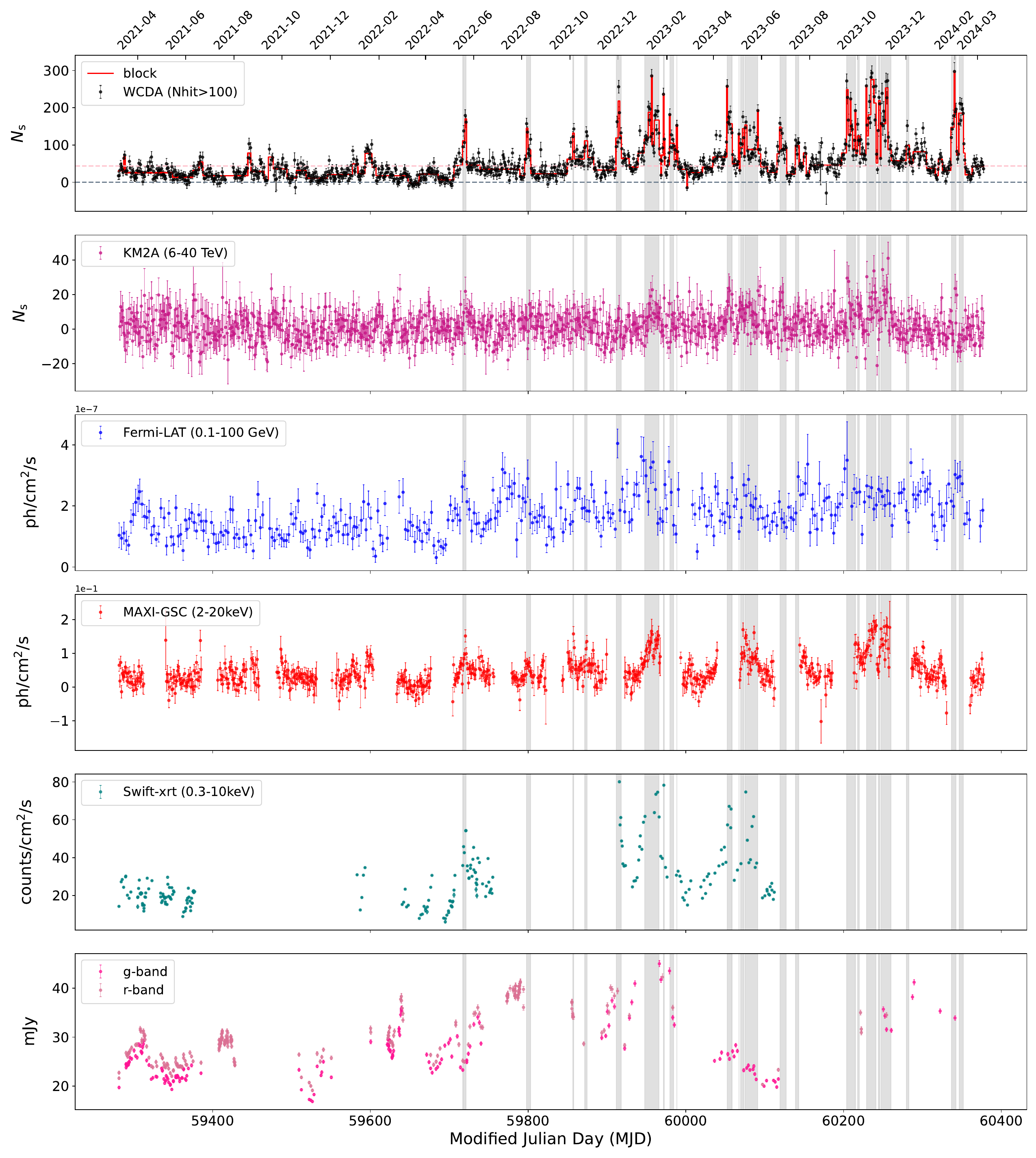}
\caption{Daily Light Curves from the Mrk~421 direction in different Energy Bands from 2021 March 8 to 2024 March 8. The panels from top to bottom refer to LHAASO-WCDA data ($N_{hit}> 100$), LHAASO-KM2A data (6 - 40 ~\rm TeV), Fermi-LAT data (0.1 - 100 ~\rm GeV), MAXI-GSC data (2 - 20 ~\rm keV), Swift-XRT data (0.3 - 10 ~\rm keV ) and optical data. In the top panel, the gray horizontal line represents zero flux level, and the pink line indicates the baseline signal (see Section~\ref{sec:dutycycle}). The red lines show the distinct $N_{s}$ states identified between change points through Bayesian blocks analysis with a 5\% false positive probability and the gray shaded areas indicate the marked outburst periods.}
\label{fig_mrk421_mul_lc}
\end{figure*}

\subsection{X-ray data from Swift-XRT}

XRT is a focusing X-ray telescope with an energy range from 0.3 to 10~\rm keV. The light curves for Mrk~421 from the Swift monitoring website\footnote{\url{https://www.swift.psu.edu/monitoring/}} were directly used in this work.
To construct the Spectral Energy Distribution (SED) for the source, all available observational data were retrieved from HEASARC. Photon events for spectral analysis were selected from a circular region with a 50-arcsecond radius centered on the source. Background data were extracted from a concentric annulus surrounding the source region, with inner and outer radii of 1.5 and 2.5 arcminutes, respectively. The extracted source spectrum was grouped to ensure a minimum of 50 counts per spectral bin for subsequent fitting. We performed X‑ray spectral fitting in XSPEC using an absorbed model. The absorption‑corrected, intrinsic flux in each of the three broad energy bands (0.5–2 keV, 2–10 keV, and 10–30 keV) was obtained by integrating the best‑fit model over the corresponding energy range. These fluxes were then used to construct the final broadband spectral energy distribution.

\subsection{Optical data from ZTF}

ZTF is a time-domain survey that had its first light at Palomar Observatory in 2017. Optical magnitudes in $g$ and $r$ bands were collected from the 20th ZTF public data release\footnote{\url{https://irsa.ipac.caltech.edu/cgi-bin/Gator/nph-scan}}. The optical data during the whole observation period were selected. After applying Galactic extinction correction using values from the NASA/IPAC Extragalactic Database (NED), the average magnitudes from each outburst were converted into data points on the SED.

\section{Results}

\begin{figure*}[!ht]
\centering
\subfigure{\includegraphics[width=0.45\textwidth]{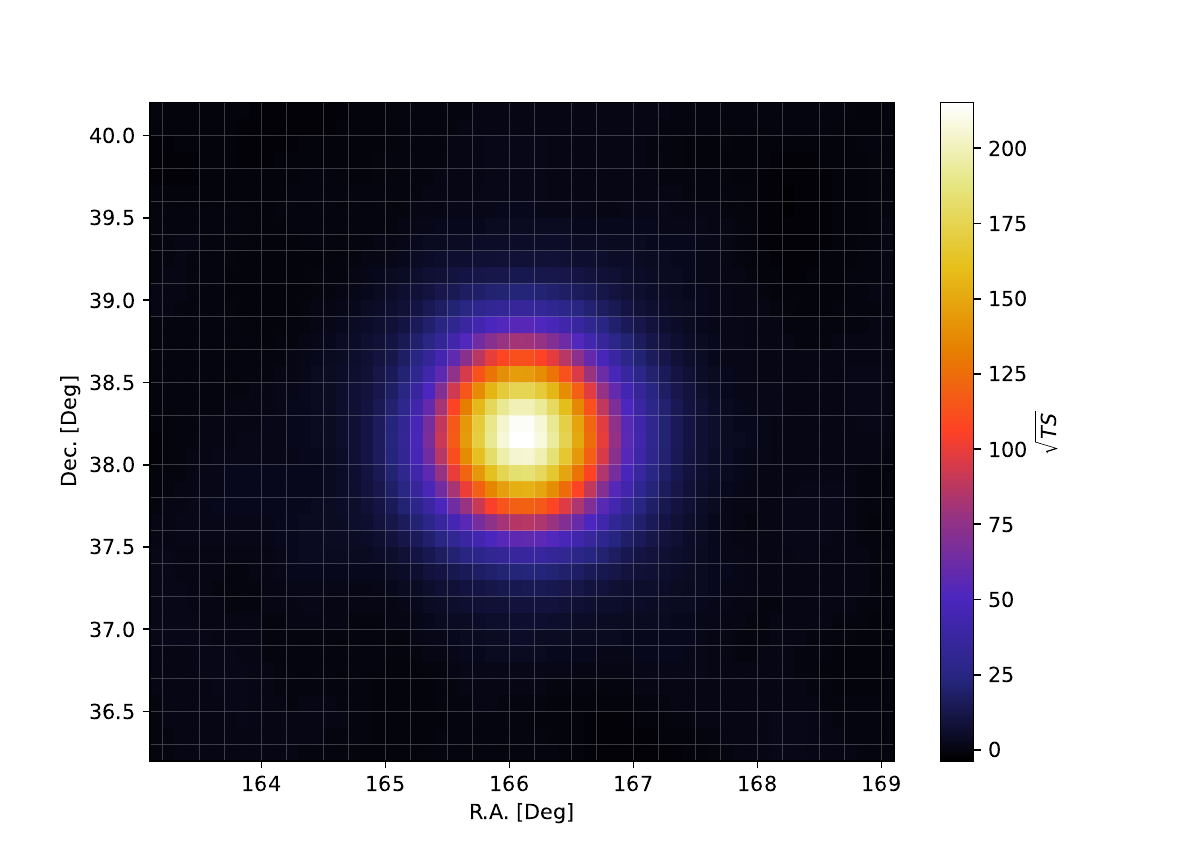}}
\subfigure{\includegraphics[width=0.45\textwidth]{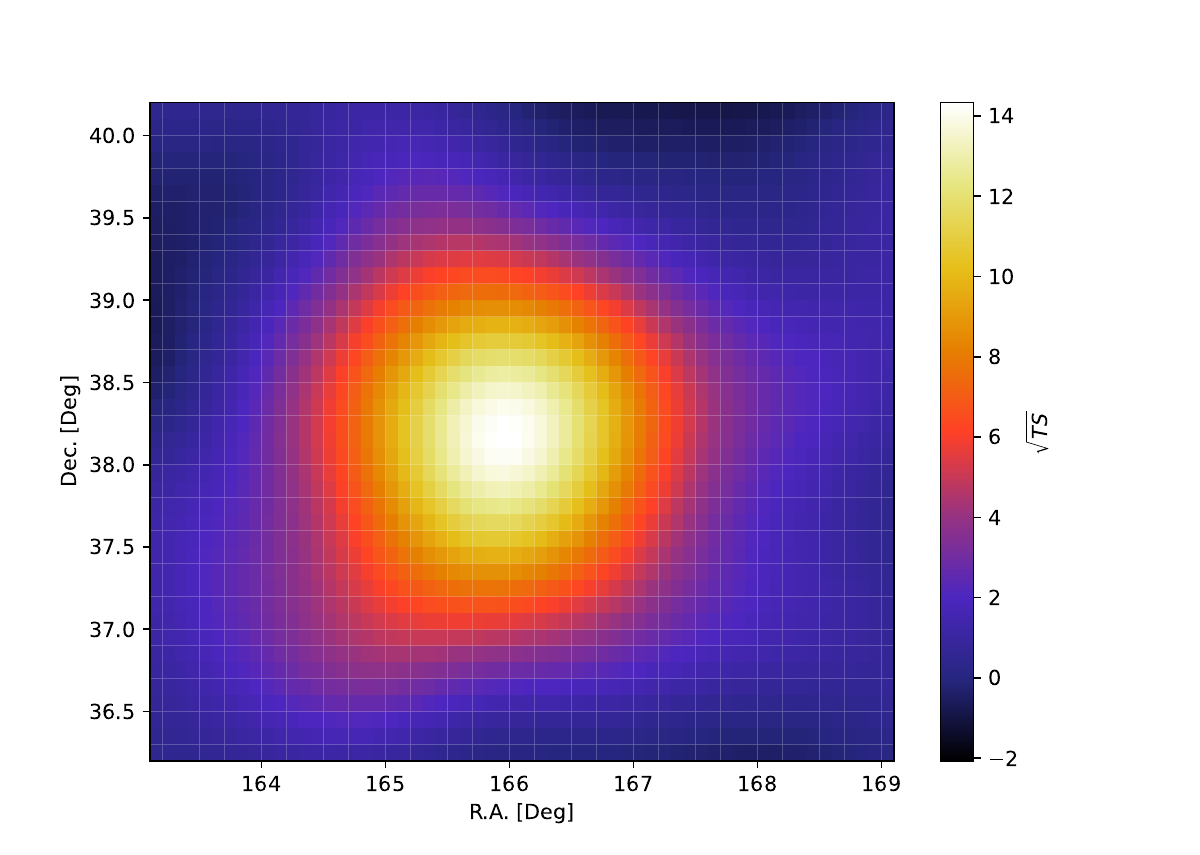}}
\caption{The left panel shows the significance sky map of Mrk~421 observed by WCDA from March 2021 to March 2024 with a maximum significance of approximately $215\ \sigma$. The right panel displays the significance sky map obtained by KM2A during the same period, with a peak significance of about $14\ \sigma$.}
\label{label:sig}
\end{figure*}

The overall detection significance of Mrk~421 by LHAASO-WCDA and LHAASO-KM2A reaches $215\ \sigma$ and $14\ \sigma$, respectively, as a point-like source (see Figure~\ref{label:sig}). The high statistics of WCDA data enable a daily-bin light curve to be constructed. In Section 4.1, we present multi-wavelength light curves spanning from optical to TeV energy bands, along with their temporal correlations. Subsequently, Section 4.2 focuses on investigating the long-term evolution of the source through cross-band correlation analysis, statistical property quantification, variability timescale characterization, and annual-timescale averaged spectral evolution features, aiming to reveal the interconnected radiation mechanisms across energy bands. Furthermore, Section 4.3 employs the Bayesian Blocks algorithm on WCDA light curves to systematically identify outburst episodes, followed by a multi-parameter investigation of each outburst event, including variability pattern decomposition and spectral modeling.

\subsection{Light Curves}


The light curves of Mrk~421, spanning the energy range from optical to TeV bands, are shown in Figure \ref{fig_mrk421_mul_lc}, with data of LHAASO-WCDA, LHAASO-KM2A, Fermi-LAT, Swift-XRT, MAXI-GSC and ZTF from top to bottom. For LHAASO-WCDA, LHAASO-KM2A, Swift-XRT and MAXI-GSC, each data point represents a daily average counting rate, while Fermi-LAT data are averaged over three-day intervals. 
During the three years of continuous monitoring covered in this study, Mrk~421 exhibited significant variability across all wavebands from optical to TeV energies, characterized by low-activity/quiescent states and high-activity/frequent outbursts. 

During 2021, LHAASO-WCDA light curve clearly indicates that Mrk~421 experienced a quiescent period, with an overall detection significance of approximately $47\ \sigma$ and a daily average significance of about $3\ \sigma$. Only occasional weak and short-lived outbursts were observed, most of which had fluxes that did not exceed twice the baseline level (see Section~\ref{sec:dutycycle} for the definition of the baseline). 
 Starting in 2022, the activity level of Mrk 421 increased significantly, entering a phase characterized by frequent outbursts with notably stronger intensities and durations ranging from days to several weeks. The maximum daily significance, higher than $20\ \sigma$, was observed on February 1, 2024, with a daily signal counts around 300, representing a factor of 7 increase compared to the quiescent. At higher energies, the LHAASO-KM2A light curve shows no significant signals from Mrk~421 due to detector sensitivity limitations in the VHE band, with only an obvious signal detected in October 2023 with a significance around 14 $\sigma$.

In the Fermi-LAT light curve, Mrk~421 exhibits a generally high frequency of activity, with enhanced fluctuations and short-term outbursts observed since 2022. Compared to the WCDA band, the activity variations in the Fermi-LAT energy range are relatively stable, though some high-intensity peaks are still apparent during specific periods. 

In the X-ray (MAXI-GSC) light curve, Mrk~421 displays frequent and relatively moderate variability, aligning well with the outbursts observed in the WCDA band. In contrast, signals in the Swift-XRT and optical bands are relatively sparse, with noticeable brightness increases mainly observed during a few outburst periods but lacking any consistent long-term trend changes. 
Between MJD 59910 and 59990, the optical light curve shows general agreement with the sparser Swift-XRT data and appears correlated with the VHE emission from LHAASO-WCDA.

\subsection{Long-term behavior}

\subsubsection{Flux distribution analysis}

The log-normal flux distributions observed in blazars may offer critical insights into outburst mechanisms and imply a potential transfer of variability signatures from accretion disks to relativistic jets, as discussed in \citep{2008bves.confE..14M} and \citep{2023ApJS..268...20K}. For instance, \citet{Sinha2016} systematically examined multi-wavelength data (radio, optical–UV, and gamma-ray) of Mrk~421 from 2009 to 2015, demonstrating that log-normal distributions provided a better fit to flux distribution across most energy bands. Such a distribution may suggest the flux variability process to be multiplicative instead of additive\citep{Uttley2005}. 

\begin{figure*}[ht]
\centering
\begin{subfigure}{}
\includegraphics[width=0.3\textwidth,height=4cm]{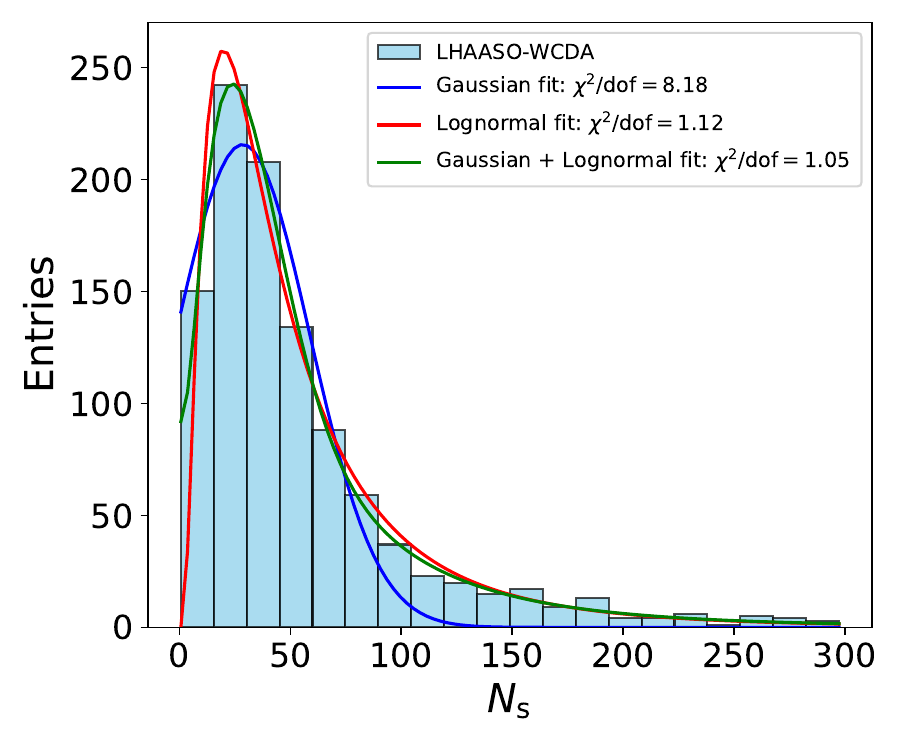}
\end{subfigure}
\begin{subfigure}{}
\includegraphics[width=0.3\textwidth,height=4cm]{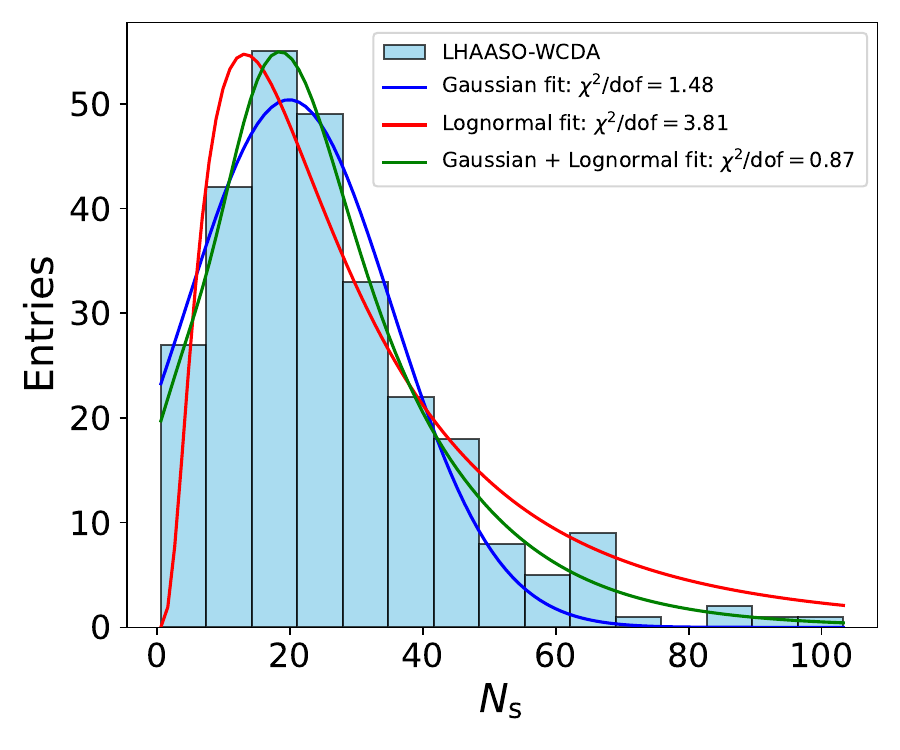}
\end{subfigure}
\caption{Histograms of the daily signal events $N_{s}$ from Mrk~421. The Left and right panels present the 3 years data and the 2021 data, respectively; in each plot, the red, blue and green corresponds to the log-normal and Gaussian and combination of Gaussian and log-normal fit.}
\label{fig: cumulative_Ns}
\end{figure*}

The left panel of Figure~\ref{fig: cumulative_Ns} shows the distribution of the daily excess event number of WCDA (roughly proportional to daily flux) from Mrk~421 over three years.
The distribution shows a long tail at high-flux end. The distribution may be described with a log-normal function, and the best-fit parameters are $\mu_{\log N}=3.23$, $\sigma_{\log N}=0.81$ with $\chi^2/dof=1.12$. We also tried to fit the distribution with the sum of a Gaussian function and a log-normal function, resulting in the best-fit parameters as $\mu_{\rm gauss}=11.76$, $\sigma_{\rm gauss}=10.35$, $\mu_{\log N}=3.44$, $\sigma_{\log N}=0.49$ with $\chi^2/dof=1.05$. The Gaussian part dominates the distribution at the low-flux end whereas the log-normal part dominates the high-flux end (i.e., the long tail). This behavior may be interpreted as the combination of a quasi-steady state (described by the Gaussian function) and a series of stochastic outburst states (described with the log-normal function), likely driven by an underlying multiplicative process. The latter scenario, if it is true, would imply two modes of energy dissipation in the blazar jet. 
Given the similar $\chi^2/dof$ values of the two fits, we cannot distinguish between these scenarios. To further test the hypothesis, we examined the flux distribution for 2021 separately, when the blazar was predominantly in a quiescent state.
As shown in the right panel of Fig.~\ref{fig: cumulative_Ns}, while a pure log-normal function cannot describe the flux distribution in the quiescent state, a pure Gaussian function results in reasonable fit. The combined fit with a Gaussian function and a log-normal function gives the best fit. It suggests that the Gaussian component plays a more important role during the quiescent state, supporting the two-mode dissipation scenario.

\begin{figure*}[!ht]
\centering
\includegraphics[width=0.99\textwidth]{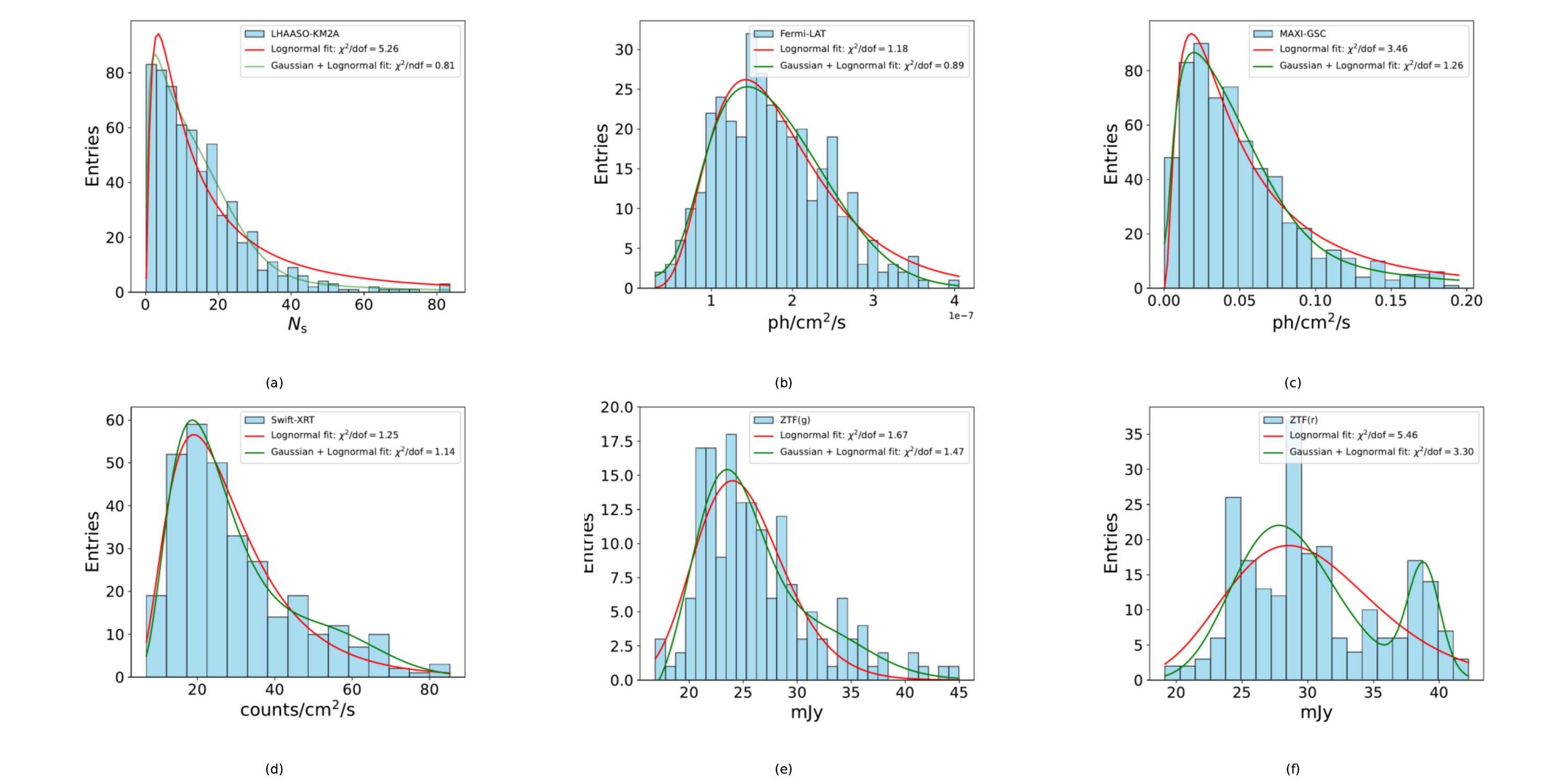}
\caption{Histograms of the MWL fluxes are shown, with (a) to (f) representing the distribution and model fits for LHAASO-KM2A, Fermi-LAT, MAXI-GSC, Swift-XRT, ZTF (g-band), and ZTF (r-band), respectively. In each plot, the red line indicates the Log-normal fit, the blue line indicates the Gaussian fit and the green line represents the combined Gaussian and Log-normal fit.} 
\label{flux_lognorm}
\end{figure*}

Following the approach for LHAASO-WCDA data, we applied log-normal and Gaussian + log-normal models to the multi-wavelength counts/flux distributions shown in Figure \ref{flux_lognorm}. For the KM2A data, the observed variability is suppressed due to the detection efficiency, (as shown in Figure \ref{fig_mrk421_mul_lc}), resulting in a signal distribution that is more readily described by a Gaussian function. For the Fermi-LAT data, the chi-square values for the two models are 1.18, 0.89, both are close to 1, indicating that two models can adequately describe the data distribution. This result may suggest that the flux variations in the energy range of the Fermi-LAT data are relatively complex, and a single model may struggle to fully capture its distribution characteristics. For the MAXI-GSC, Swift-XRT, and ZTF (g-band) data, log-normal distribution can better describe their distribution characteristics. However, for the r-band data, due to insufficient data or low statistical significance, two models could not yield a satisfactory result.

\subsubsection{Fractional variability analysis}

\begin{figure*}[ht!]
\centering
\includegraphics[width=0.9\textwidth,height=6cm]{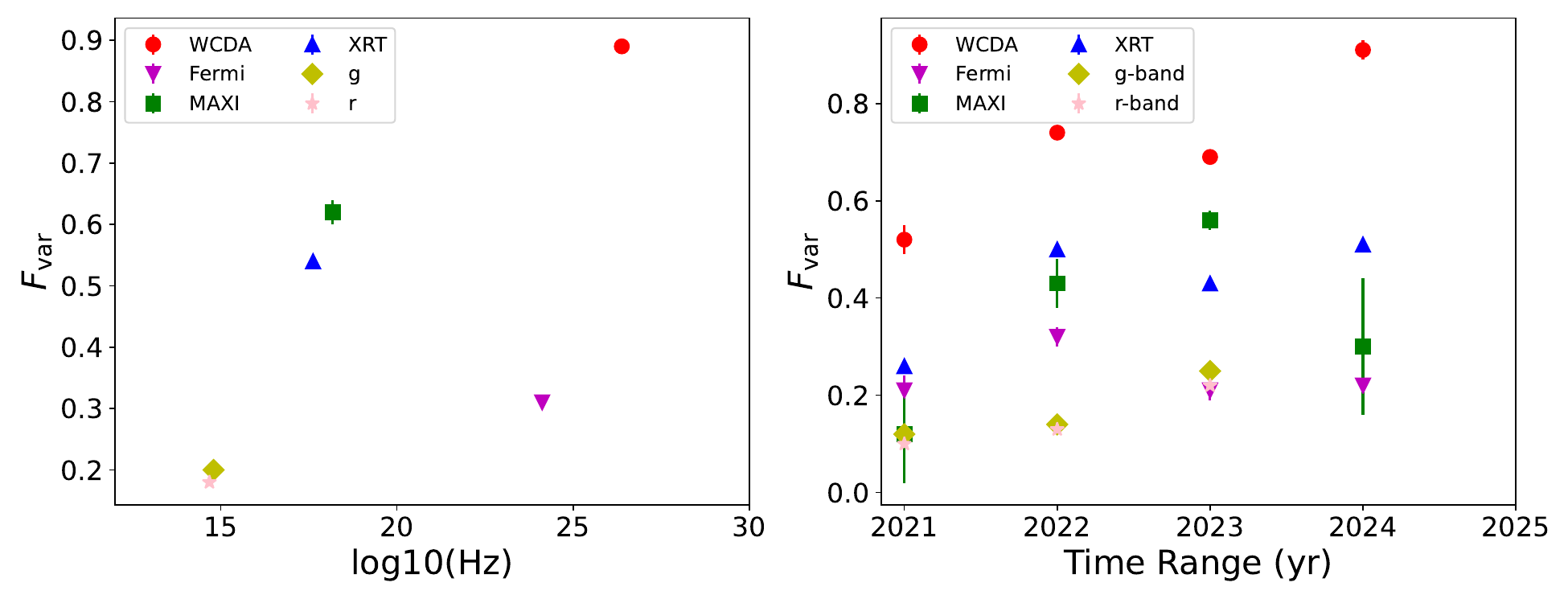}
\caption{Left panel: Multi-instrument fractional variability $F_\mathrm{{var}}$ as a function of the energy over the entire observation period. Right panel: The variation of $F_\mathrm{{var}}$ in LHAASO-WCDA over , with each vertical dashed line representing a time range.}
1\label{Fvar}
\end{figure*}

As demonstrated by the long-term multi-wavelength light curves in Figure \ref{fig_mrk421_mul_lc}, Mrk~421 exhibited low- and high-activity phases across wide wavebands during the three-year observational period analyzed in this work. To systematically quantify the variability amplitudes across energy bands, we calculated the fractional variability parameter ($F_\mathrm{var}$) \citep{1996ApJ...470..364E}, 
with results presented in the left panel of Figure \ref{Fvar}. The definition of $F_\mathrm{var}$ is:
\begin{equation}
  F_\mathrm{var} = \sqrt{\left(\frac{S^{2}-<\sigma_\mathrm{err}^{2}>}{<x>^{2}}\right)}
\end{equation}
$<x>$ is the average signal, S is the standard deviation of the measurements, and $<\sigma_{err}^{2}>$ is the mean squared error, all determined for a given instrument and energy band.

In left panel of Figure \ref{Fvar}, the $F_\mathrm{{var}}$ value increases from optical to X-ray, then declines in HE band, and reaches the maximum value in the VHE band. This trend is consistent with the behaviour shown in the light curves Figure \ref{fig_mrk421_mul_lc}, the variability plot shows the typical double bump structure, whereas the most variation observed in the TeV and keV energy bands. 
In detail, the variability amplitude in the hard X-ray bands rises from 54\% in the 0.3 - 10~\rm keV range to 62\% in the 2 - 20 keV range.  In the GeV-TeV gamma-ray band, it increases from 31\%, to 89\% in the WCDA TeV energies. Due to large fluctuations and  low signal-to-noise ratio in the LHAASO-KM2A data, a reliable $F_\mathrm{{var}}$ value could not be determined. 

Complementing the three-year $F_\mathrm{{var}}$ analysis, we also investigate the distribution of $F_\mathrm{var}$ on annual timescales, the result is shown in the right panel of Figure \ref{Fvar}. Overall, the annually binned $F_\mathrm{{var}}$ values are systematically lower than the three-year values in their corresponding energy bands, as expected given the reduced variability amplitude on shorter timescales.
Focusing on the TeV band from 2021 to 2024, the annual $F_\mathrm{var}$ reveals a clear transition from quiescence to high activity:
it rises sharply from ~52\% in 2021 to 74\% in 2022, remains elevated at 69\% in 2023, and peaks at 91\% in 2024,the highest variability recorded in the observing period. This sustained increase in annual $F_\mathrm{var}$ systematically traces the rising activity level of the source.

It is worth noting that observations of Fermi-LAT have shown an increasing trend in the variability since 2022. Moreover, the X-ray and optical bands also display an increase in $F_{\rm var}$ after 2022, indicating that Mrk~421 entered a more active phase across multi-bands at that time. It suggests that emissions in these energy bands may share the same origin.

During the 2021 quiescent VHE period, the variability amplitude in the X-ray band dropped significantly from 0.61 to 0.10, approaching the optical value.
The fractional variability plot in this year shows only a single bump structure in the VHE band. 
During the 2021 quiescent VHE period, the variability amplitude in the X-ray band dropped significantly, from 0.61 to 0.10, approaching the optical value. $F_{\rm var}$
in that year shows only a single bump in the VHE band. In contrast, during the 2022–2024 active period, most of the variation is concentrated in the X-ray and VHE bands. The distinct fractional variability-versus-energy patterns in these two periods imply that different physical mechanisms drive the broadband variability in quiescent and active states.

\subsubsection{Correlation analysis}
 
The light curves show several prominent X-ray, GeV and VHE gamma-ray outbursts observed by Swift-XRT, MAXI-GSC and Fermi-LAT. To quantify the correlation more rigorously, we used the discrete correlation function (DCF)\citep{1988ApJ...333..646E} between the LHAASO-WCDA light curve and those from other instruments. The DCF provides an assumption-free representation of correlation without the need for time interpolation, making it particularly suitable for uneven light curves caused by differences in sampling and varying uncertainties.

\begin{figure}[ht]
    \centering
    \begin{minipage}[b]{0.31\linewidth}
        \centering
        \includegraphics[width=\linewidth]{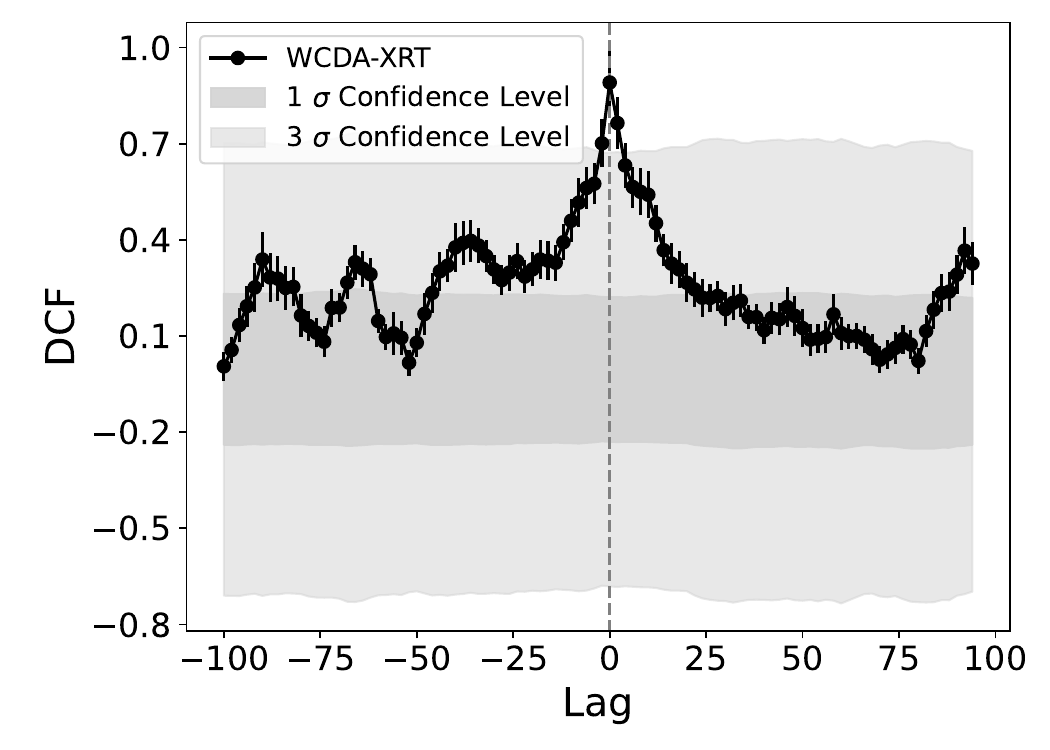} 
        \textbf{(a)} WCDA-XRT
    \end{minipage}
    \begin{minipage}[b]{0.31\linewidth}
        \centering
        \includegraphics[width=\linewidth]{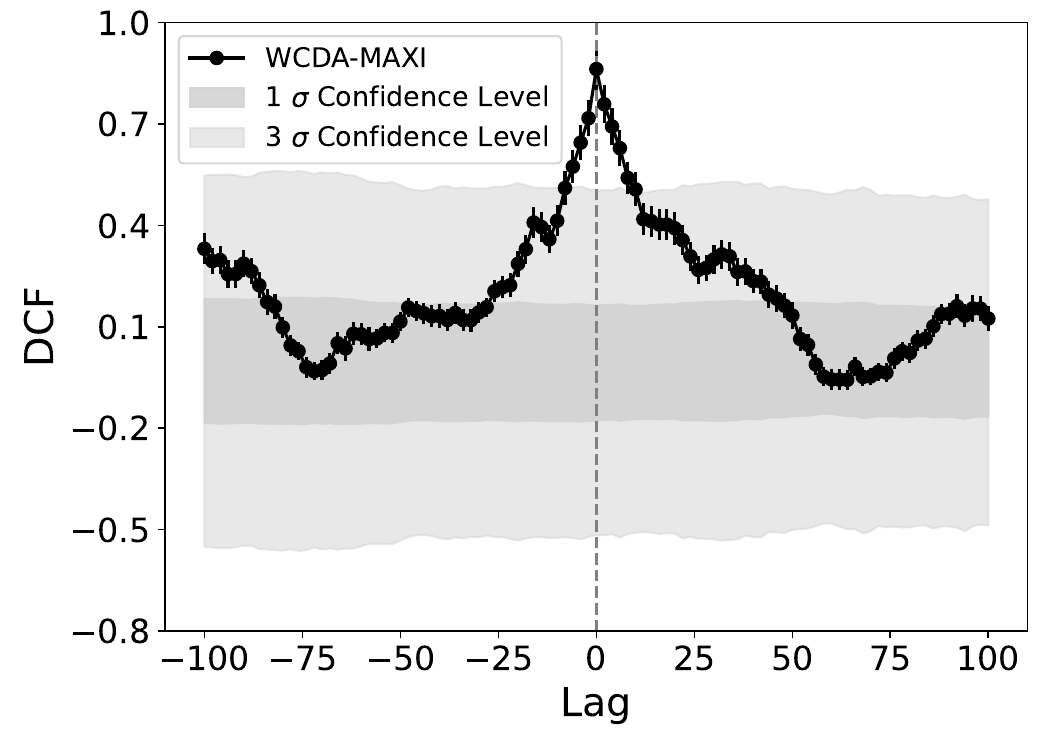} 
        \textbf{(b)} WCDA-MAXI
    \end{minipage}      
    \begin{minipage}[b]{0.31\linewidth}
        \centering
        \includegraphics[width=\linewidth]{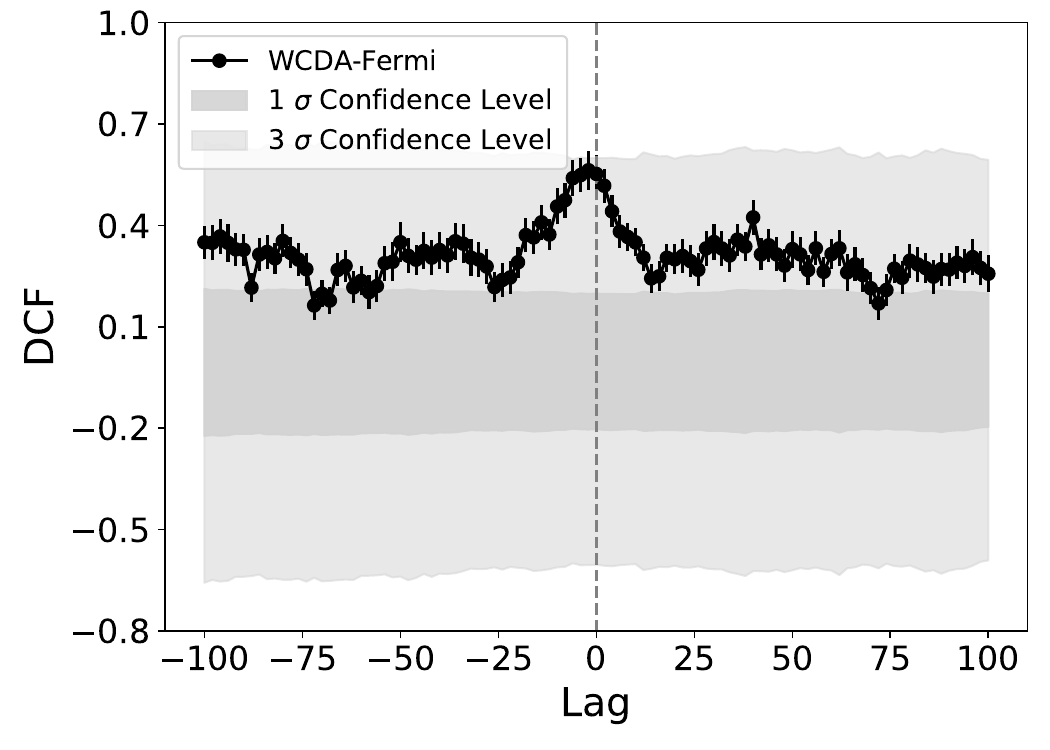} 
        \textbf{(c)} WCDA-Fermi
    \end{minipage}
      
    \caption{The DCF of the LHAASO-WCDA light curve, correlated with the Swift-XRT, MAXI-GSC and Fermi-LAT light curve. The black error bars represent the uncertainties as derived from \citep{1988ApJ...333..646E}. The gray area represents the $1\sigma$ and $3\sigma$ extremes of the DCF distribution of simulated WCDA light curves when correlated with the measured light curves from other energy bands.}
    \label{fig: DCF}
\end{figure}

\begin{table}[ht]
  \centering
    \caption{Discrete correlation analysis between different energy bands}
   \begin{tabular}{lccccccc}
    \hline
    \hline
    Energy Band & Time Lag(d) & DCF  \\
    \hline
    WCDA vs. XRT & 0 & $0.89\pm0.10$ \\
    WCDA vs. GSC & 0 & $0.86\pm0.06$ \\
    WCDA vs. LAT & $-2\pm 1.9$ &  $0.56\pm0.06$ \\
    \hline
    \end{tabular}%
    \label{tab:DCF_coeficient}
\end{table}

To evaluate the statistical significance of the measured DCF, we simulated 1,000 uncorrelated light curves in the LHAASO‑WCDA energy band using the DELCgen package \citep{2013MNRAS.433..907E}. The simulations were based on the power spectral density (PSD) and probability density function (PDF) of the original observed light curve. The PSD was modeled using the PSRESP method \citet{2002MNRAS.332..231U} to reproduce the intrinsic variability of the source, while the PDF was described by a log‑normal plus Gaussian mixture distribution. Through iterative optimization, the amplitude distribution of the simulated light curves was adjusted to ensure that their statistical properties matched those of the original data.
For each simulated flux point, the measurement error was randomly drawn from the actual error distribution of the observations. The same binning, time‑shifting, and correlation calculation procedures as applied to the real light curves were performed on each simulated curve, yielding a corresponding DCF distribution. Based on this distribution, we determined the confidence levels at which random correlations can be excluded and derived significance contours. This approach fully accounts for the specific sampling patterns and flux‑measurement uncertainties of the two light curves, thereby providing a robust assessment of the statistical significance of the observed correlations.

The cross-correlation between LHAASO-WCDA and other energy bands was  analyzed and the results of the flux variation correlation are presented in Figure \ref{fig: DCF} and Table \ref{tab:DCF_coeficient}. Over the time lag range of $-100$ to $+100$ days, significant positive correlations, $> 3 \sigma$, were found between LHAASO-WCDA and X-ray data, indicating strong synchronization. Notably, the correlation between WCDA and the MAXI band showed a strong positive correlation with a coefficient of $0.86\pm0.06$ at zero time lag, while the correlation with Swift-XRT was even more pronounced, with a coefficient of $0.89\pm0.10$ at zero lag, both exceeding the $3\ \sigma$ confidence level - consistent with results from past decades of observations. These results suggest that the physical processes in these energy bands may share a common origin or mechanism, and the high correlation likely reflects simultaneous outbursts or intense activity across these bands. However, the KM2A light curve suffers from high noise levels, leading to unstable correlation of weak statistical significance. Consequently, no clear conclusion can be drawn regarding the correlation between KM2A and WCDA, and this part of the analysis should be interpreted with caution. 

The correlation between WCDA and Fermi-LAT was weaker. Although a correlation coefficient of 0.56 was found at a 2-day time lag, the DCF peak values were generally low and mostly within the $3\ \sigma$ confidence interval, suggesting that the correlation between these two energy bands is not statistically significant. This may imply that the activities detected by Fermi-LAT and WCDA are less synchronized in time, or that the part of the emission observed by the two instruments may have different origins, leading to nonidentical activity patterns. To further assess the statistical significance of temporal lags, we generated 1,000 simulated TeV light curves using Poisson statistics based on daily background event counts from WCDA observations. For each realization, we calculated the DCF between the simulated TeV light curve and the Fermi-LAT GeV light curve, recording the time lag corresponding to the maximum DCF value. The standard deviation ($\sigma = 1.9$ days) of the derived lag distribution was used to quantify uncertainties. Statistical analysis reveals that the observed 2-day lag remains consistent with statistical noise within uncertainties, failing to reject the null hypothesis of zero-lag correlation.  

Note that, the optical band (g/r) data is not available for the correlation analysis due to the limitation of statistics.

\subsubsection{Time average spectrum}

In this section, we present a detailed analysis of the long-term SED of Mrk~421, with a focus on its characteristics across different years. In this work, the observed differential energy spectrum spanning from 400~\rm GeV to 20~\rm TeV is well described by a power-law function with a high-energy exponential cutoff function (ECPL), defined as
$dN/dE = N_{0}\times(E/E_{0})^{-\alpha}\times \exp(-E/E_{\rm cut})$, where the reference energy $E_{0}$ is set at 3 ~\rm TeV.  Additionally, for energy bins with the significance below 2~$\sigma$, 95\% confidence level flux upper limits are calculated using the Feldman–Cousin method. For KM2A upper limit points, the spectral index is fixed as the value measured by WCDA.

\begin{figure*}[!ht]
    \subfigure[]{\includegraphics[width=0.46\textwidth]{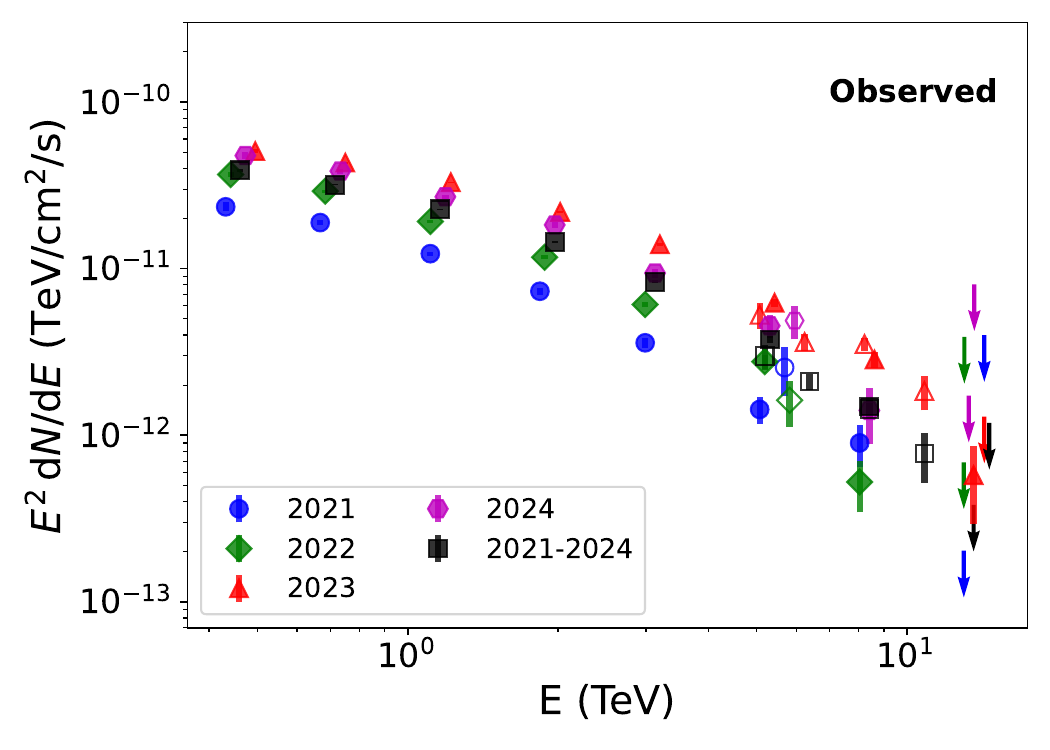}}
    \subfigure[]{\includegraphics[width=0.46\textwidth]{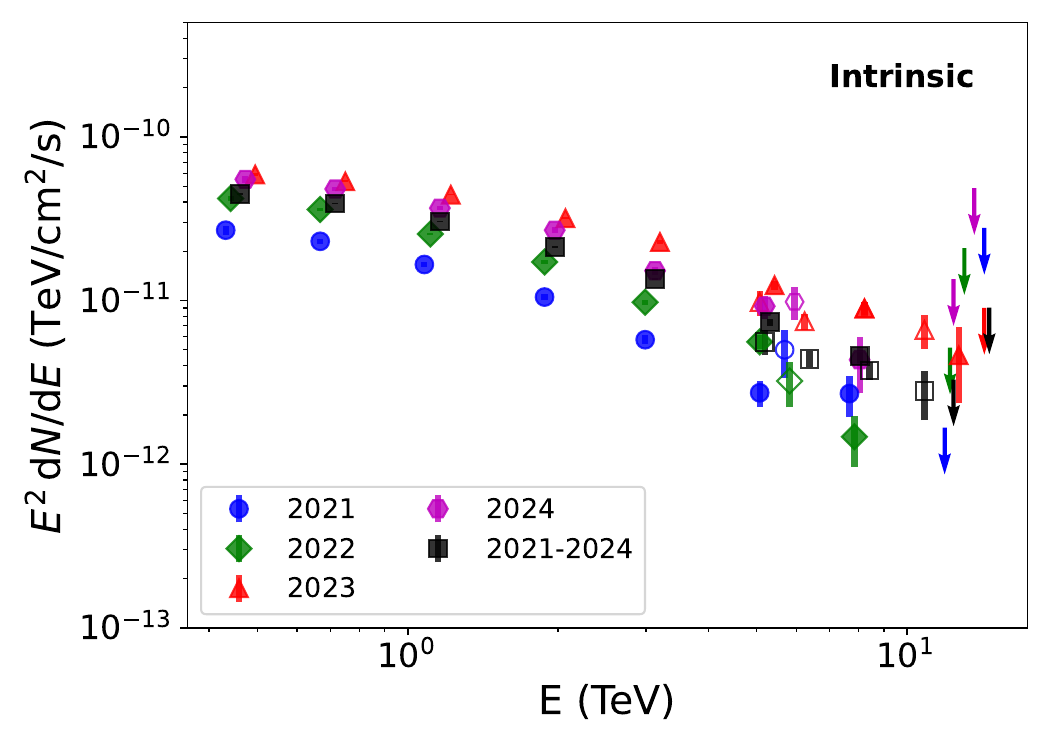}} 
    \caption {The energy spectrum measurement results of LHAASO for Mrk~421: The left figure is the observed energy spectrum, the right figure is the intrinsic energy spectrum. Among them, the solid points represent the observation results of WCDA, the hollow points represent the observation results of KM2A, and the upper limit is the energy limit given by KM2A. Different colors represent different time intervals, respectively.} 
    \label{average_sed}
\end{figure*}

The time-averaged spectral SED over the entire observational time and the average SED of each year are shown in the left panel of Figure~\ref{average_sed}, only statistical uncertainties are displayed. Regarding the systematic uncertainty of the flux, we followed the method described in paper\citep{2023Sci...380.1390L}, which yields an energy-dependent uncertainty ranging from approximately 8\% to 24\%.

Given its distance of more than 100\,Mpc, the observed spectrum of Mrk~421 at multi-TeV energies
is affected by the absorption from the EBL. The intrinsic spectrum is derived from the observed spectrum by $(dN/dE)_{\rm int} = (dN/dE)_{\rm obs} \times \exp(-\tau(E))$, 
where $\tau(E)$ is the energy-dependent optical depth calculated using the EBL model provided by \citet{2021MNRAS.507.5144S}. 
After EBL correction, the spectrum becomes harder:  the power-law index decreases by about 0.12 and the normalization increases by 30\%, as shown in the right panel of  Figure~\ref{average_sed}. The best-fit parameters of both observed spectrum and intrinsic spectrum are listed in Table~\ref{tab_average_spectrum_params}. 

\begin{figure*}[!ht]
\centering
\includegraphics[width=0.9\textwidth]{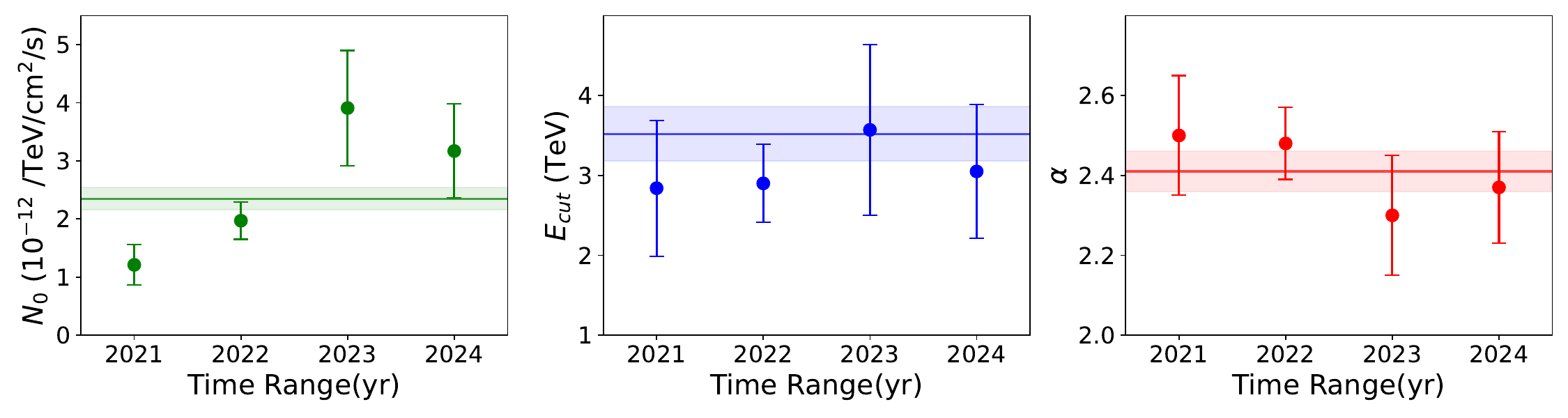}
\caption{Temporal evolution of ECPL model parameters, showing from left to right the normalized flux $N_{0}$, cutoff energy $E_{cut}$, and spectral index $\alpha$. Each solid line represents the best-fit results during the 2021-2024 observation period, with the shaded region indicating the $1\ \sigma$ confidence interval.}
\label{fig:params_years}
\end{figure*}

Figure~\ref{fig:params_years} shows the temporal evolution of ECPL model parameters: the normalization $N_{0}$, the cutoff energy $E_{\rm cut}$ and the spectral index $\alpha$. $N_{0}$ exhibits a 
clear upward trend starting in 2021, peaks in 2023, and then declines slightly in 2024, 
indicating a significant brightening phase during the observed period. 
Similarly, the cutoff energy $E_{\rm cut}$, follows a comparable pattern with relative large uncertainties: it gradually increases from 2021 to a maximum in 2023 before decreasing gently in the following year. 
As expected, the spectral index $\alpha$ exhibits an opposite tendency, becoming smaller in 2023. 
A comparison of the annual spectra reveals a ‘harder-when-brighter’ tendency:
that is, a higher normalization $N_0$ generally coincides with a harder spectrum (smaller the spectral index $\alpha$) and a larger cutoff energy $E_{\rm cut}$. 
In summary, the annual integrated energy fluxes in the 0.4–20 TeV range from 2021 to 2024 correspond to 0.27, 0.44, 0.78, and 0.63 CU, respectively. The year 2021 represents the baseline quiescent state, whereas 2022–2024 exhibited significant flux enhancements, reaching 1.60, 2.85, and 2.33 times the 2021 baseline level  ($2.82\times 10^{-11} \rm TeV/cm^{2}/s$), respectively.

\begin{table}[!ht]
\centering
\caption{Optimization of the fitting parameters for LHAASO SEDs across different year-time ranges.}
\begin{tabular}{lcccc}
\hline
\hline
Observed & Epoch  & $N_{\rm 0}$ & $\alpha$ & $E_{\rm cut}$ \\ 
& & $(10^{-12} \rm /TeV/cm^{2}/s)$ & & (TeV) \\
\hline
&2021  & $1.21\pm0.35$ & $2.50\pm0.15$ & $2.84\pm0.85$\\
&2022  & $1.97\pm0.32$ & $2.48\pm0.09$ & $2.90\pm0.49$\\
&2023  & $3.91\pm0.99$ & $2.30\pm0.15$ & $3.57\pm1.07$\\
&2024  & $3.17\pm0.81$ & $2.37\pm0.14$ & $3.05\pm0.84$\\
&2021-2024 & $2.35\pm0.19$ & $2.41\pm0.05$ & $3.52\pm0.34$\\
\hline
\hline
Intinsic & Epoch  & $N_{\rm 0}$ & $\alpha$ & $E_{\rm cut}$ \\
& & $(10^{-12} \rm /TeV/cm^{2}/s)$ & & (TeV) \\
\hline
&2021  & $1.76\pm0.50$ & $2.37\pm0.15$ & $3.16\pm1.05$\\
&2022  & $2.79\pm0.46$ & $2.36\pm0.09$ & $3.32\pm0.66$\\
&2023  & $5.48\pm0.79$ & $2.18\pm0.08$ & $4.33\pm0.92$\\
&2024  & $4.42\pm1.13$ & $2.26\pm0.14$ & $3.60\pm1.18$\\
&2021-2024  & $3.34\pm0.27$ & $2.29\pm0.04$ & $4.17\pm0.49$\\
\hline 
\hline
\end{tabular}
\label{tab_average_spectrum_params}
\end{table}

\begin{figure*}[!ht]
    \centering
    \includegraphics[width=0.6\textwidth]{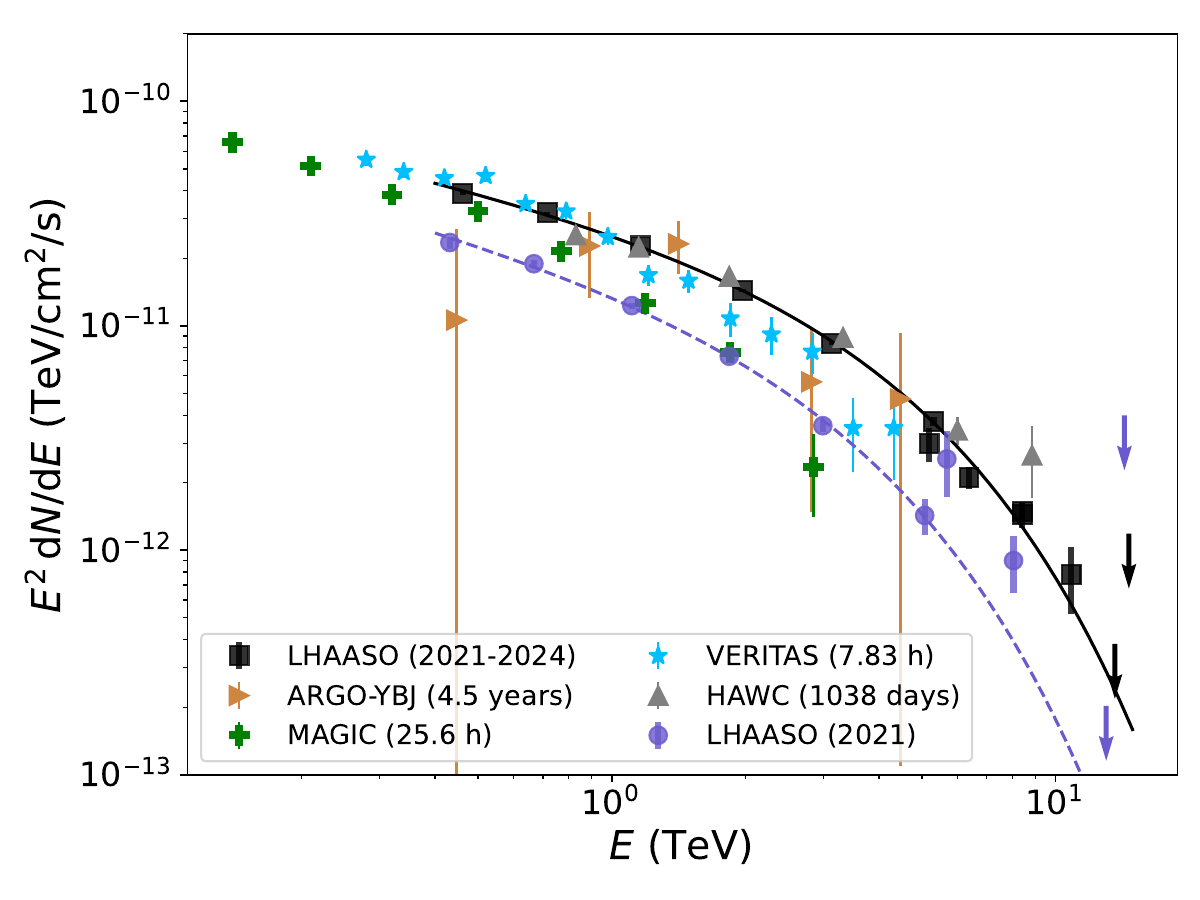}
    \caption {VHE spectrum of Mrk~421. Black circles correspond to data from LHAASO (2021 - 2024), the green crosses are the MAGIC data for the 25.6 hours observation campaign in 2004 November and 2005 April \citep{2013PhDT.......475B}. The light brown triangles are the ARGO-YBJ for a 4.5-year period between 2008 and 2012 \citep{2016ApJS..222....6B}. The grey diamonds are HAWC results form June 2015 and July 2018 \citep{2022ApJ...929..125A}. The blue stars correspond to observations performed by VERITAS telescopes (7.83 h) respectively \citep{2011ApJ...738...25A}} 
    \label{long_term_sed_check}
\end{figure*}

Several long-term and low-state SED from MAGIC, VERITAS, ARGO-YBJ and HAWC are compared with our 2021-2024 and baseline result (2021) in the Figure~\ref{long_term_sed_check}.
Notably, the long-term results from the large field-of-view detectors are in good agreement, even though their observation epochs do not overlap. Our measured quiescent baseline is also consistent with the low-state spectra obtained by IACTs, while extending the coverage to higher energies, reaching up to 8~TeV. At energies above 10~TeV, a more stringent upper limit is established.

\subsubsection{HE-to-VHE joint SED}

\begin{figure*}[!ht]
\centering
\includegraphics[width=0.9\textwidth]{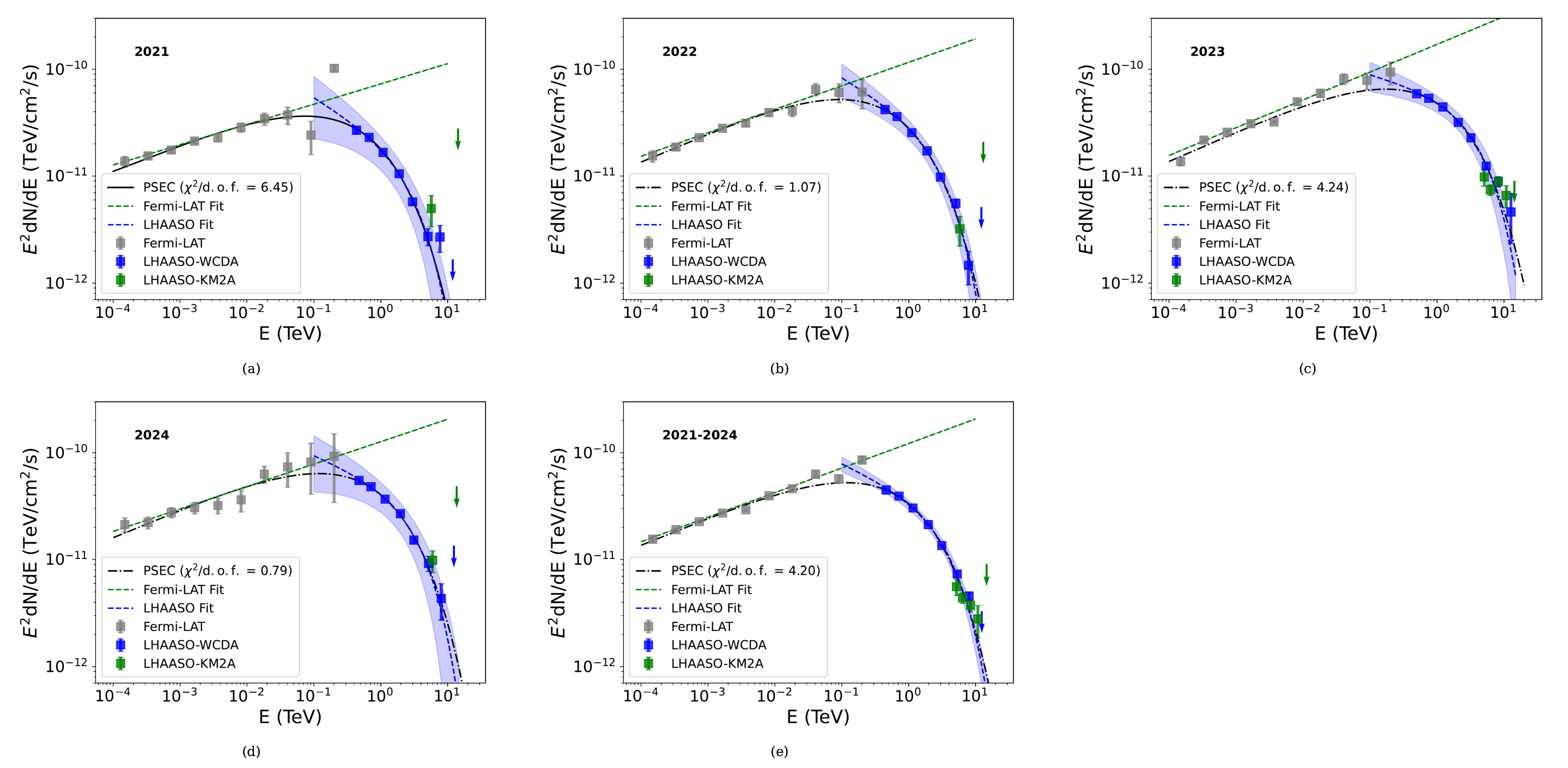}
\caption{The long-term SED of Mrk 421 in the HE-to-VHE range, measured using LHAASO data and Fermi-LAT, along with fitted functions of the data (see parameters in Table 3).}
\label{fig:average_sed_fit}
\end{figure*}

\begin{table*}
\centering
\caption{Parameters of the joint spectral fit to the gamma-ray SEDs obtained with LHAASO-WCDA and Fermi-LAT data. }

\begin{tabular}{lcccc}
\hline
\hline 
Epoch & $N_{0}$ & $\alpha$ & $E_{c}$ & $\chi^2/dof $ \\
& $(\times(10^{-12} \rm /TeV/cm^{2}/s))$&&(TeV)&\\
\hline
2021 & $1.18\pm0.10$ & $1.74\pm0.01$ & $0.28\pm0.02$ & 6.45 \\
2022 & $1.76\pm0.09$ & $1.72\pm0.01$ & $0.30\pm0.01$ & 1.07 \\
2023 & $1.89\pm0.05$ & $1.72\pm0.01$ & $0.54\pm0.02$ & 4.24 \\
2024 & $1.94\pm0.15$ & $1.73\pm0.01$ & $0.41\pm0.04$ & 0.79 \\
2021-2024 & $1.57\pm0.04$ & $1.74\pm0.01$ & $0.42\pm0.01$ & 4.20 \\
\hline
\hline 
\end{tabular}
\label{tab:Time_averge_fit}
\end{table*}

Figure \ref{fig:average_sed_fit} compares the intrinsic SED measured by LHAASO between 0.4 and 20~TeV with the simultaneous Fermi-LAT energy spectrum. The comparison shows a clear spectral steepening at high energies. To describe the transition between the HE (Fermi-LAT) and VHE (LHAASO) regimes, a power law with a subexponential cutoff (PSEC) model of the form $dN/dE = N_{0} \times (\frac{E}{1\mathrm{TeV}})^{-\alpha}\times \exp(-\sqrt{\frac{E}{E_{c}}})$, 
was fitted to the gamma‑ray SED using only the statistical uncertainties of each data point, as shown in Figure \ref{fig:average_sed_fit} with the best‑fit parameters, summarised in Table \ref{tab:Time_averge_fit}.
The PSEC model captures the spectral variation well across the full energy range and the cutoff energy $E_{cut}$ reaches its highest value, about 0.54~TeV, in 2023. Together with 
the flux increase by a factor of about 1.7 from 2021 to 2023, this suggests an ‘$E_{cut}$-higher‑when‑brighter’ behavior, while the photon index remains almost constant. 

In the Fermi-LAT band (0.1–200~GeV), the relative annual integrated energy fluxes for 2021–2024 are 0.56, 0.80, 1.00, and 0.90, respectively (normalized to the 2023 value of 1.00), corresponding to a maximum year‑to‑year variation of a factor of ~1.8. In the VHE band, as discussed in Section 4.2.4, the corresponding variation reaches a factor of ~3.4. The markedly larger variability amplitude in the VHE band compared to the GeV band further suggests that the GeV and VHE gamma‑ray emissions originate from distinct physical processes or spatial regions. 

Another interesting result from this joint fit is that the position of the IC peak also shifts in correlation with the energy flux, shown in left panel of Figure \ref{SED_index_F0}. 

\subsection{TeV Outbursts}

This section systematically characterizes the properties of outburst activity observed throughout the three-year observational campaign, with particular focus on multi-timescale variability patterns and phenomenological interpretations enabled by a parsimonious multi-zone model requiring only essential free parameters.

\subsubsection{Source State Definition and outburst duty cycle}\label{sec:dutycycle}

In this subsection, we focus on the remarkable VHE gamma-ray outbursts, aiming to investigate the spectral variations across different periods, with particular emphasis on comparisons with low-activity phases. The activity states of Mrk~421 are primarily defined using the LHAASO-WCDA light curve. To achieve this, we applied the Bayesian Block algorithm\citep{1998ApJ...504..405S,2013ApJ...764..167S}, which detects significant changes in data series independently of gaps or variations in exposure time, producing a block-wise constant representation of sequential data. Unlike methods which bin data into predefined temporal intervals, the Bayesian Block algorithm preserves critical flux changes by adapting to actual variability in the data. This approach remains highly effective in characterizing local variability in astronomical light curves, even with uneven data sampling, as it provides optimal temporal divisions based on genuine fluctuations. Thus, it serves as an efficient method for detecting outbursts. 

The optimal segmentation, defined by its change points, maximizes the goodness of fit with a certain model for the data within each block. The point measures fitness function is used for the Bayesian Blocks and applied to the daily signal points to find the change points at the transition from one signal state to the next. The algorithm requires the initial choice of a Bayesian prior, called $\mathrm{ncp}_\mathrm{{prior}}$, for the probability of finding a new change of flux states. A false-positive rate($p_\mathrm{{0}}$) is associated with the choice $\mathrm{ncp}_\mathrm{{prior}}$. The false-positive rate was chosen to be $p_\mathrm{{0}}$ = 0.05. The resulting segmentation yields 129 blocks, represented by the red line overlaid on the WCDA flux points (black dots) in Figure \ref{fig_mrk421_mul_lc}.
The height of each block, calculated as the weighted average of all signals within it, is used to identify outburst periods in the light curve across different observation epochs.

In the following, VHE gamma-ray outbursts are defined as instances where the flux increases by at least a factor of two compared to the quiescent phase. To determine the quiescent phase signal, the weighted average of the input light curve is calculated. Data points that are more than twice the average or less than half the average are iteratively removed, and this process is repeated until the average value stabilizes.
The result of the final iteration is taken as the baseline signal, which is 43.47 photon counts per day. An outburst may consist of multiple consecutive rising steps, collectively forming a local maximum in flux. After reaching this peak, the flux typically decreases to a lower level, which can occur over a day or a longer timescale. The duration of the outburst period is estimated by calculating the time between the first data point of a block and the last point of the last block, providing a conservative measure of the outburst's duration. These outbursts, detected using the Bayesian Block algorithm, are highlighted in Figure 1 with shaded gray areas to facilitate comparison with features in the light curves across other wavelengths. 
In total 23 outbursts have been identified in the three-year observation. The details of each outburst, i.e., starting/ending time, signal counts, and spectral properties, are reported in Table \ref{tab:source_state}. The detailed multi-band light curve in the outburst periods is shown in Figure \ref{fig:multi_lc_outburst} of Appendix-A
The brightest day was observed between MJD 60340.6 and 60341.6 (1st Feb. 2024), with a significance of approximately $20\ \sigma$. During this period, the fitted spectral index was about $-2.27\pm0.12$, and the integral flux above $E > 0.4 \mathrm{TeV}$ was $(3.49 \pm 0.45) \times 10^{-10} \ \mathrm{TeV/cm^{2}/s}$, corresponding to 3.3 CU.

\begin{table}[!ht]
    \centering
    \caption{The table summarizes the key parameters of 23 outburst episodes from the TeV light curve of Mrk421, including temporal characteristics and spectral features. Parameters include: start/end time (MJD Start/Stop), duration (in units of day), signals of outburst periods ($N_{s}$), normalized flux at 3 TeV ($N_{0} \times 10^{-12} \rm /TeV/cm^{2}/s$), spectral index ($\alpha$), rising/decay timescales ($t_{r}$,$t_{d}$ in days), and significance (Sig).}
   \begin{tabular}{lccccccccc}
    \hline
    \hline
    Epoch & MJD & MJD & Duration & $N_{s}$ & $N_{0}$ & $\alpha$ & $t_{r}$ & $t_{d}$ & Sig. \\
     & Start & Stop & (day) & & & & (day) & (day) \\
    \hline
    O1 & 59717.4 & 59722.3 & 5 & $131.69\pm6.18$ &$4.82\pm0.30$ &$2.38\pm0.06$ & \makecell[l]{$2.7\pm0.5$} & \makecell[l]{$0.3\pm0.1$} & 30.81\\
    O2 & 59798.1 & 59804.1 & 6 & $104.23\pm5.46$ & $2.98\pm0.22$ &$2.64\pm0.06$ & \makecell[l]{$0.3\pm0.1$} & \makecell[l]{$2.2\pm0.4$} & 26.27\\
    O3 & 59856.9 & 59858.9 & 2 & $132.33\pm9.83$ & $4.70\pm0.50$ &$2.34\pm0.10$ & \makecell[l]{$0.6\pm0.1$} & \makecell[l]{$0.7\pm0.1$} & 18.79\\
    O4 & 59871.9 & 59875.9 & 4 & $111.45\pm6.72$ & $3.63\pm0.32$ &$2.52\pm0.08$ & \makecell[l]{$1.1\pm0.2$} & \makecell[l]{$1.2\pm0.2$} & 22.38\\
    O5 & 59911.8 & 59918.8 & 7 & $142.17\pm5.32$ & $5.67\pm0.28$ &$2.39\pm0.05$ & \makecell[l]{$1.1\pm0.2$} & \makecell[l]{$1.8\pm0.3$} & 38.96 \\
    O6 & 59947.7 & 59966.7 & 19 & $142.04\pm3.10$ &$5.25\pm0.15$ &$2.54\pm0.02$ & - & - & 67.46 \\
    O7 & 59971.7 & 59973.7 & 2 & $162.01\pm10.27$ & $7.09\pm0.53$ &$2.25\pm0.08$ & \makecell[l]{$0.6\pm0.2$} & \makecell[l]{$0.5\pm0.2$}  & 24.44\\
    O8 & 59979.6 & 59985.6 & 6 & $113.60\pm5.14$ & $4.01\pm0.25$ &$2.61\pm0.05$ & \makecell[l]{$0.2\pm0.1$} & \makecell[l]{$3.8\pm0.7$} & 30.77\\
    O9 & 59988.6 & 59989.6 & 1 & $152.80\pm14.04$ & $6.13\pm0.82$ &$2.34\pm0.13$ & \makecell[l]{$0.5\pm0.1$} & \makecell[l]{$0.5\pm0.1$} & 15.14\\
    O10 & 60052.4 & 60059.4 & 7 & $169.17\pm5.88$ & $5.69\pm0.28$ &$2.48\pm0.04$ & \makecell[l]{$0.4\pm0.2$} & \makecell[l]{$4.7\pm0.6$} & 42.78\\
    O11 & 60067.4 & 60068.4 & 1 & $129.68\pm13.15$ & $5.25\pm0.76$&$2.30\pm0.15$  & - & -  & 14.02\\
    O12 & 60069.4 & 60074.4 & 5 & $111.17\pm5.60$ & $4.01\pm0.28$ &$2.49\pm0.06$  & - & - & 28.66\\
    O13 & 60075.4 & 60092.3 & 16 & $98.60\pm3.15$ & $3.49\pm0.15$ &$2.51\pm0.04$  & - & - & 44.22\\
    O14 & 60119.3 & 60128.2 & 9 & $100.50\pm4.33$ & $3.18\pm0.19$ &$2.47\pm0.05$ & - & - & 31.80 \\
    O15 & 60139.2 & 60144.2 & 5 & $98.40\pm5.98$ & $2.74\pm0.24$ & $2.61\pm0.07$ & - & - & 21.62\\
    O16 & 60204.0 & 60215.9 & 12 & $146.33\pm4.49$ & $4.90\pm0.19$&$2.50\pm0.03$ & - & - & 49.26\\
    O17 & 60217.9 & 60220.9 & 3 & $124.62\pm8.27$ & $3.75\pm0.34$ &$2.42\pm0.08$ & \makecell[l]{$0.6\pm0.3$} & \makecell[l]{$2.7\pm0.9$} & 20.82\\
    O18 & 60228.9 & 60241.9 & 12 & $199.37\pm4.77$ & $6.91\pm0.21$&$2.32\pm0.03$ & \makecell[l]{$5.9\pm0.7$} & \makecell[l]{$1.5\pm0.2$} & 65.43\\
    O19 & 60243.9 & 60246.9 & 3 & $186.76\pm9.36$ & $6.67\pm0.49$ &$2.37\pm0.07$ & - & - & 29.84 \\
    O20 & 60247.9 & 60260.8 & 13 & $135.31\pm4.66$ & $5.35\pm0.22$&$2.38\pm0.04$ & - & - & 46.29\\
    O21 & 60279.8 & 60283.8 & 4 & $99.83\pm6.36$ & $3.39\pm0.29$ & $2.45\pm0.08$ & \makecell[l]{$1.2\pm0.3$} & \makecell[l]{$1.0\pm0.4$} & 21.31\\
    O22 & 60336.7 & 60343.6 & 7 & $168.85\pm5.78$ & $6.23\pm0.30$ &$2.45\pm0.04$ & \makecell[l]{$2.1\pm0.3$} & \makecell[l]{$1.2\pm0.2$} & 43.62\\
    O23 & 60346.6 & 60352.6 & 6 & $185.58\pm6.27$ &$6.17\pm0.30$ &$2.54\pm0.04$ & - & - & 42.97\\
    \hline
    \end{tabular}%
    \label{tab:source_state}
\end{table}

As previously mentioned, blazars exhibit significant variability on short time scales. The lowest stable flux level is referred to as the baseline state. The activity level of a source can be characterized by the fraction of time it spends in outburst states, commonly known as the duty cycle (DC), and is expressed as follows \citep{2014ApJ...782..110A}:
\begin{equation}
    DC = \frac{\sum_{i}t_{i}}{T_{obs}}
\end{equation}
where $t_{i}$ is the time that the source has experienced during the $i$th outburst, with $i$ running over all identified outburst in the observation period $T_{\rm obs}$. To determine the duty cycle, a flux threshold must firstly be established to differentiate between outburst and non-outburst states. We utilize the method described in Section 4.3.1 to identify the outburst states. 
 Based on three years continuous monitoring and our definition of VHE outbursts, the total duration time of outbursts identified by LHAASO lasts 147.85 days, indicating a duty cycle of approximately 14.6\%. This value is significantly lower than the $(43\%\pm13\%)$ reported by IACTs \citep{2007JPhCS..60..318T}. 
Such discrepancy is expected, as the duty cycle derived from IACT observations is likely an overestimate due to their follow-up strategy, which preferentially targets the source when it is in a high state as indicated by other instruments.

\begin{figure*}[!ht]
\centering
\includegraphics[width=0.7\textwidth]{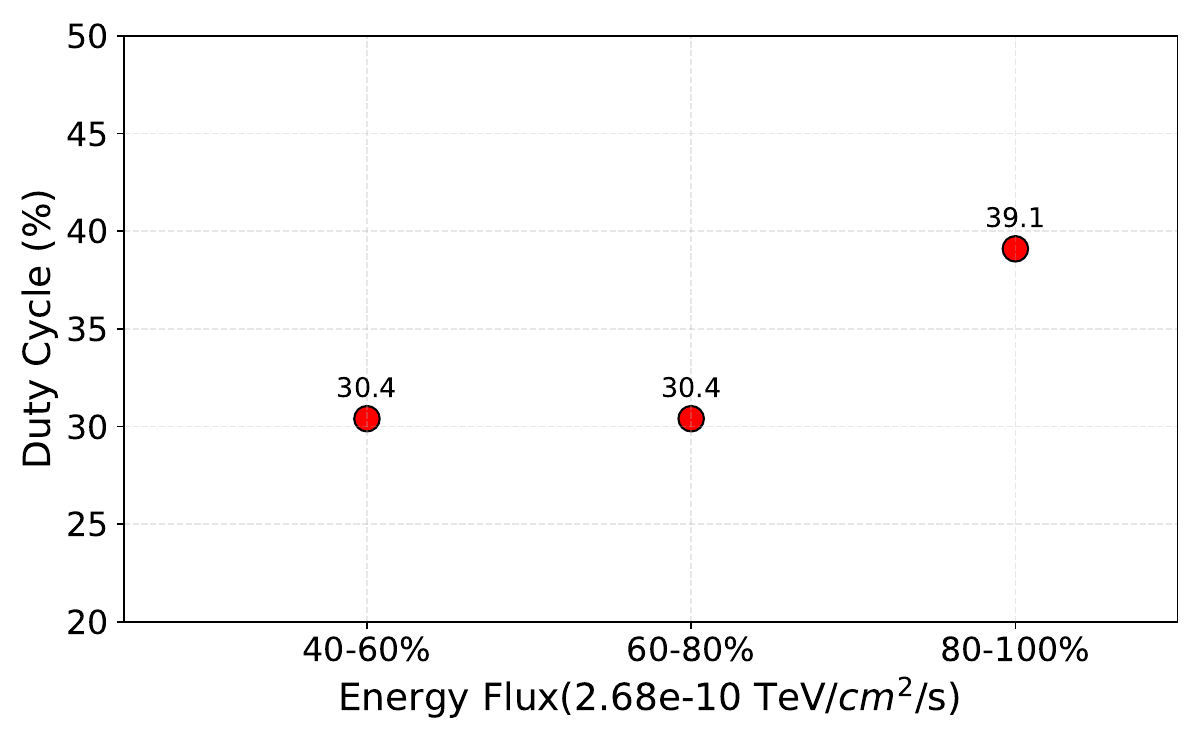}
\caption{The distribution of the duty cycle relative to different energy flux threshold for 23 outbursts}
\label{fig:duty_cycle_flaring}
\end{figure*}

We also examine the duty cycle relative to higher flux thresholds. Figure \ref{fig:duty_cycle_flaring} shows the distribution of the VHE outburst duty cycle as a function of different integrated energy flux thresholds from this work. 
The highest measured flux in the 0.4–20 TeV range among all 23 outbursts is
 $2.68\ \times10^{-10} \mathrm{TeV}/\mathrm{cm^{2}}/\mathrm{s}$ which we adopt as the reference value. The outbursts are then divided into three intensity levels: 
 40\%–60\%, 60\%–80\%, and 80\%–100\% of this reference. The corresponding duty cycles for these levels are 4.5\%, 3.9\%, and 6.2\%, respectively. Expressed in terms of event frequency, these three categories account for 30.4\%, 30.4\%, and 39.1\% of the 23 outbursts. Notably, the duty‑cycle distribution across intensity levels appears relatively flat.

More detailed analysis of these outburst periods, including timescale and energy spectra, is provided in Sections 4.3.2 and 4.3.3. 

\subsubsection{VHE multi-timescale variability}

This section focuses on the properties of the VHE gamma-ray variability of Mrk~421 during the outbursts detected by LHAASO. We employ the Fred function to fit the light curve:
\begin{equation}
    F(t) = F_{\rm c} + F_{0} \times [e^{(t_{0}-t)/T_{\rm r}}+e^{(t-t_{0})/T_{\rm d}}] ^{-1},
\label{Formula:F(t)}
\end{equation}
where $F_{\rm c}$ represents an assumed constant level underlying the outburst, $F_{0}$ measures the amplitude of the outbursts, $t_{0}$ describes approximately the time of the peak, $T_{\rm r}$ and $T_{\rm d}$ measure the rise and decay timescale.

\begin{figure}[ht]
    \centering
    \subfigure[O1]{\includegraphics[width=0.24\textwidth,height=2.6cm]{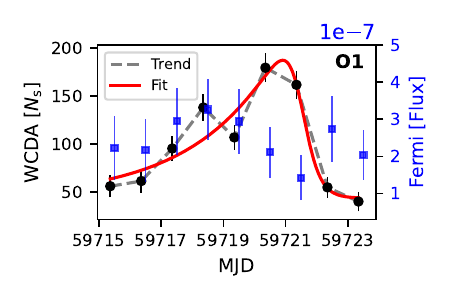}}
    \subfigure[O18]{\includegraphics[width=0.24\textwidth,height=2.6cm]{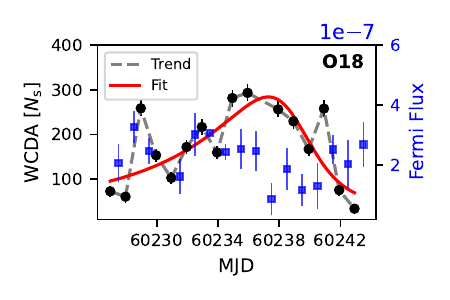}} 
    \subfigure[O22]{\includegraphics[width=0.24\textwidth,height=2.6cm]{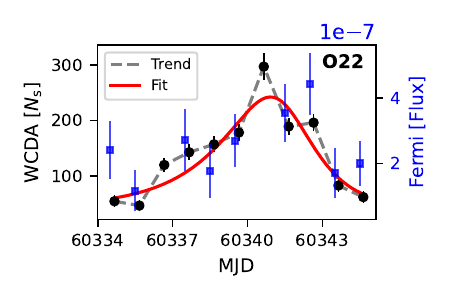}}
    \subfigure[O2]{\includegraphics[width=0.24\textwidth,height=2.6cm]{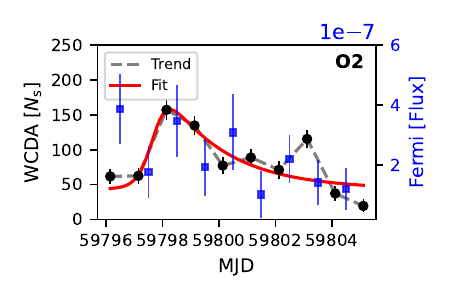}} \\

    \subfigure[O5]{\includegraphics[width=0.24\textwidth,height=2.6cm]{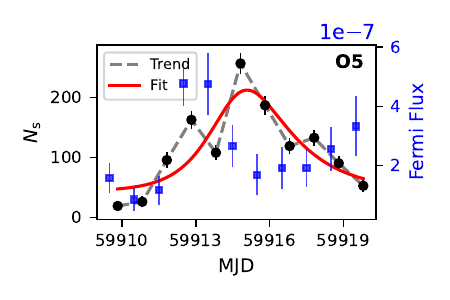}} 
    \subfigure[O8]{\includegraphics[width=0.24\textwidth,height=2.6cm]{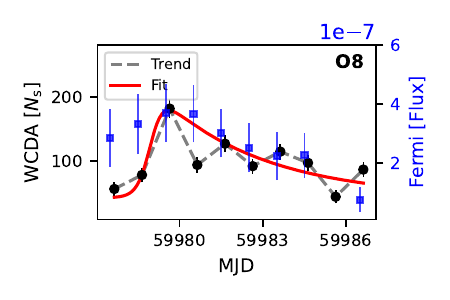}}
    \subfigure[O10]{\includegraphics[width=0.24\textwidth,height=2.6cm]{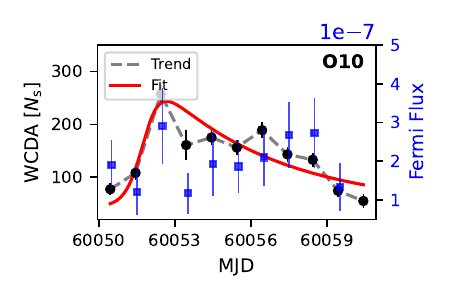}}
    \subfigure[O17]{\includegraphics[width=0.24\textwidth,height=2.6cm]{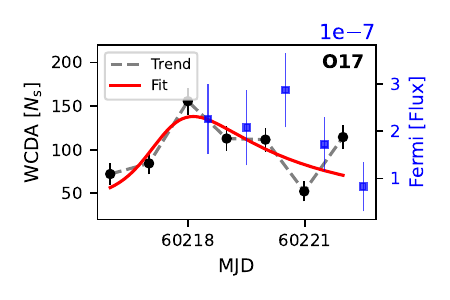}}  

    \subfigure[O3]{\includegraphics[width=0.24\textwidth,height=2.6cm]{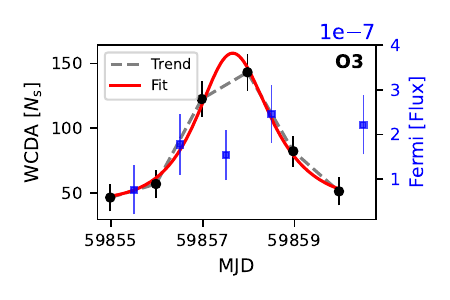}} 
    \subfigure[O4]{\includegraphics[width=0.24\textwidth,height=2.6cm]{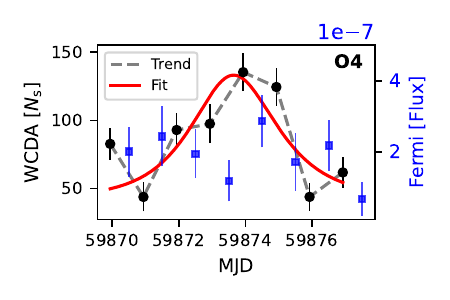}} 
    \subfigure[O7]{\includegraphics[width=0.24\textwidth,height=2.6cm]{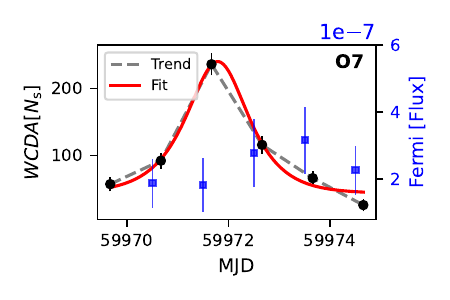}}
    \subfigure[O9]{\includegraphics[width=0.24\textwidth,height=2.6cm]{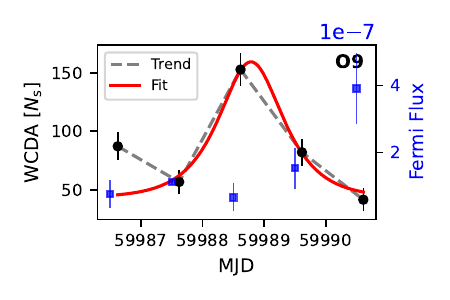}}\\
    
    \subfigure[O21]{\includegraphics[width=0.24\textwidth,height=2.6cm]{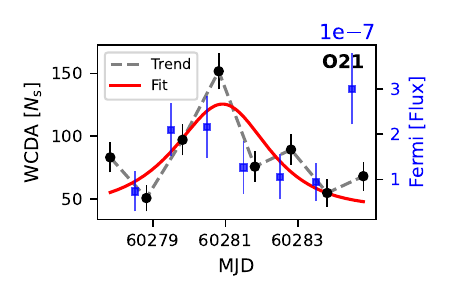}}  
    \subfigure[O6]{\includegraphics[width=0.24\textwidth,height=2.6cm]{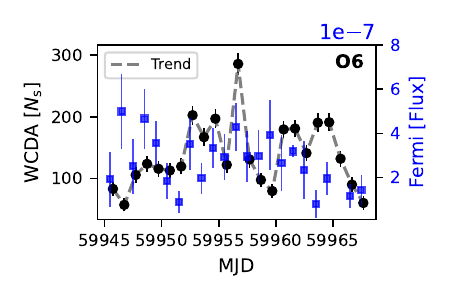}} 
    \subfigure[O11\_O13]{\includegraphics[width=0.24\textwidth,height=2.6cm]{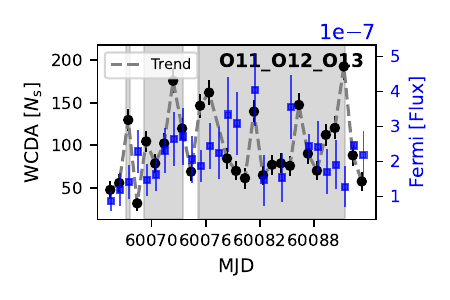}} 
    \subfigure[O14]{\includegraphics[width=0.24\textwidth,height=2.6cm]{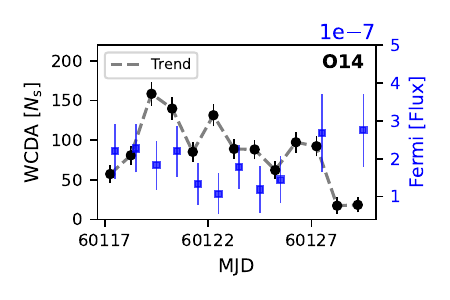}}\\
 
    \subfigure[O15]{\includegraphics[width=0.24\textwidth,height=2.6cm]{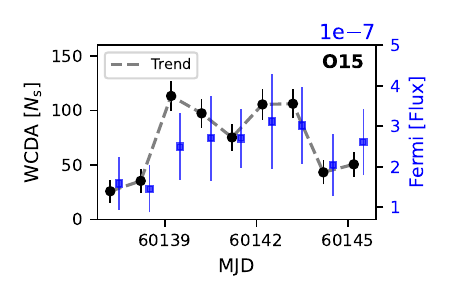}}
    \subfigure[O16]{\includegraphics[width=0.24\textwidth,height=2.6cm]{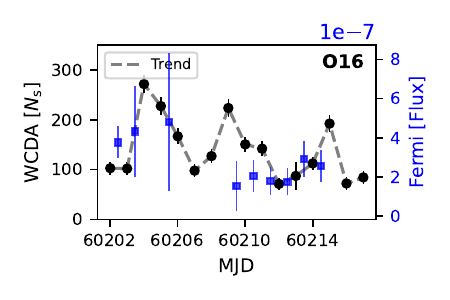}} 
    \subfigure[O19\_O20]{\includegraphics[width=0.24\textwidth,height=2.6cm]{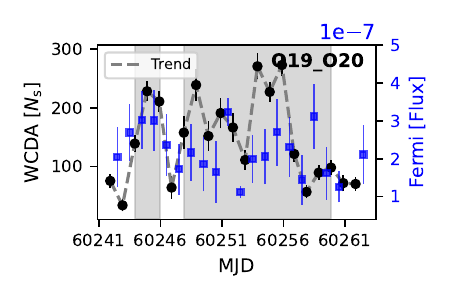}} 
    \subfigure[O23]{\includegraphics[width=0.24\textwidth,height=2.6cm]{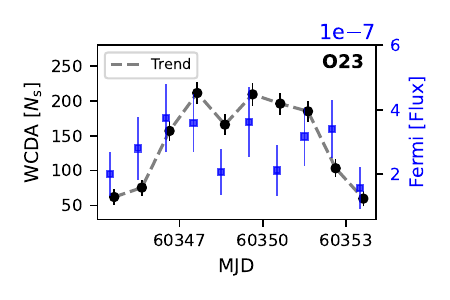}}  
    \caption{The detailed 23 outbursts light curves from LHASO-WCDA : panel (a)-(c) is for Group-A with longer rising timescale, panel (d)-(h) with longer decay timecale, panel (i)-(m) with roughly symmetric timescale and panel (n)-(t) for ‘outburst forest’. Here, the black points represent the light curves from WCDA, the blue points represent the light curves from Fermi, and the red curves in Groups A, B, and C show the fitting results obtained using the Fred function.}
    \label{fig:groupD}
\end{figure}

In the fitting process, the constant background level underlying the outburst $F_{\rm c}$, defined as the long-term weighted average signal value, is fixed to the baseline value. For each outburst, $T_{\rm r}$, $T_{\rm d}$ and $F_0$ are free parameters. Equation \ref{Formula:F(t)} provides a rough description of the light curve, but it may not perfectly fit all data points. For example, in strong and long-duration outbursts (such as O11-O13, etc.), the light curves have complex substructures, which cannot be reproduced by Fred function.

\begin{figure*}[ht]
    \centering
    \includegraphics[width=0.8\textwidth]{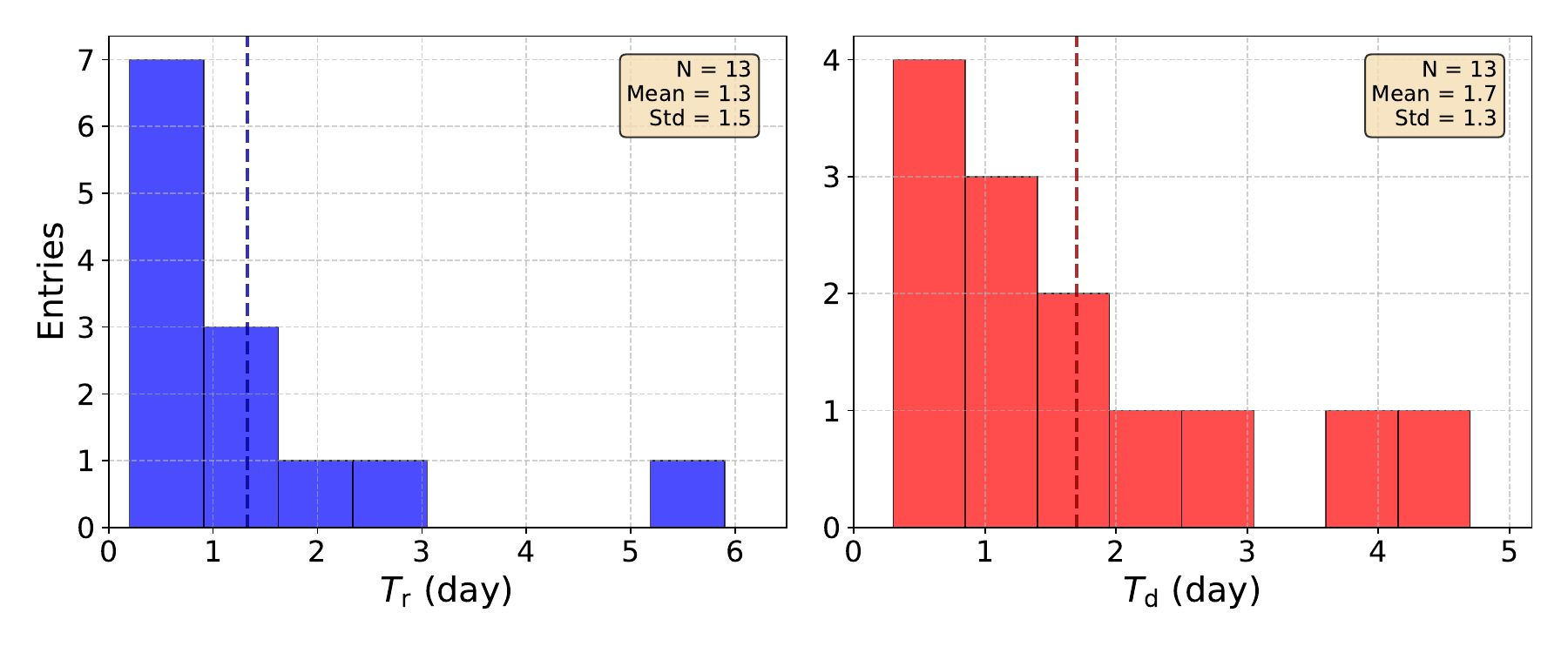}
    \caption{Histogram distribution of rise ($T_{r}$ (day)) and decay ($T_{d}$ (day)) timescales during outburst periods.}
    \label{tr_td}
\end{figure*}

We performed the fitting procedure to all 23 identified outbursts, and best-fit parameters are listed in Table~\ref{tab:source_state} and the corresponding light curves are presented in detail in Appendix A Figure~\ref{fig:groupD}. Based on the profile of the light curves, the outbursts can be categorized into four distinct groups:

\begin{itemize}
\item Group-A: the light curve can be fitted with a Fred function exhibiting longer rising timescale than the decay timescale. O1, O18, and O22 belong to this group, and their light curves are shown in panel (a) - (c) of Figure~\ref{fig:groupD}. Among them, the rising timescale ranges from 1.2 days to 5.9 days, while the decay timescale are typically on the order of a  day level, ranging from 0.3 to 1.5 days. 

\item Group-B: similar to the Group-A, the light curve can be fitted with a Fred function, but shows a shorter rising timescale than the decay timescale. O2, O5, O8, O10, and O17
belong this group as shown in panel (d) - (h) of Figure~\ref{fig:groupD}. The magnitude of the decay timescale is within a few days level, whereas all rising timescale in this group are less than one day. 
ermined by the emission region’s spatial size, as theorized in studies like {\color{red}Zhang et al. (2002)}.

\item Group-C: the light curve can be fitted with a Fred function.
In this case, the outbursts exhibit an almost symmetric light curve with comparable rising and decay timescales. There are five outburst events in this category, namely O3, O4, O7, O9, and O21 as shown in panel (i) - (m) of Figure~\ref{fig:groupD}. It is interesting to see that all of these events show variability within the intra-day scope.

\item Group-D: these outbursts exhibits very complex temporal behaviours, featuring with two or more spikes being found in their light curves, which cannot be adequately described by a single Fred function. 
Their light curves may be fitted with the superposition of a few consecutive Fred functions but our statistics and time resolution does not allow us to claim so. Here total 10 outbursts belong to this group as shown in panel (n) - (t) of Figure~\ref{fig:groupD}. We call them ‘outburst forest’.

\end{itemize}

Figure~\ref{tr_td} shows the distributions of the rising timescale and the decay timescale of fitted outbursts (excluding Group-D). 
During the outburst period presented here, 13 outbursts occurred on intra-day time scales and generally showed larger fractional variability than the outburst detected with longer duration. Note that, the rising timescale depends on particle acceleration/injection timescale and the decay timescale depends on particle cooling/escape timescale. In addition, if these timescales are shorter than the light crossing time of the radiation zone, the rising or decay timescale will be determined by the light crossing time. Therefore, in addition to the value of the rising and decay timescales, different shapes of the light curves also shed light on the dissipation and radiation mechanism of the outburst.
To be more precise, we also find that the fitting may not be perfect due to the complex profile of the light curve and daily data point. 
During certain outbursts, like shown in panel (a),(b) (d) of Figure~\ref{fig:groupD}, a small, fast sub-flare has been observed superimposed on the global/general outburst behaviour.  However, these fitting results still provide a good indication of the convolution of the outburst events. 
Detailed multi-wavelength light curves for each of the 23 outbursts are shown individually in Figure \ref{fig:multi_lc_outburst} in Appendix A.

\subsubsection{Outburst spectral behavior}

\begin{figure}[htbp]
\centering
\includegraphics[width=0.9\textwidth,height=10cm]{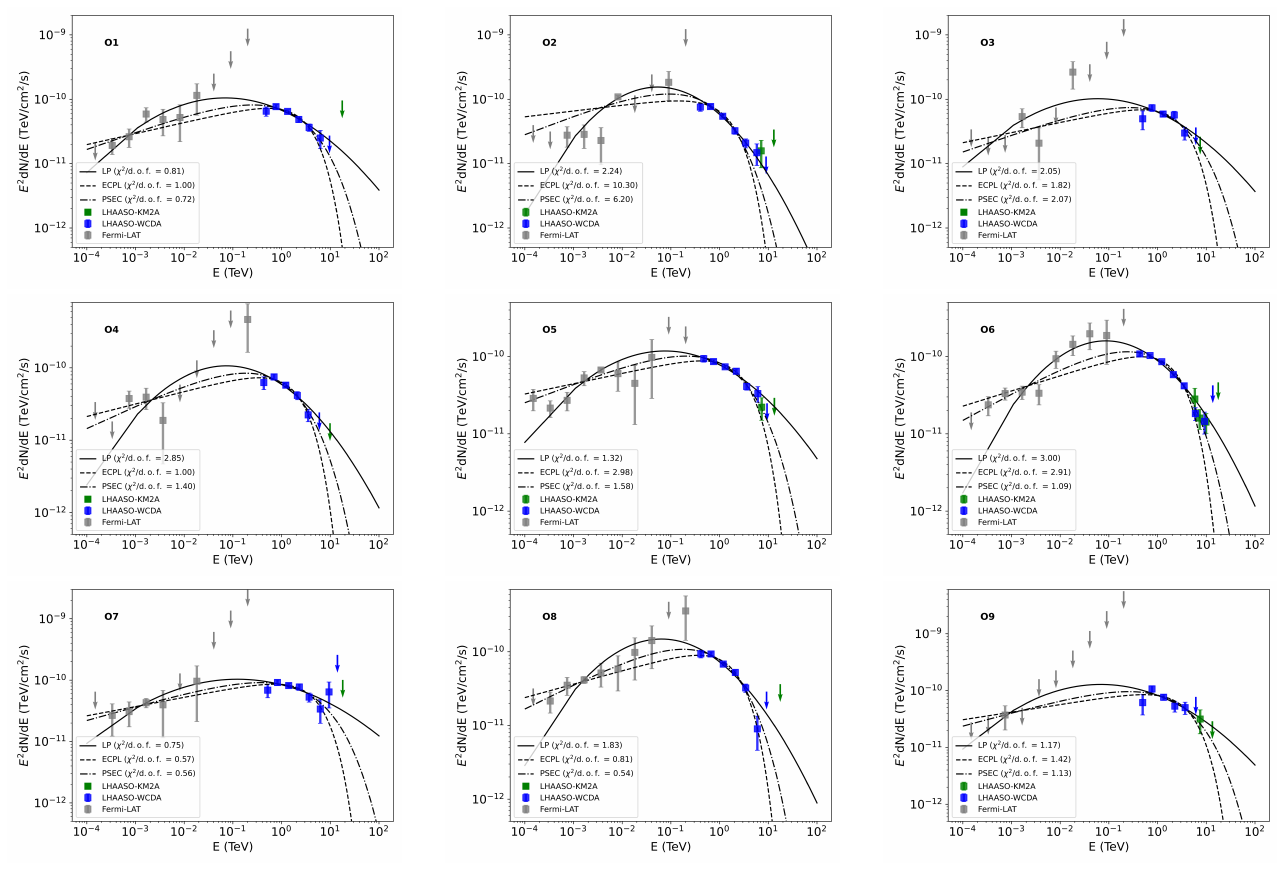}
\caption{Energy spectrum of Mrk~421 during 23 outbursts, simultaneous SED from Fermi-Lat is included. Nine panels are relative to outburst O1 to O9. based on joint observational data from LHAASO-WCDA, LHAASO-KM2A, and Fermi-LAT during observation periods O1 to O9. The solid gray, blue, and green squares represent the observational results from Fermi-LAT, LHAASO (WCDA and KM2A), with arrows indicating the upper limits of flux. The solid line in the figure represents the fitting result of the LP model, the dashed line represents the fitting result of the ECPL model, and the dotted line represents the fitting result of the PSEC model.}
\label{Fit_gamma_ray_part1}
\end{figure}

To enable a direct comparison between outbursts, 
the SEDs of these 23 outbursts were derived following the procedure described in Section 4.2.4, but with a simplified power-law model to accommodate the varying durations and limited statistics of individual events. The best‑fit SED parameters for the 23 outbursts are listed in Table \ref{tab:source_state}. 
Figure \ref{Fit_gamma_ray_part1} show the individual SED, which include simultaneous GeV data. Among the outbursts, the hardest spectrum has an index of $2.25\pm 0.08$ and normalization $N_{0} =7.09 \pm 0.53$, corresponding to an integral flux of 2 CU in the energy range of 0.4-20 TeV region. While the softest spectrum has an index of $2.64 \pm 0.06$ and $N_{0} = 2.98 \pm 0.22$, corresponding to an integral flux of 1.38 CU. Compared to the 2021 averaged SED, the flux during these outbursts increased by a factor of 5–7. 
 
\begin{figure}[htbp]
\centering
\includegraphics[width=0.9\textwidth,height=10cm]{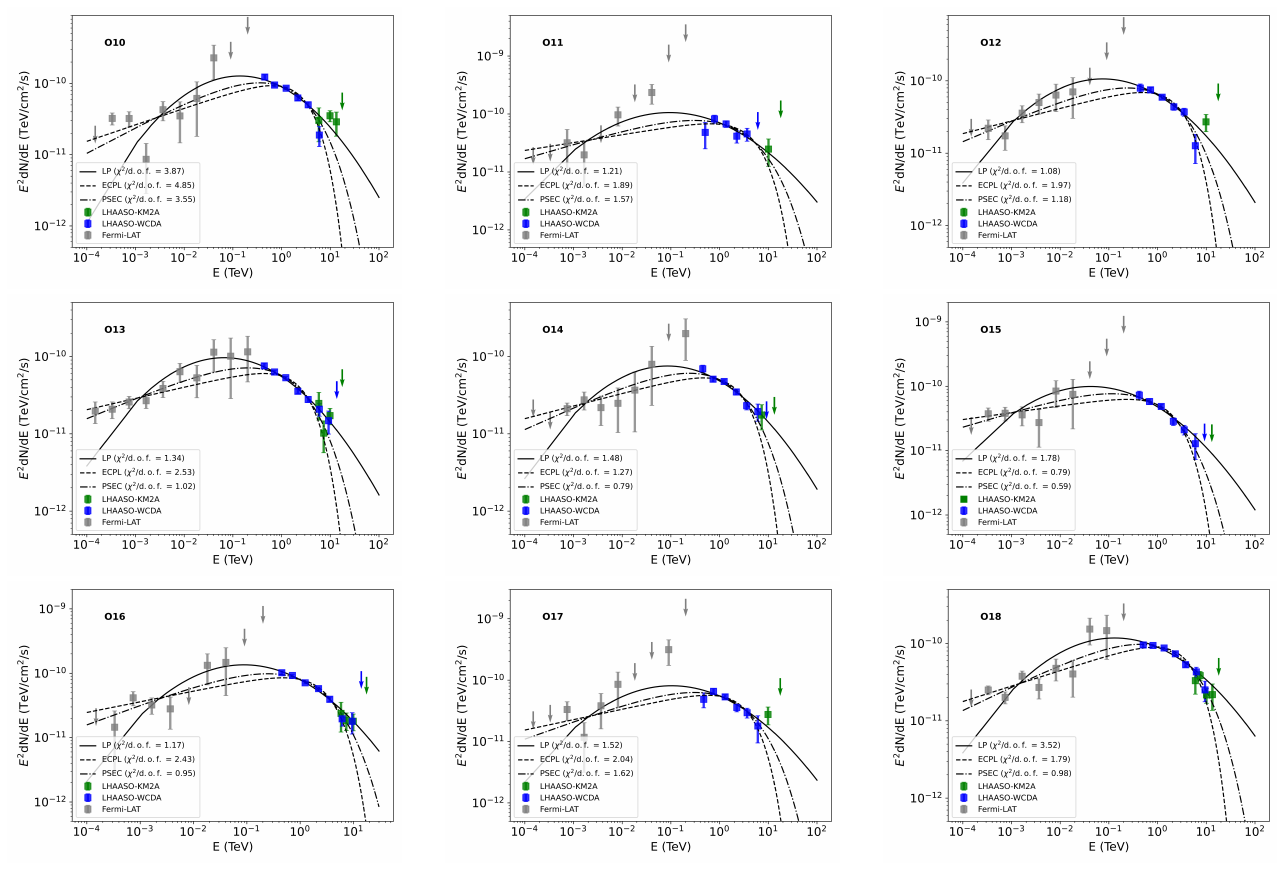}
\center{Figure \ref{Fit_gamma_ray_part1} -- Continued for outburst O10 to O18}
\end{figure}

\begin{figure}[htbp]
\centering
\includegraphics[width=0.9\textwidth,height=6.2cm]{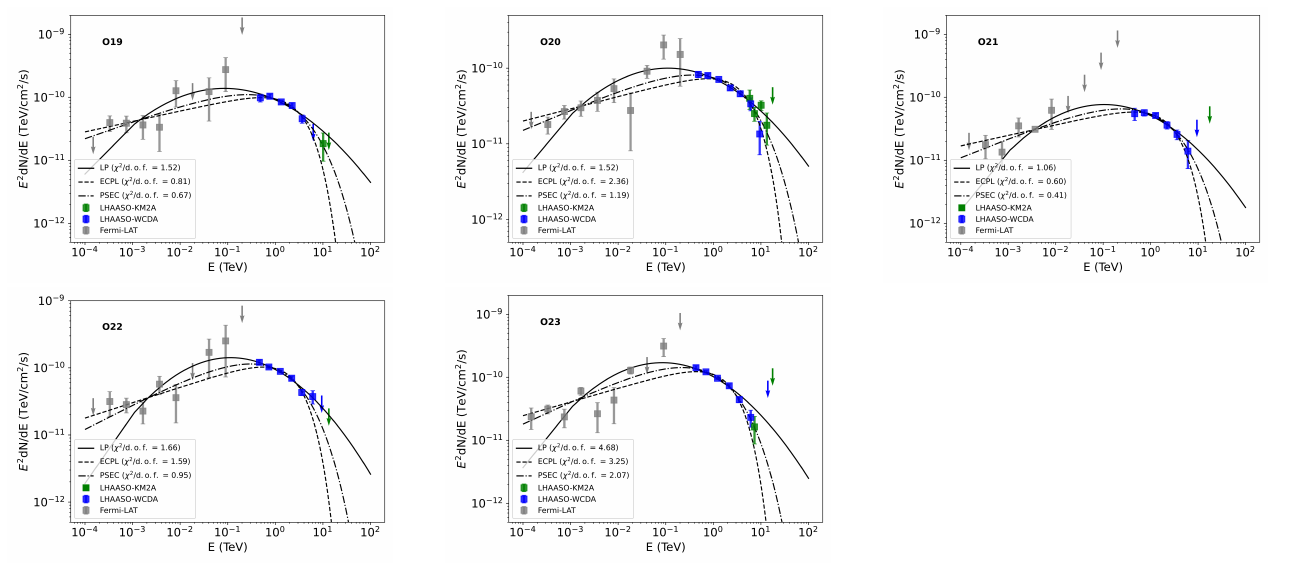}
\center{Figure \ref{Fit_gamma_ray_part1} -- Continued for outburst O19 to O23}
\end{figure}

In Figure.\ref{Fit_gamma_ray_part1} we also show the simultaneous SED in the GeV band from Fermi-LAT for the 23 active states. 
Three functions are adopted to fit GeV-to-VHE gamma-ray SEDs, here taking into account both the statistical and systematic errors of each data point. The three functions are the log-parabola (LP), ECPL, and PSEC models. In the figure, the solid, dashed, and dotted lines represent the LP, ECPL, and PSEC fit results, respectively.
Based on  chi-square value distributions, shown in Appendix-B Figure.~\ref{chi2_distribution}, and GeV-to-TeV fitting results, we observed significant differences among the three models: The ECPL model not only exhibited substantial deviation in overall chi-square distribution but also showed clear discrepancies in low-energy band fitting, failing to achieve smooth spectral transitions. Although the LP model outperformed the ECPL model overall, it still demonstrated systematic deviations in high-energy bands (GeV/TeV bands), with predicted fluxes exceeding observational upper limits. In contrast, the PSEC model demonstrated optimal fitting performance—its chi-square values were concentrated in the range of $\chi^{2}/dof < 2$, and it exhibited better explanatory power for the high-energy gamma-ray SEDs observed by LHAASO, Fermi-LAT. This model not only captured spectral variations across both high and low-energy bands but also maintained good adaptability throughout the entire energy range. The PSEC model is therefore used in the subsequent IC peak estimation.

\begin{figure}[htpb]
\centering
\subfigure{\includegraphics[width=0.45\textwidth]{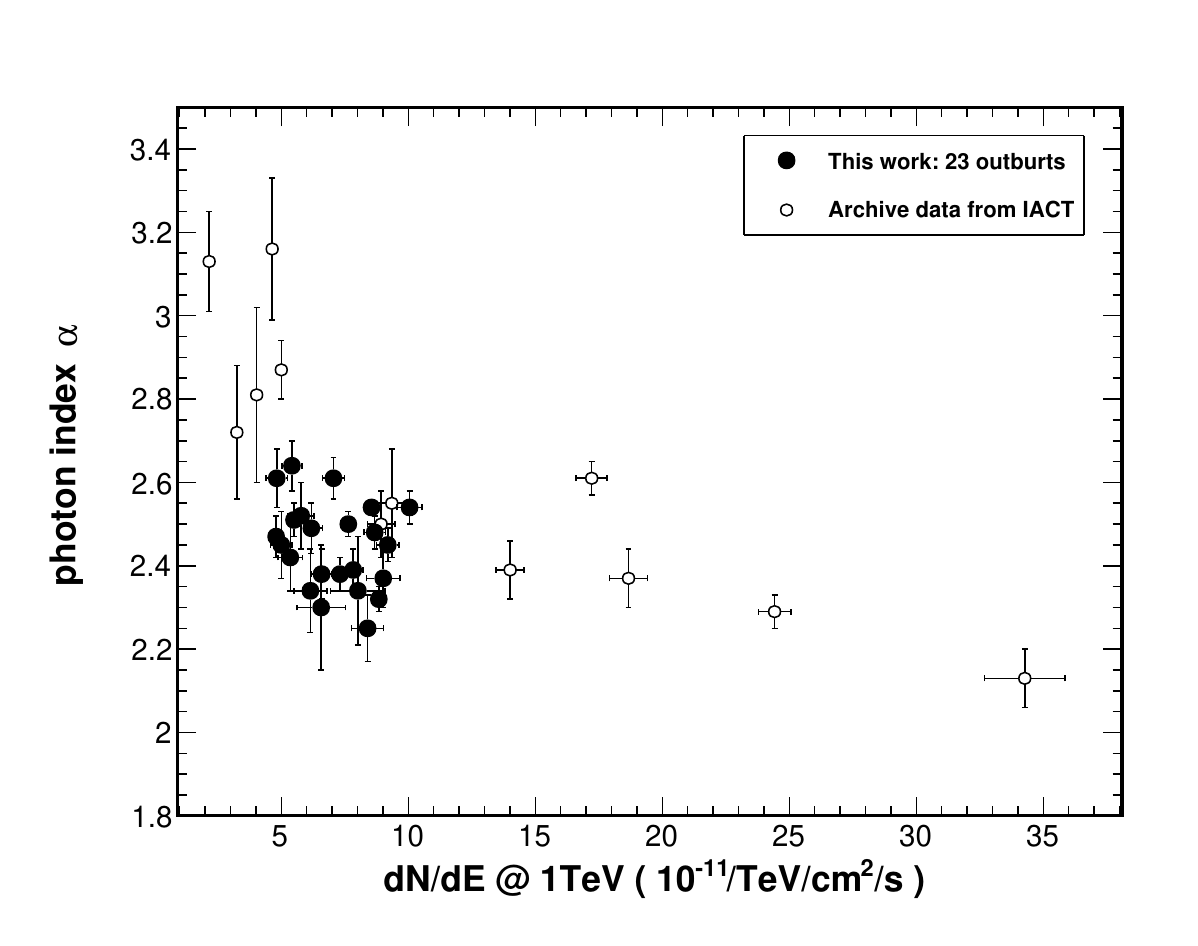}}
\subfigure{\includegraphics[width=0.45\textwidth]{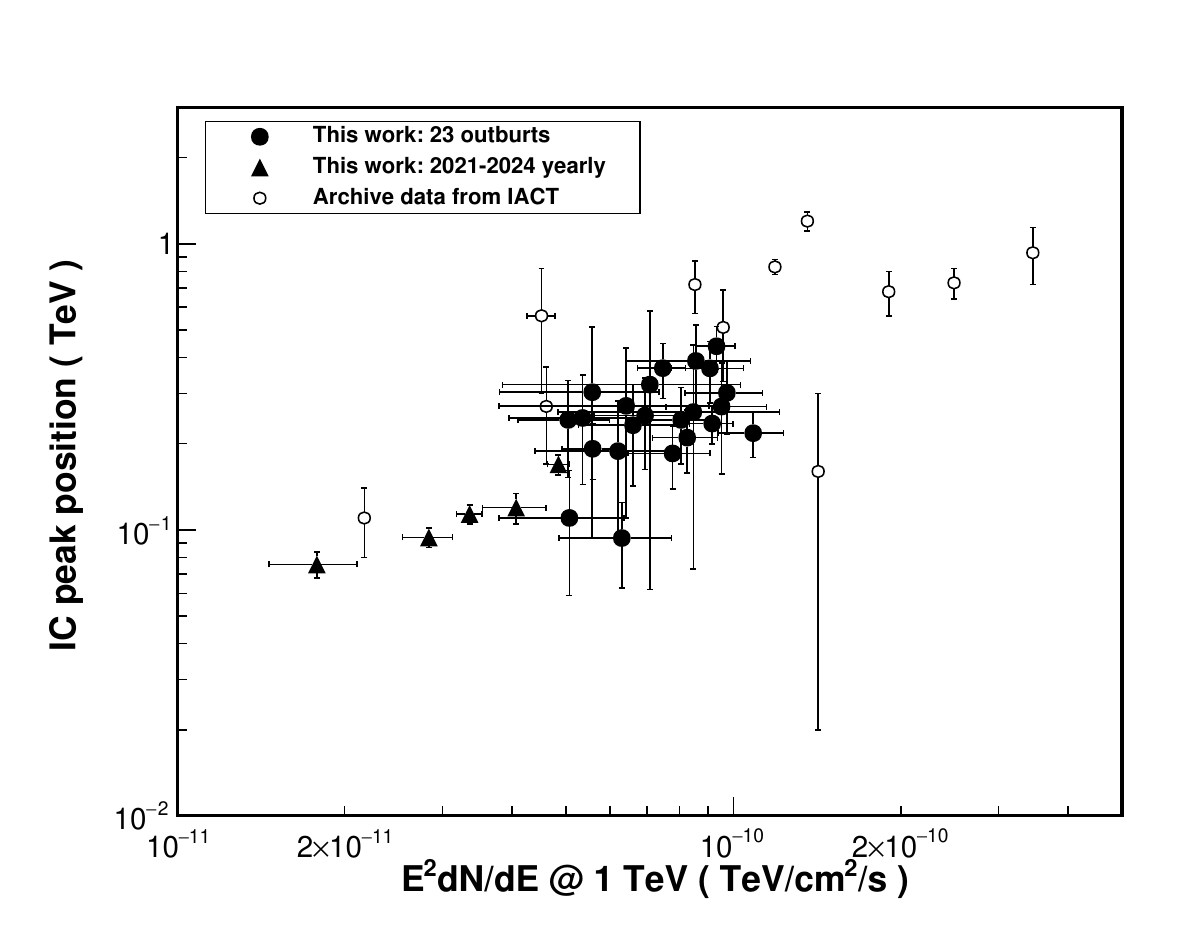}}
\caption{Left panel: relationship between the intrinsic spectral index and the 1 TeV flux for the 23 outburst periods. Right panel: relationship between the fitted SED peak position and the 1 TeV flux.  Data are shown for the 23 outbursts from this work (filled black circles), the annual SED results from this work (black triangles), and archival IACT measurements (open circles, \citep{2007ApJ...663...125} ).}
\label{SED_index_F0}
\end{figure}

As noted earlier, the observations analyzed here were obtained during the high-activity state of the source, which exhibits not only an elevated flux but also a harder spectral shape. This is consistent with the established ‘harder-when-brighter’ trend illustrated in left panel of  Figure \ref{SED_index_F0}. The figure plots the spectral index against the 1~TeV flux for various datasets, including archival IACT measurements taken during different source states(\citep[]{2007ApJ...663...125}).

To quantify the correlation between spectral index and energy flux, 
we performed a linear fit ($y = \alpha x + \beta$) to the 23 outbursts detected by LHAASO. The best-fit slope is $\alpha=-0.013\pm0.008$ and $\beta = 2.58\pm0.06$. This slope is smaller than the value of -0.027 reported from archival IACTs observations (\citep{2007ApJ...663..125A}). Overall, the LHAASO outburst sample shows a consistent behaviour with earlier IACT results, displaying a clear negative correlation. Whether this correlation is strictly linear or follows a more complex pattern remains an open question for discussion.

Another notable trend is the ‘$E_{\rm cut}$-higher-when-brighter’ behaviour. In right panel of Figure \ref{SED_index_F0}, we compare the positions of IC peak derived from the joint GeV-to-VHE SED and fluxes $(E^{2}dN/dE)$ at 1 TeV reported by different experiments. The data show that the peak systematically shifts toward higher energies as the flux increases. Furthermore, the LHAASO measurements-which predominantly sample low to moderate flux states—provide substantial new information that significantly enriches our understanding of the source's behaviour in the low-energy regime.

\begin{figure*}[!ht]
    \centering
    \includegraphics[width=0.8\textwidth]{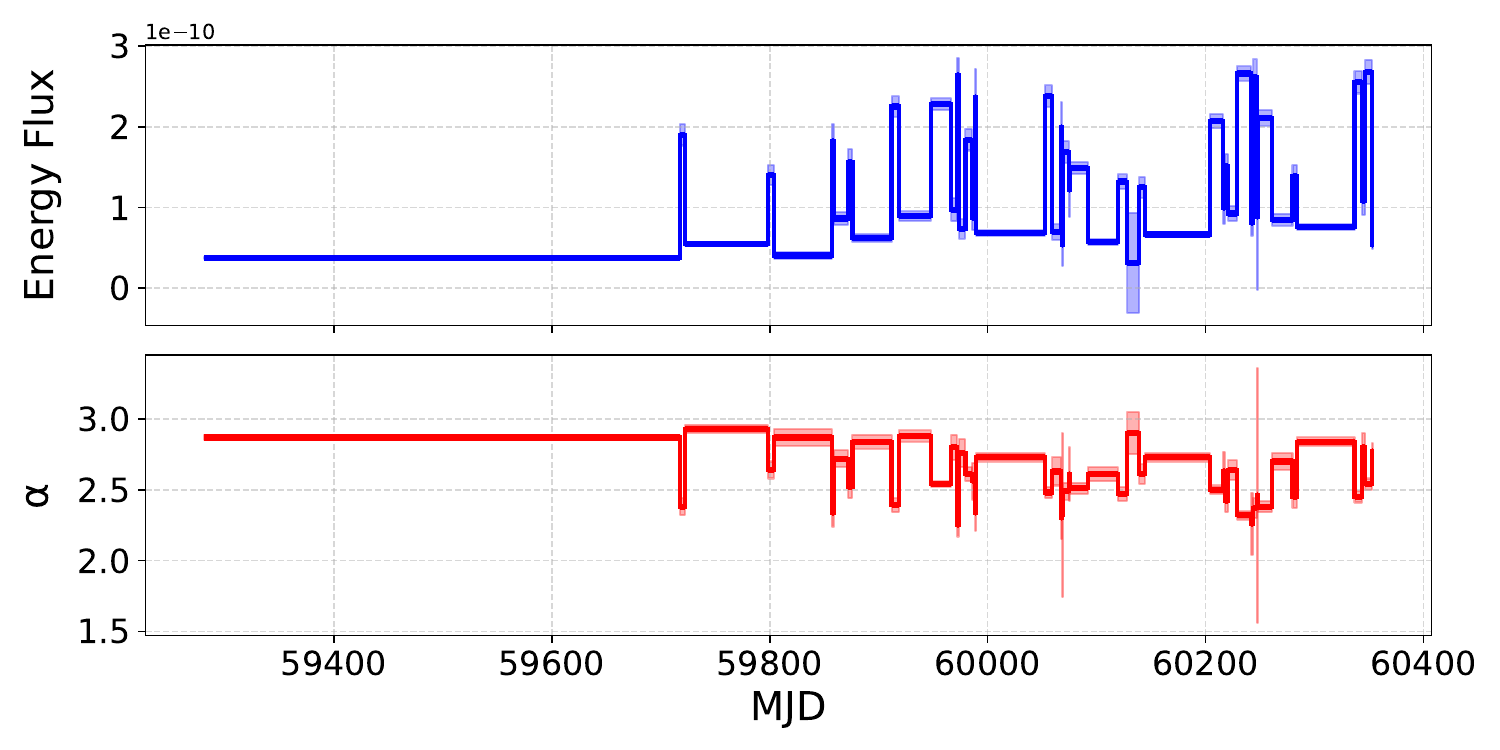}
    \caption{Temporal distribution of flux and spectral index during 3 years observation. Upper panel shows the variation of the integrated energy flux over time for 23 outburst periods and the 24 quiescent intervals in between, while the lower panel presents the evolution of the photon index during the same periods. The blue (upper) and red (lower) curves represent the fitted results for each time segment, with the shaded areas indicating the $1 \sigma$ statistical uncertainty bands.}
    \label{Index_flux_evolution}
\end{figure*}

Additionally, Figure \ref{Index_flux_evolution} illustrates the spectral index and normalized flux across various emission states (from quiet to active) observed over a three-year observation period. The observed anti-correlation between these two parameters suggests underlying changes in the physical mechanisms of the source during different activity states. This may indicate variations in the energy distribution of the radiation region or modifications in particle acceleration processes. Further analysis and interpretation of this behavior are provided in the SED modeling section.

\section{Discussion}

\subsection{different EBL model effects}

\begin{figure*}[!ht]
\centering
\includegraphics[width=0.60\textwidth]{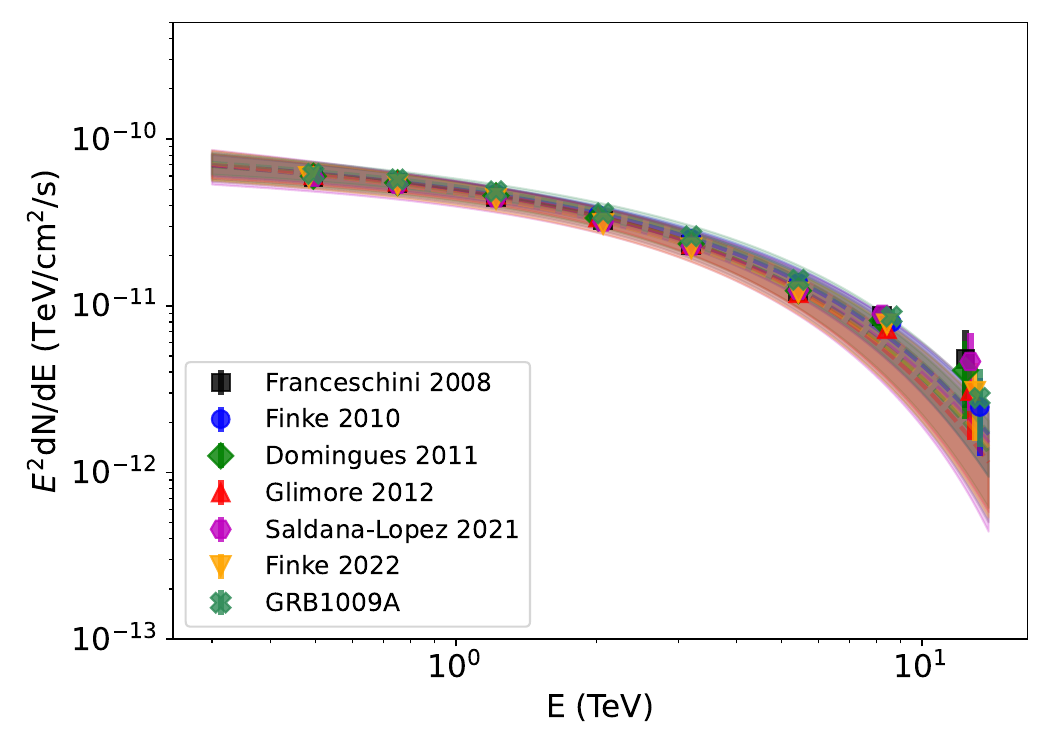}
\caption{Different intrinsic SEDs in the year of 2023 (high state) derived from various EBL models.}
\label{EBL}
\end{figure*}

Nonetheless, we do compare the effects of different EBL models in this study. 
The intrinsic SEDs for Mrk 421, obtained under different EBL models, are compared in Figure \ref{EBL}. All SEDs are modeled with a simple power-law function scaled by an EBL attenuation factor. Relative to the Saldana-Lopez et al. (2021) model, the flux at 10 TeV varies across the models by a factor of 0.85 to 1.15, with the Gilmore et al. (2012) and the LHAASO-GRB221009A models corresponding to the minimum and maximum values, respectively. The intrinsic spectrum indices are in consistent, with the value from $-2.13 \pm 0.05$ to   $-2.20 \pm 0.07$.
Gamma-ray astronomy offers a means to indirectly set limits on the EBL by studying the effects of gamma-ray absorption in the VHE SED of blazars. In fact, there is ongoing work aimed at using all the aboved 23 flares of Mrk~421 to constrain the EBL in our collaboration; however this is beyond the scope of the current paper. 

\subsection{X-ray .vs. VHE and HE vs. VHE flux-flux Correlations}

\begin{figure}[!ht]
   \centering
    \subfigure[MAXI-WCDA]{\includegraphics[width=0.48\textwidth]{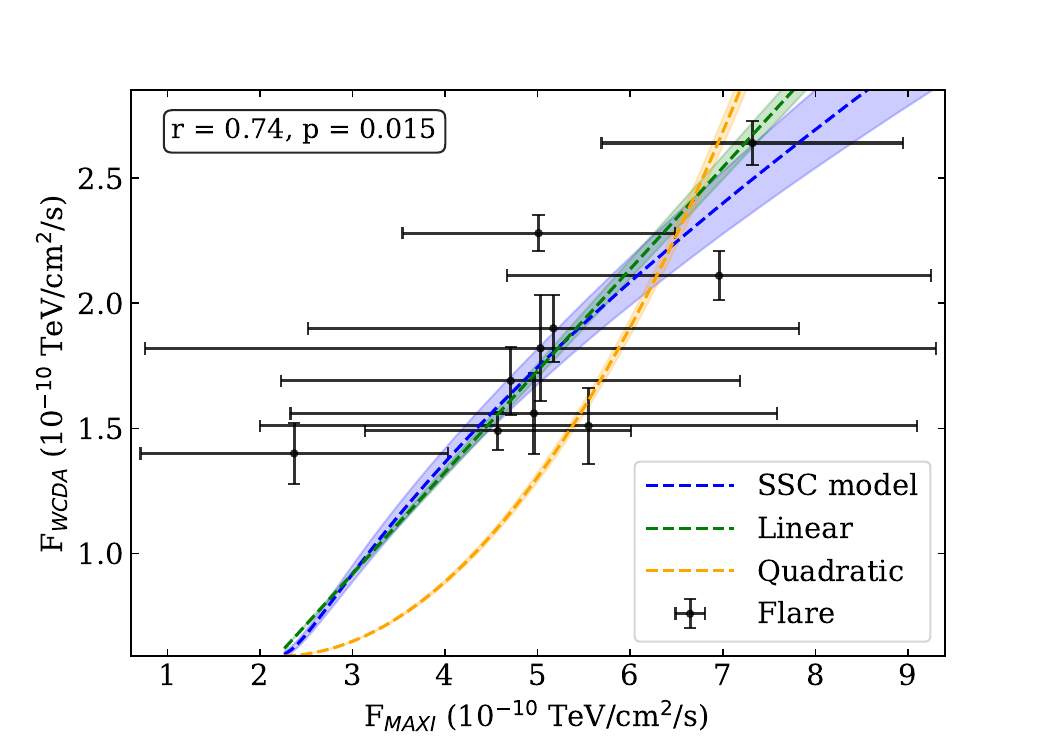}}
    \subfigure[Fermi-WCDA]{\includegraphics[width=0.48\textwidth]{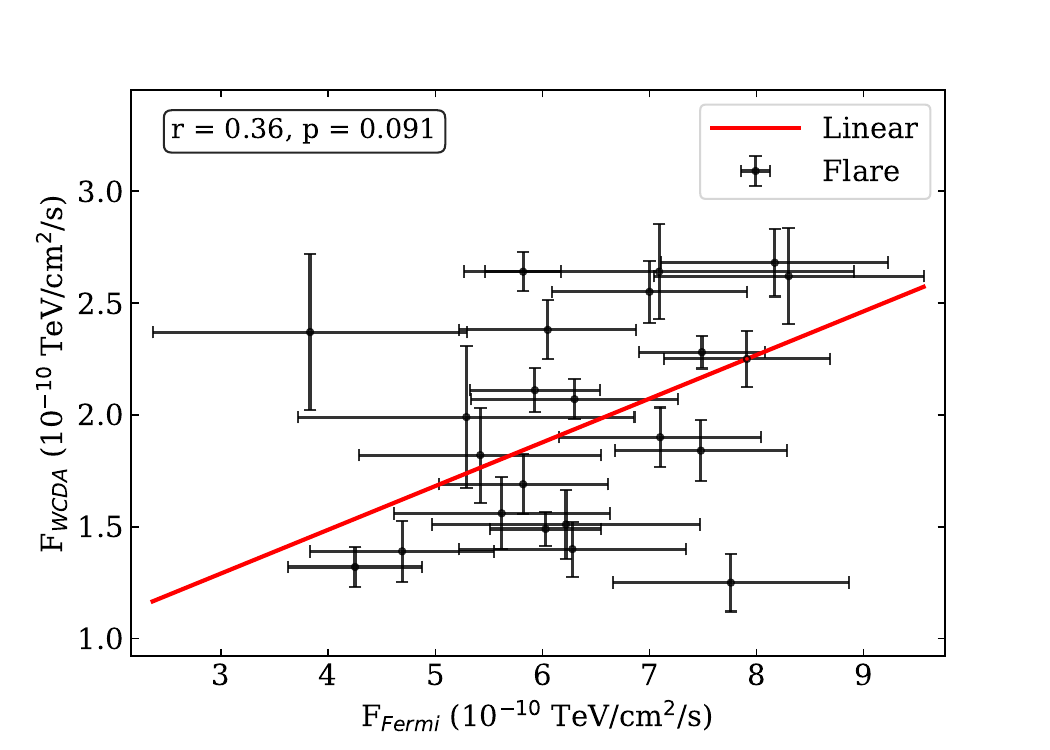}}

    \caption{Left panel: VHE $\gamma$-ray WCDA flux vs. X-ray Maxi flux. The blue line is the flux relationship derived from SSC scenario with Klein-Nishina effect considered, the green and orange lines are the fitted results with linear and quadratic function, respectively. Right panel: VHE $\gamma$-ray WCDA flux vs. HE $\gamma$-ray Fermi flux. The red line is the fitted result with linear function. Each point represents the average radiation flux during an outburst period.}
    \label{fig:flux-flux}
\end{figure}

Flux–flux correlations have been a target for many multi-wavelength campaigns for blazars in both active and low states, as they provide crucial information for distinguishing between different emission models.
In this study, we present the flux–flux correlations of each flare. Specifically, we examine these correlations between the VHE flux (from WCDA, for E$>$1TeV) and the X-ray flux (from MAXI, in the 2–20 keV band),  as shown in panel (a) of Figure \ref{fig:flux-flux}. Our findings reveal an overall strong correlation, suggesting that emissions are likely co-spatial, which is consistent with the positive correlation in the long-term observation as discussed in section 4.2.3. This indicates that electrons with approximately the same energy drive these emissions through synchrotron radiation in X-rays and inverse Compton scattering (SSC) processes producing gamma rays. 

The analysis of the daily data reveals a quadratic fit resulting in $\chi^{2}/dof$ = 1.49, whereas a linear fit yields $\chi^{2}/dof$= 1.21. This suggests that the observed correlation between gamma-ray and X-ray fluxes is better described by a linear relationship rather than a quadratic one.

In the SSC scenario, TeV $\gamma$-ray emission and X-ray emission likely originate from the same population of electrons. Considering that there is a stable emission $F_s$ in the steady state and the outburst state is attributed to the contribution $F_f$ from a separate outburst region. The overall emission is $F=F_s+F_f$. The X-ray emission scales as $F_{f,X}\propto K_e B^2$, while the TeV $\gamma$-ray emission scales as $F_{f,\gamma} \propto K^2_e B^2 R^{-2}$. $K_e$ represents the total number of electron and $B$ and $R$ are the magnetic field and radius of the emission region, respectively. For a jet with a conserved magnetic energy density, one obtais $F_{f,\gamma} \propto F^2_{f,X}$, corresponding to a quadratic relationship. Taking Klein-Nishina effect into modifies the relation to $F_{\gamma}=Cf_{KN}(F_X-F_{s,X})^2 + F_{s,\gamma}$. Higher X-ray emission is accompanied by higher-energy X-ray photons, which leads to a more significant Klein-Nishina effect. Therefore, $f_{KN}$ is negatively correlated to $F_{f,X}$, resulting in a deviation from the expected quadratic relation. As shown in panel (a) of Figure \ref{fig:flux-flux}, the blue line represents the flux relationship in a SSC scenario with Klein-Nishina corrected. The data from MAXI and WCDA during flares exhibit a strong and statistically significant correlation ($r=0.74,\, p=0.015$), which supports the scenario that the X-ray and TeV flaring emissions share a common origin.

Panel (b) of Figure \ref{fig:flux-flux} shows the correlation between WCDA (VHE) and Fermi-LAT (HE) data. Given the small Pearson correlation coefficient and the large p-value ($r=0.36,\, p=0.091$), no statistically significant correlation is found between the observed flaring fluxes in the HE and VHE bands.

\subsection{Modeling of the Temporal Evolution of the multi-wavelength SEDs}

Mrk~421 flux variations can be associated with the intrinsic astrophysical mechanisms of the emission. In this section, we will investigate the major parameters correlated to these variations in the framework of both the one-zone SSC model and the multi-zone model.

A widely used  model to study the spectrum and variability of blazars is the one-zone leptonic model \citep[e.g.,][]{1996ApJ...463..555I, 1997A&A...320...19M}. It is assumed that all the non-thermal radiation of a blazar is generated from a compact spherical plasma blob filled with high-energy electrons inside the jet. The radius of the blob is denoted by $R$. Due to the light travel time effect, the variability timescale produced by the blob cannot be shorter than $(1+z)R/c\delta_D$. The variability timescale is also influenced by the particle injection timescale. When the injection timescale is relatively long, it can lead to flare durations that significantly exceed the light-crossing time of the emission region. Therefore the radius is constrained by the variability timescale $t_{\rm var}$ as $R < ct_{\rm var}\delta_D/(1+z) = 7.5\times10^{16} ({t_{\rm var}}/{1 \rm day}) ({\delta_D}/{30}) {\rm cm}$. The steady electron energy distribution is described by a broken power-law with indices $p_1$ and $p_2$ below and above the break energy $\gamma_b m_e c^2$. The minimum and maximum electron Lorentz factors are fixed to be 10$^2$ and 2$\times$10$^6$, respectively, which are not crucial to the fitting results. Due to relativistic Doppler boosting, the observed emission will be significantly enhanced, characterized by the Doppler factor ${\rm \delta_D} \approx \Gamma$, where $\Gamma$ is the bulk Lorentz factor of the blob. 

\begin{figure}[ht!]
    \subfigure{\includegraphics[width=0.32\textwidth]{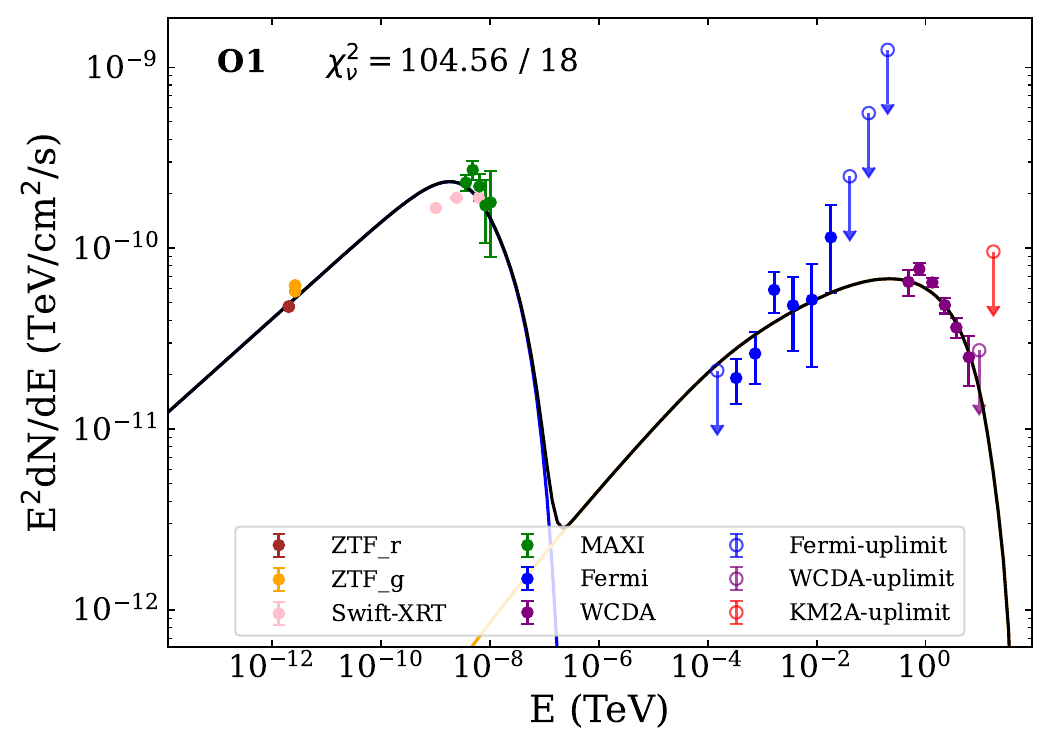}}
    \subfigure{\includegraphics[width=0.32\textwidth]{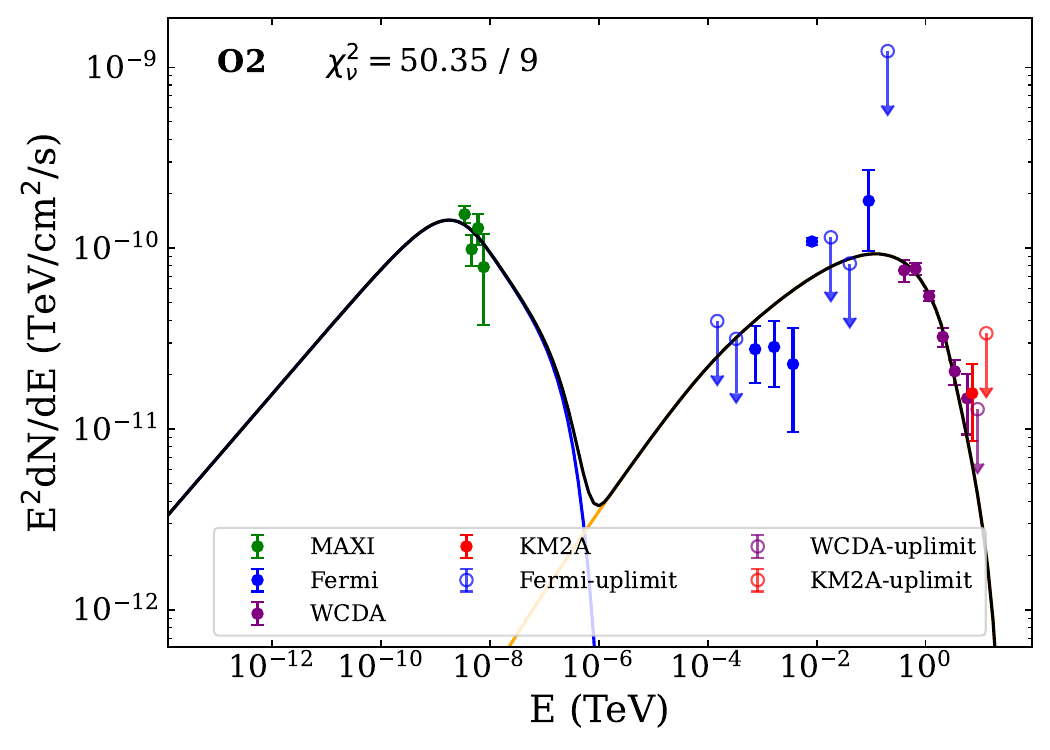}} 
    \subfigure{\includegraphics[width=0.32\textwidth]{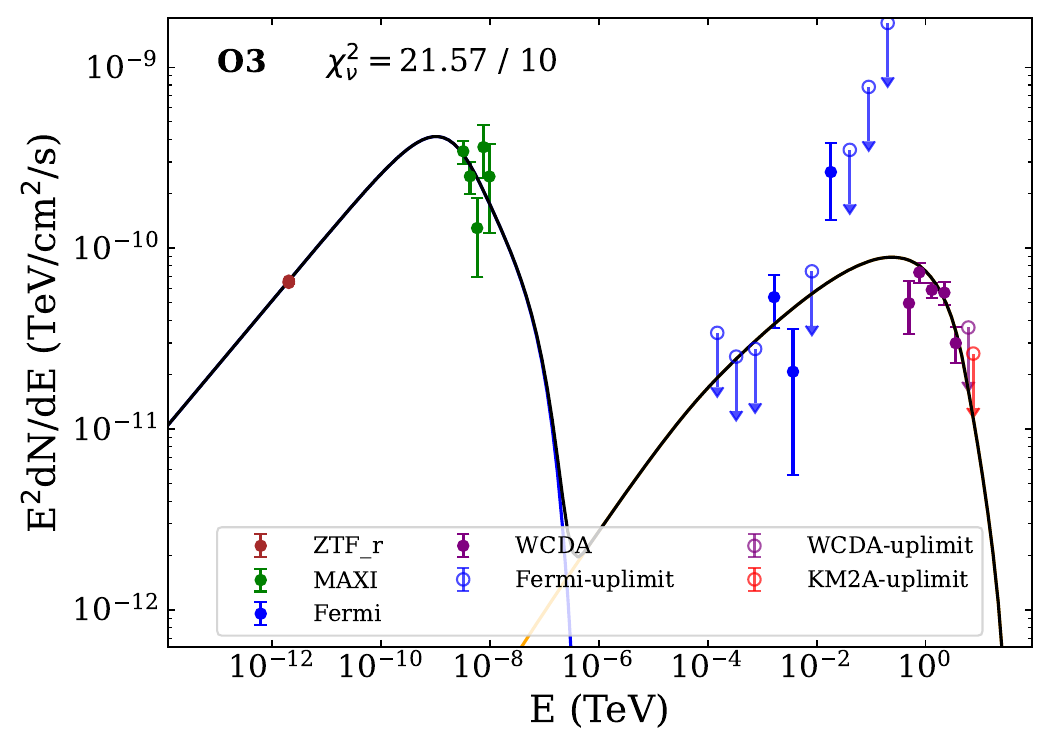}}

    \subfigure{\includegraphics[width=0.32\textwidth]{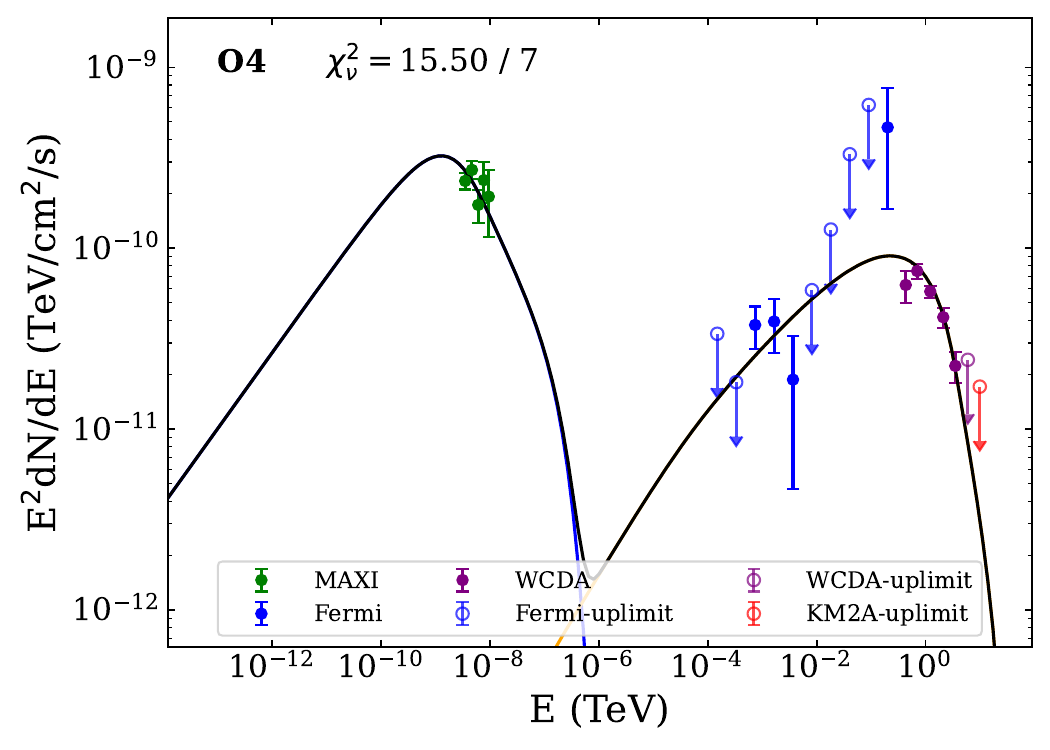}} 
    \subfigure{\includegraphics[width=0.32\textwidth]{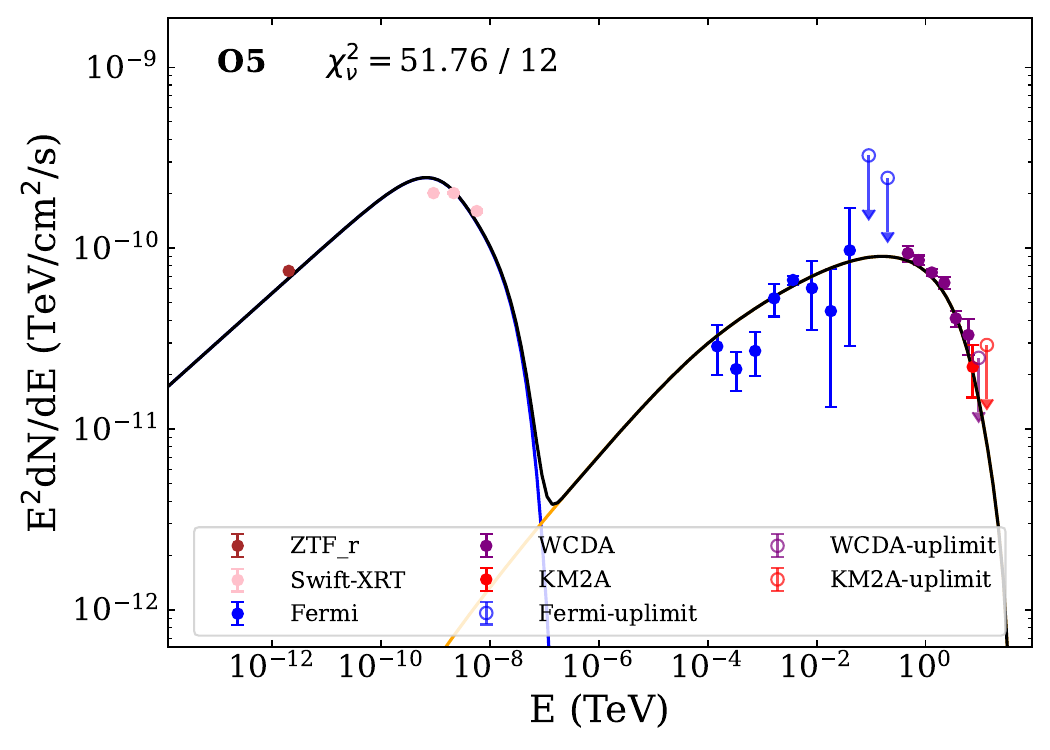}} 
    \subfigure{\includegraphics[width=0.32\textwidth]{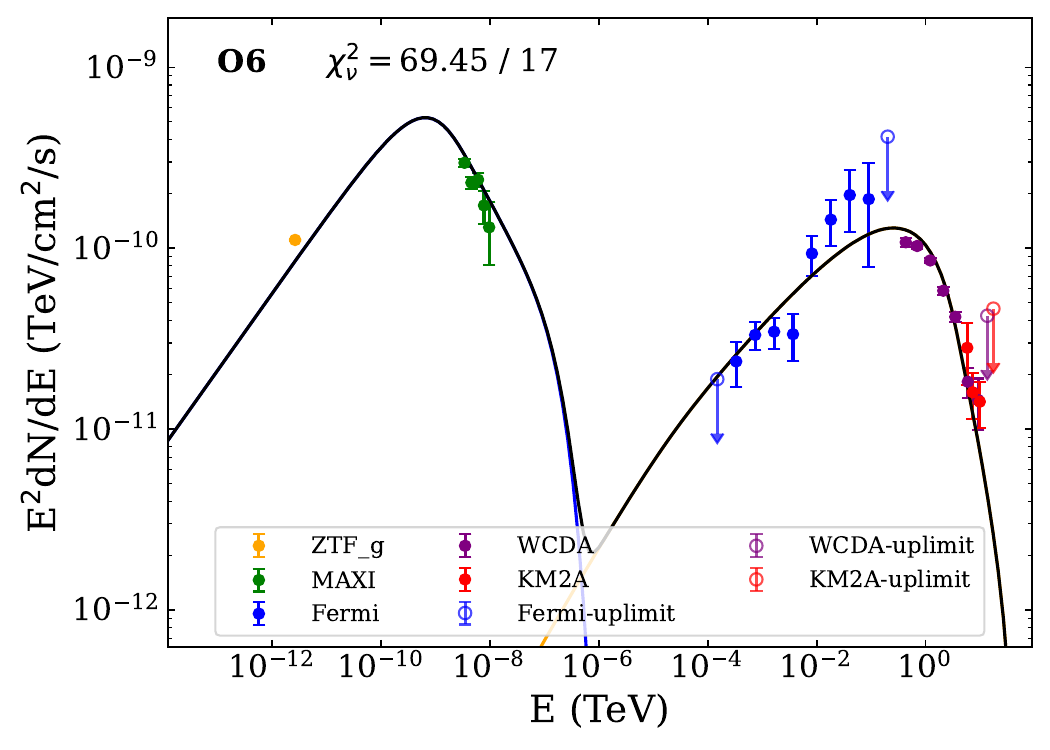}}\\
   
    \subfigure{\includegraphics[width=0.32\textwidth]{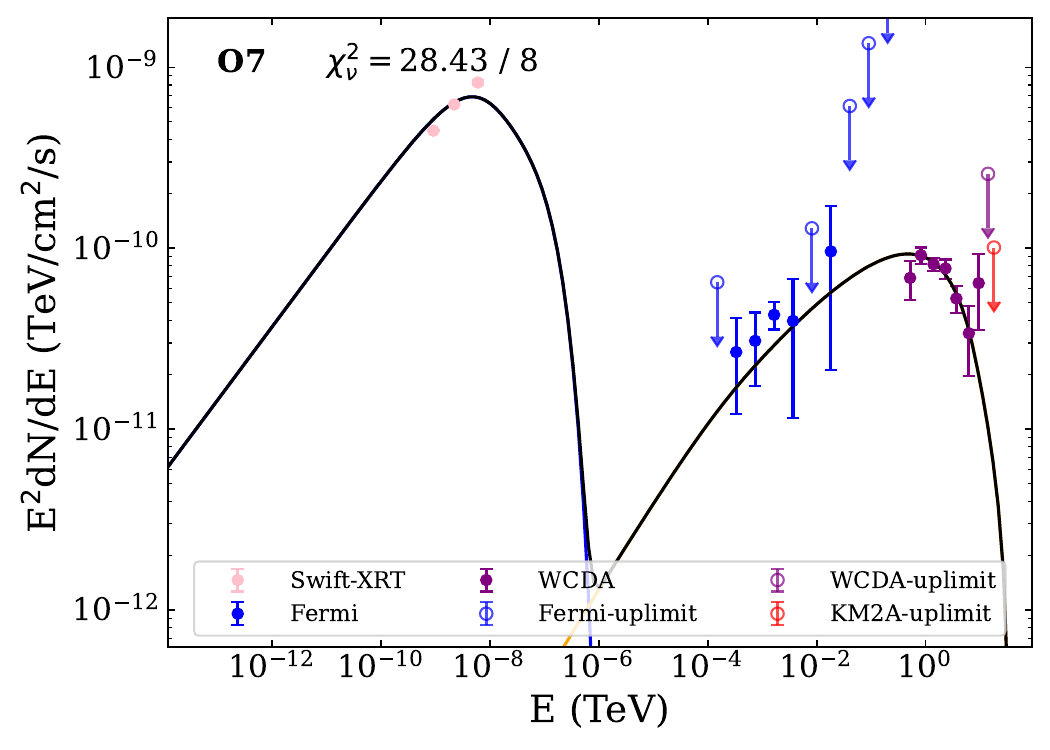}} 
    \subfigure{\includegraphics[width=0.32\textwidth]{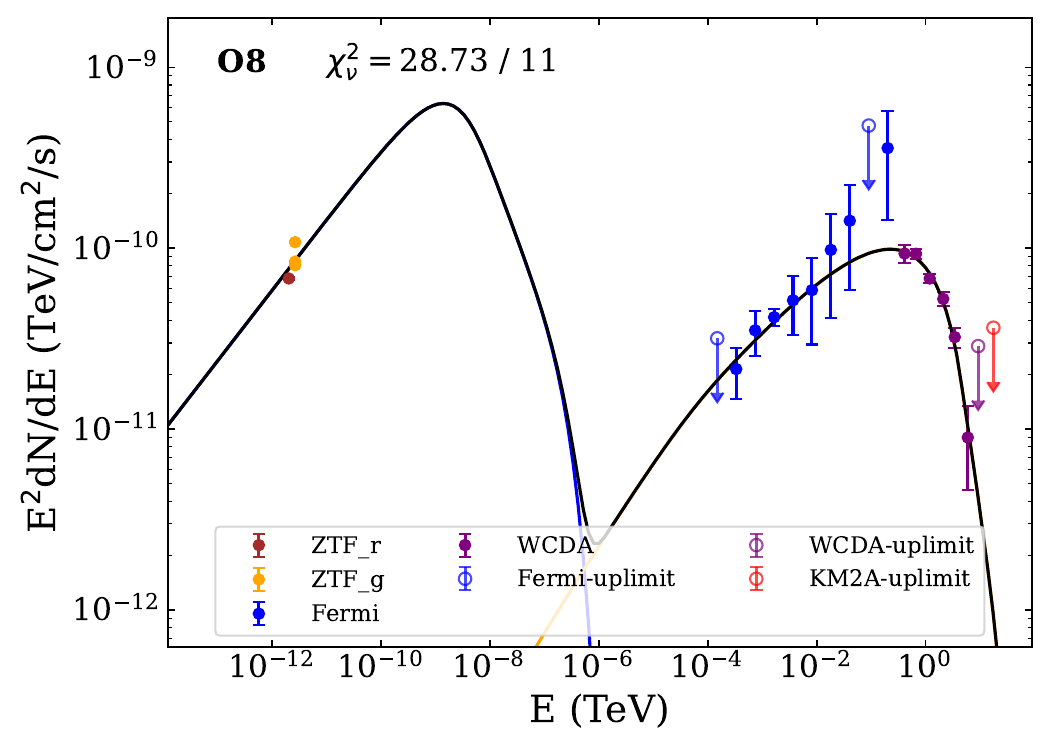}} 
    \subfigure{\includegraphics[width=0.32\textwidth]{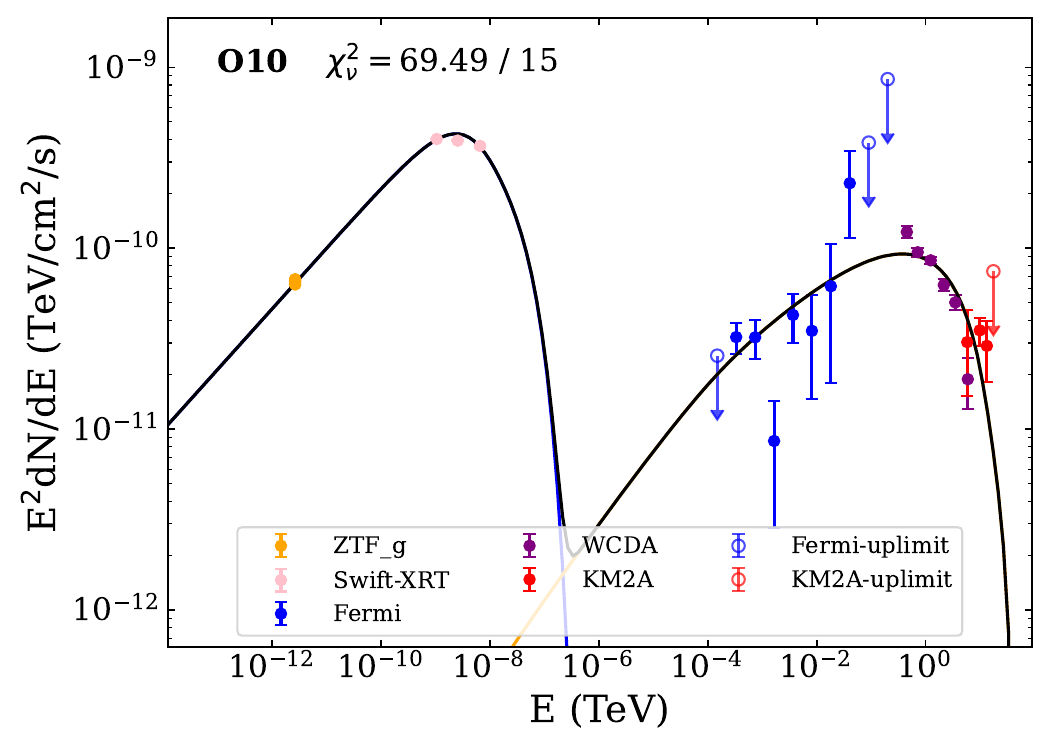}}

    \caption{The fitting SEDs based on the one-zone SSC model during each outburst. The dashed orange and blue lines represent synchrotron radiation and SSC radiation, respectively. The black solid line is the total radiation in the outburst state. The corresponding model parameters are reported in Table \ref{tab:one_zone model_params_list}. 
    }
    \label{fig:one_zone_ssc_model_figure_part1}
\end{figure}

The model parameters are reported in Table \ref{tab:one_zone model_params_list}, and the SEDs are depicted in Figure~\ref{fig:one_zone_ssc_model_figure_part1}. Among the 23 outbursts, 6 of them (O9, O11, O14, 015, 016, and O21) do not have sufficient multiwavelength data and hence model parameters could not be well constrained. 
The results show that the magnetic field is highly associated with the size of the emission region, as shown in the left panel of Figure~\ref{fig:one_zone_model_params}, which can be described by $B\propto R^{-0.97}$. 
 Taking into account the configuration of magnetic field along the jet as $B\propto r^{-1}$ \citep{2021ApJ...906..105C}, the ratio between the blob radius and the transverse radius of the jet is likely a constant, which supports a conical jet structure. For all outbursts, synchrotron spectra seem to peak at the keV energy band and the peak frequency is estimated as $\nu_{\rm p,syn} \propto \delta_D B \gamma^2_b$. As the magnetic field decreases with increasing radius, the break electron energy should be enhanced with a larger radius as shown in the middle panel of Figure~\ref{fig:one_zone_model_params}.
The magnetic power, $P_B = \Gamma^2 \pi R^2 c u_B$, is therefore considered to be conserved for all outbursts, displayed in the right panel of Figure~\ref{fig:one_zone_model_params}. 
Additionally, considering an electron-proton jet, the kinetic luminosity $P_k = \Gamma^2 \pi R^2 c (u_{\rm e} + u_{\rm p}) \propto R^{1.63}$ is supposed to be much higher in bigger blobs, which are likely to be generated in a farther region inside the jet. These may reflect that the capability of electron acceleration is promoted  and energy dissipation becomes more efficient with increasing distance from the central engine in a certain distance range.

In addition, the steady state SED has also been fitted. Compared with outburst states, the fitted Doppler factor is significantly smaller. This suggests that the enhanced radiation flux during outburst states may primarily result from an increase of the Doppler factor. At the same time, a larger Doppler factor in the outburst state can suppress the Klein-Nishina effect in the production of TeV emission, leading to a harder observed TeV spectrum.

Though the one-zone model performs well in explaining the spectra of many blazars, it is a highly simplified model. In fact, blazar jets are extended structures, reaching scales of up to hundreds of parsecs or even longer distance. Numerous bright structures can be observed along the jet in radio, optical, and X-ray bands \citep{2002ApJ...564..683M}, indicating existence of multiple energy dissipation zones and radiation zones along blazar jets. Therefore, the variety in the observed TeV emission among different outbursts may be better understood in the multi-zone framework (i.e., the stochastic dissipation model \citealt{2022PhRvD.105b3005W, 2023MNRAS.526.5054L}; turbulent extreme multi-zone model \citealt{Marscher2014}), where a single outburst may be generated for different reasons, such as the leading blob scenario\citep{Lefa2011}, the jet-in-jet scenario\citep{Giannios2009} and etc.

\begin{figure}[ht!]
    \subfigure{\includegraphics[width=0.32\textwidth]{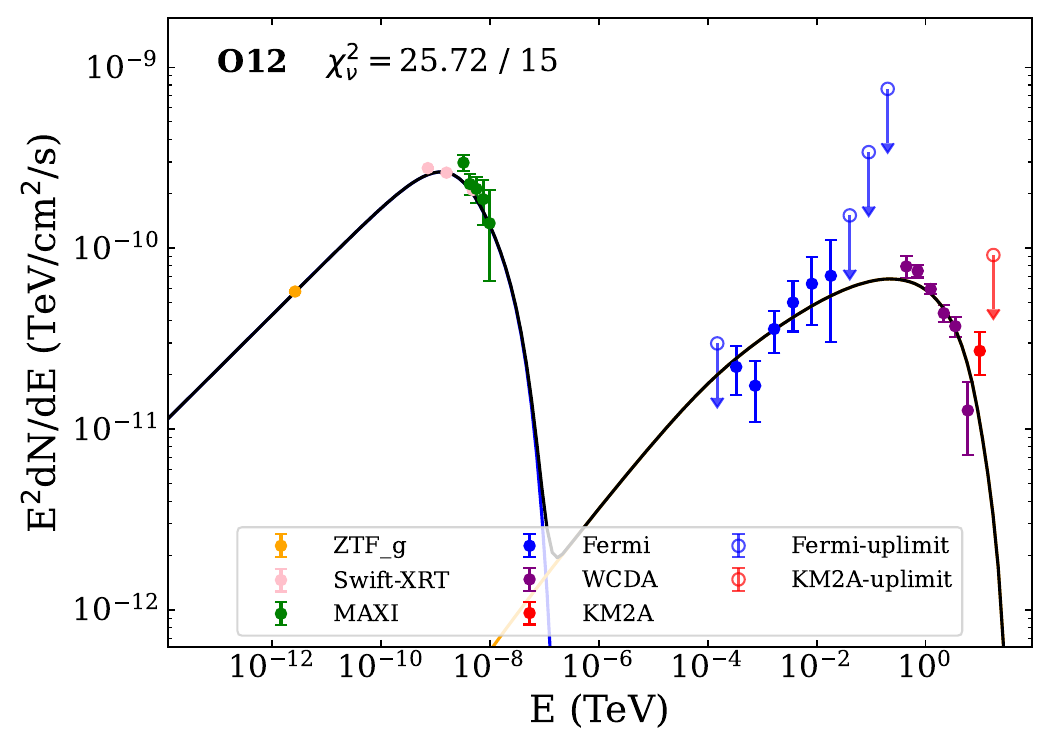}} 
    \subfigure{\includegraphics[width=0.32\textwidth]{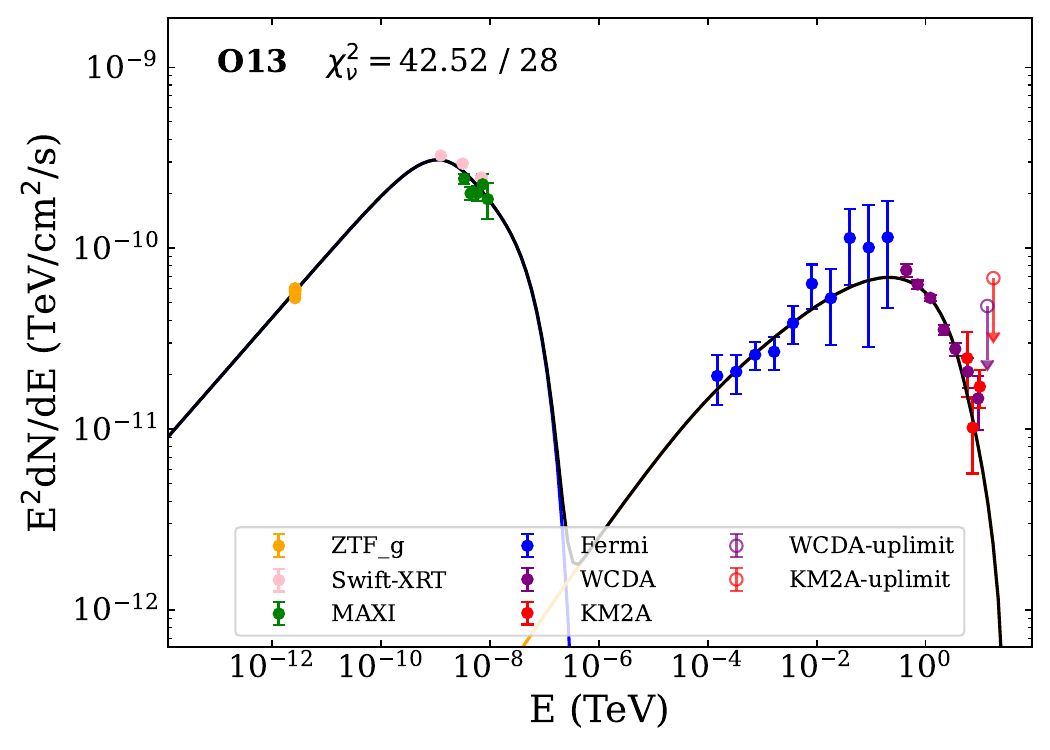}} 
    \subfigure{\includegraphics[width=0.32\textwidth]{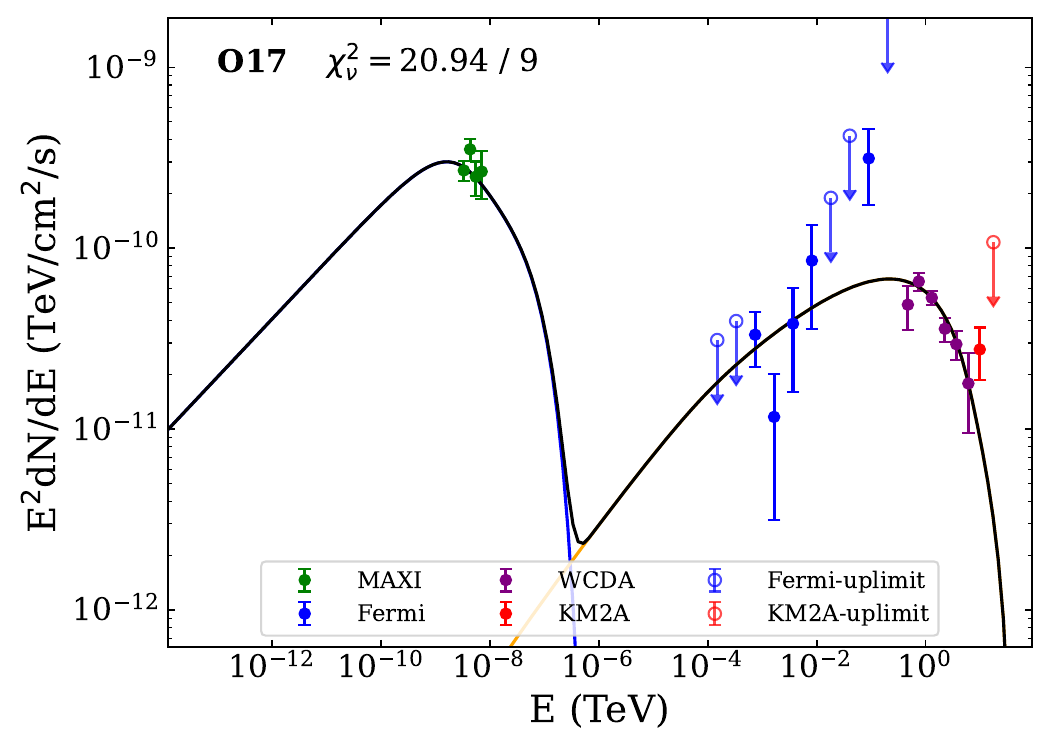}}

    \subfigure{\includegraphics[width=0.32\textwidth]{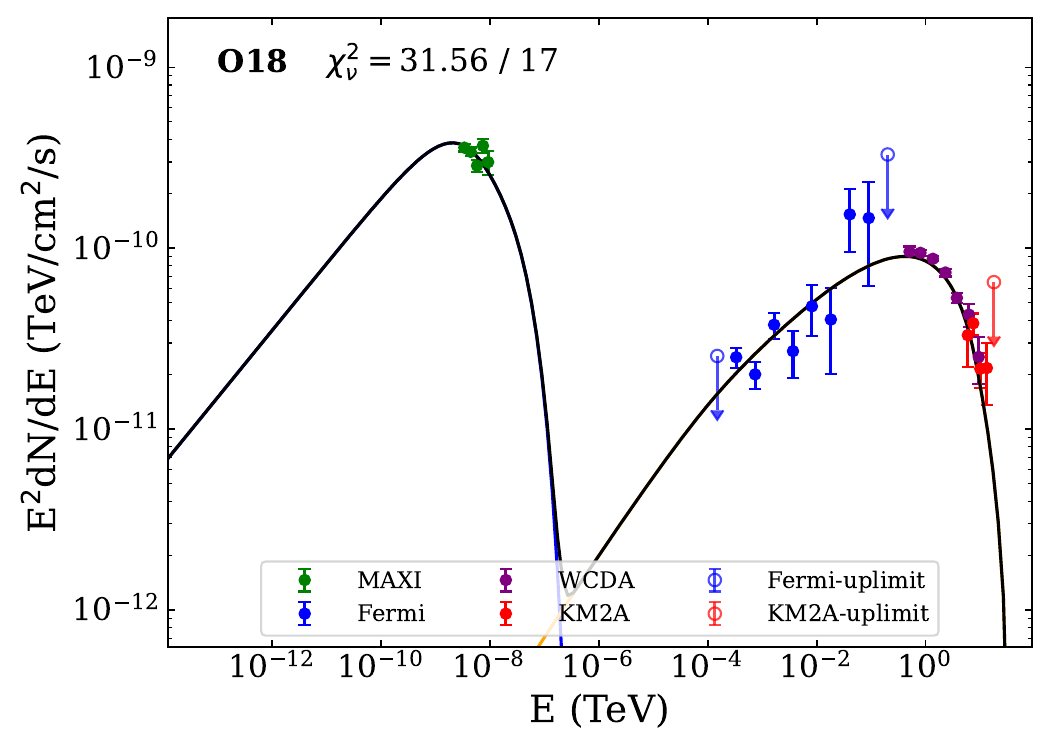}} 
    \subfigure{\includegraphics[width=0.32\textwidth]{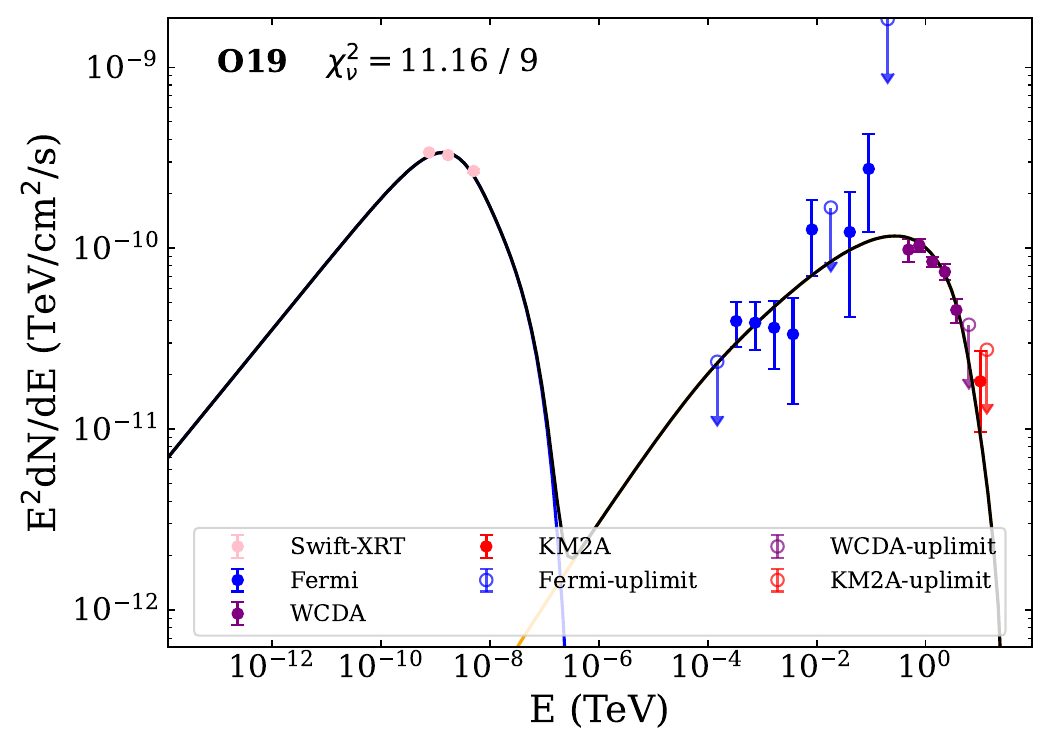}} 
    \subfigure{\includegraphics[width=0.32\textwidth]{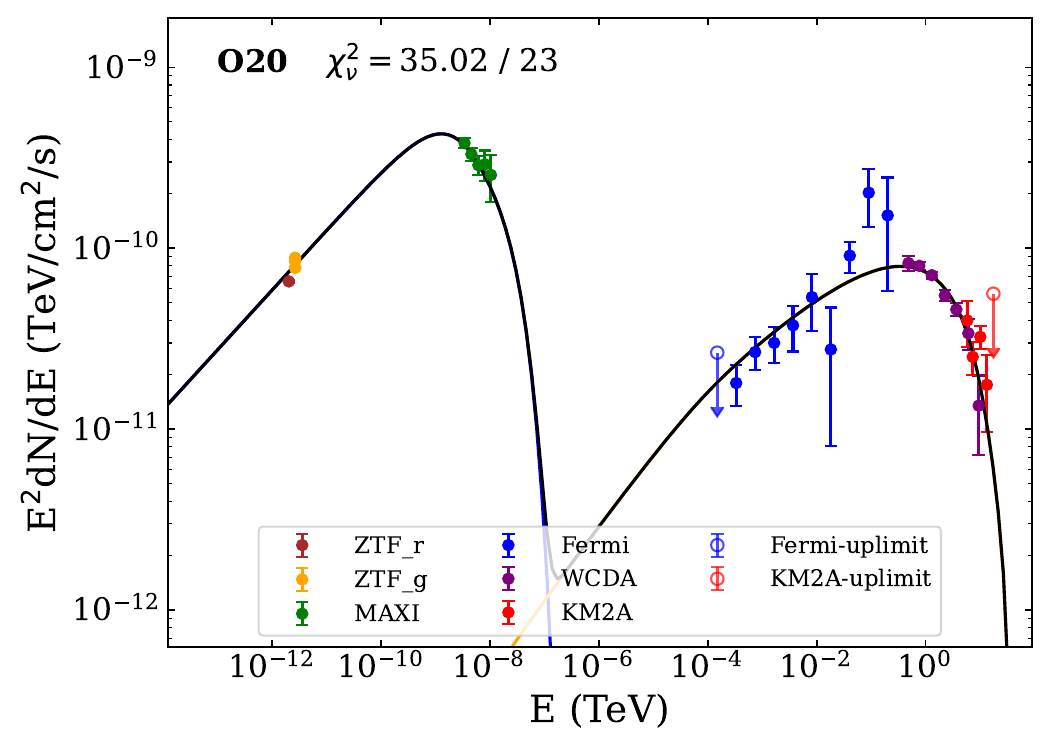}}

    \subfigure{\includegraphics[width=0.32\textwidth]{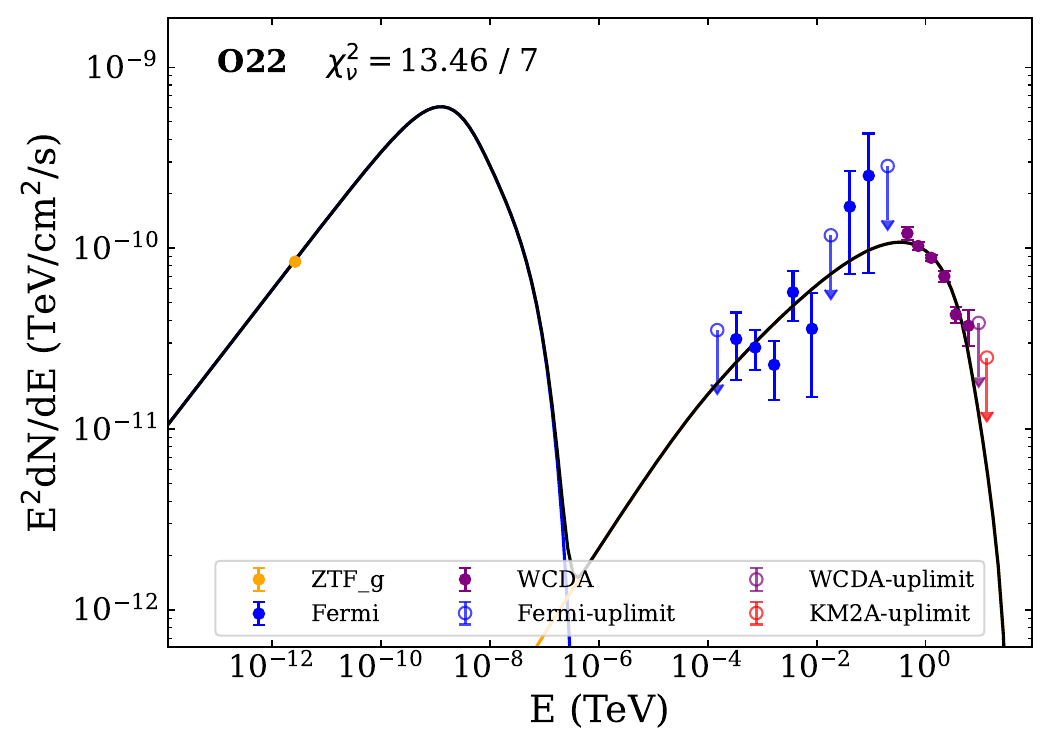}} 
    \subfigure{\includegraphics[width=0.32\textwidth]{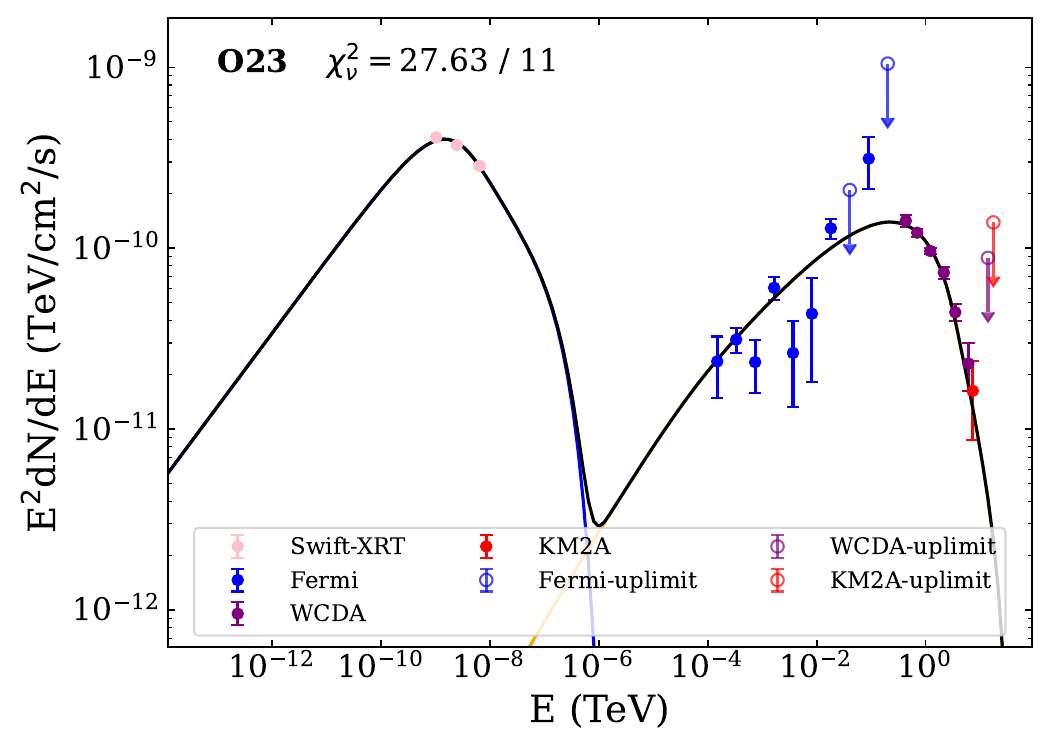}} 
    \subfigure{\includegraphics[width=0.32\textwidth]{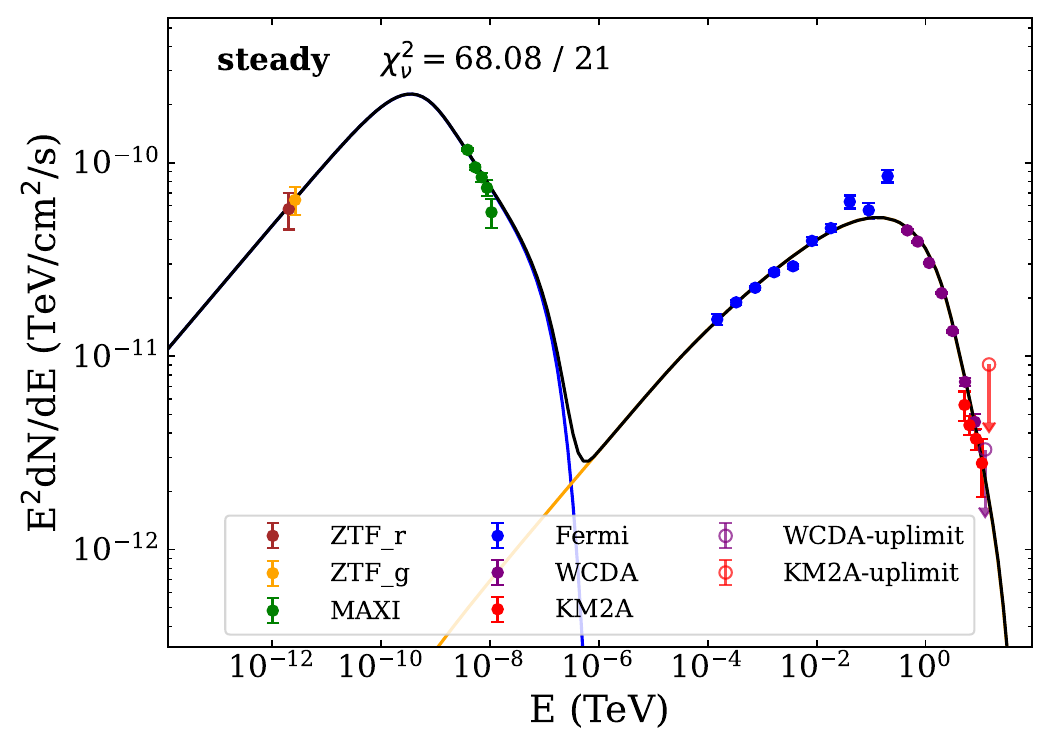}} 
\center{Figure \ref{fig:one_zone_ssc_model_figure_part1} -- Continued}
\end{figure}

\begin{figure}[htpb]
   \centering
    \subfigure{\includegraphics[width=0.32\textwidth]{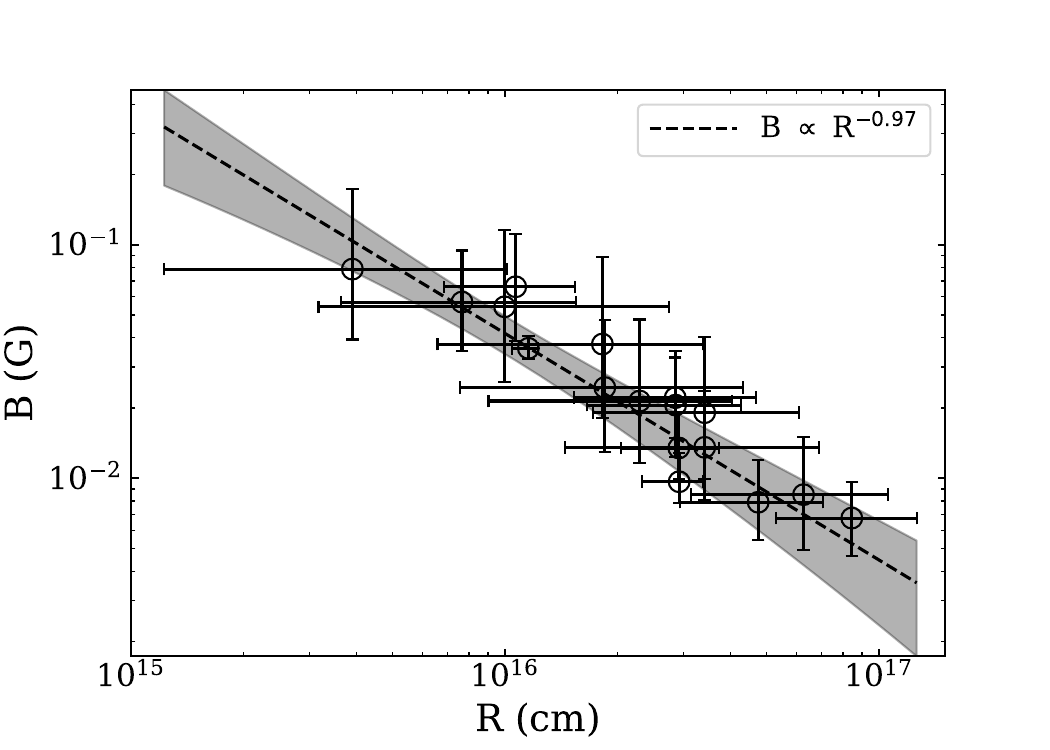}}
    \subfigure{\includegraphics[width=0.32\textwidth]{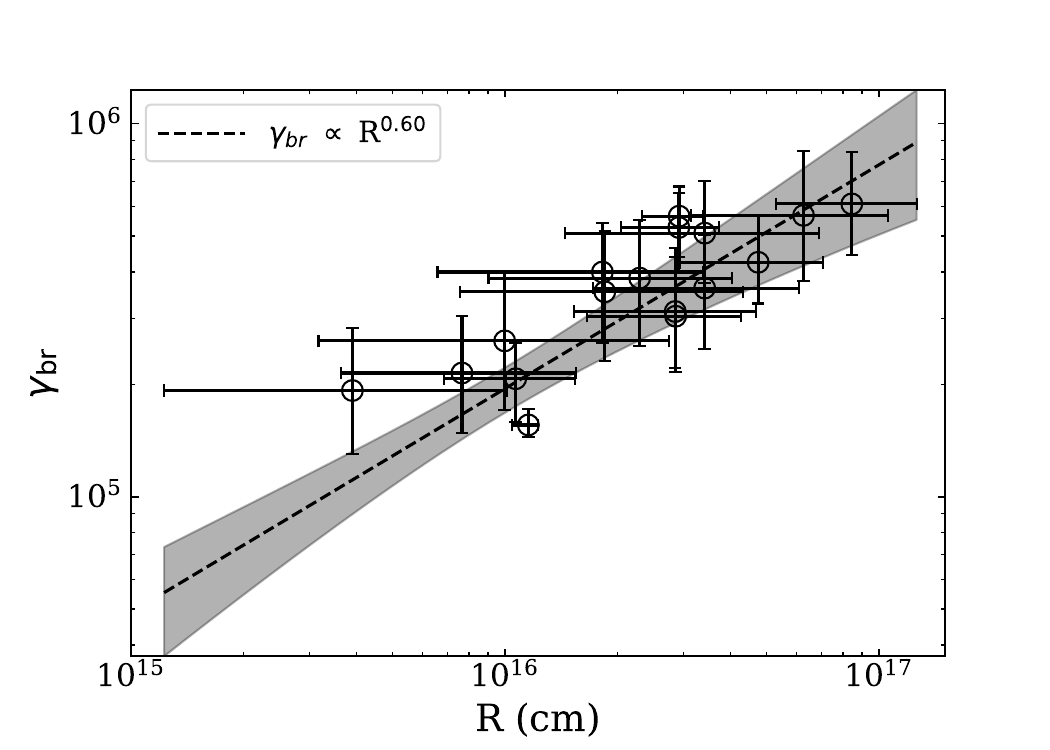}} 
    \subfigure{\includegraphics[width=0.32\textwidth]{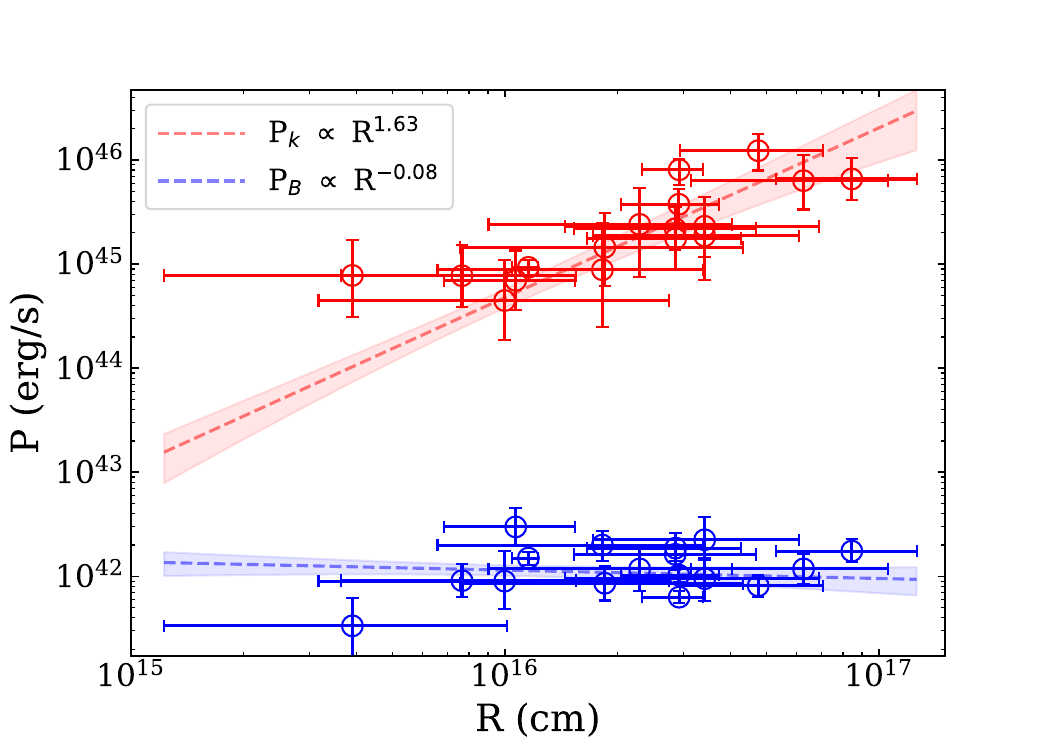}}

    \caption{The correlation between fitting parameters of the one-zone SSC model. Left panel: the blob size ($R$) versus the magnetic field ($B$), $B\propto R^{-0.97}$. Middle Panel: the blob size ($R$) versus the break Lorentz factor of electron ($\gamma_b$), $\gamma_b \propto R^{0.60}$. Right panel: the blob size ($R$) versus kinetic and magnetic luminosity ($P_B$, $P_k$), $P_B \propto R^{-0.08}$, $P_k \propto R^{1.63}$.}
    \label{fig:one_zone_model_params}
\end{figure}

\begin{table}[htpb]
\centering
\caption{The fitting parameters of the one-zone SSC model. Parameters include: spectral indices of the electron energy distribution $p_1$ and $p_2$, break Lorentz factor of electron $\gamma_{b}$, magnetic field of the emission region $B$, radius of the region $R$, Dopplor factor $\delta_{\rm D}$, magnetic power $P_{B}$,  kinetic power $P_{k}$ and the ratio between proton and electron energy density $u_{p}/u_{e}$.}
\resizebox{\textwidth}{!}{
\begin{tabular}{lccccccccc}
\hline
\hline
Epoch & $p_{1}$ & $p_{2}$ & $\gamma_{b}$ & $B$~[G] & $R$~[cm] & $\delta_{\rm D}$ & $P_{B}$~[erg~s$^{-1}$] & $P_{k}$~[erg~s$^{-1}$] & $u_{p}/u_{e}$\\
\hline
O1 & 2.45$_{-0.04}^{+0.02}$ & 4.18$_{-0.33}^{+0.47}$ & 5.65$_{-1.26}^{+1.14}$ $\times$ 10$^{5}$ & 9.67$_{-1.83}^{+3.16}$ $\times$ 10$^{-3}$ &2.92$_{-0.60}^{+0.46}$ $\times$ 10$^{16}$ &47.15$_{-2.57}^{+1.77}$ &6.32$_{-0.72}^{+0.95}$ $\times$ 10$^{41}$ &8.07$_{-2.26}^{+2.13}$ $\times$ 10$^{45}$ &5.83\\ 
O2 & 2.29$_{-0.08}^{+0.06}$ & 3.96$_{-0.26}^{+0.42}$ & 1.93$_{-0.62}^{+0.91}$ $\times$ 10$^{5}$ & 7.87$_{-3.94}^{+9.49}$ $\times$ 10$^{-2}$ &3.90$_{-2.68}^{+6.20}$ $\times$ 10$^{15}$ &33.90$_{-10.06}^{+9.79}$ &3.35$_{-1.64}^{+2.89}$ $\times$ 10$^{41}$ &7.79$_{-4.71}^{+9.22}$ $\times$ 10$^{44}$ &4.61\\ 
O3 & 2.26$_{-0.10}^{+0.08}$ & 4.38$_{-0.48}^{+0.59}$ & 3.05$_{-0.88}^{+1.34}$ $\times$ 10$^{5}$ & 2.06$_{-0.82}^{+1.45}$ $\times$ 10$^{-2}$ &2.86$_{-1.20}^{+1.41}$ $\times$ 10$^{16}$ &41.64$_{-6.17}^{+5.09}$ &1.87$_{-0.45}^{+0.75}$ $\times$ 10$^{42}$ &1.76$_{-0.89}^{+1.81}$ $\times$ 10$^{45}$ &4.25\\ 
O4 & 2.14$_{-0.10}^{+0.09}$ & 4.33$_{-0.41}^{+0.56}$ & 2.62$_{-0.91}^{+1.38}$ $\times$ 10$^{5}$ & 5.43$_{-2.84}^{+6.12}$ $\times$ 10$^{-2}$ &9.97$_{-6.80}^{+17.53}$ $\times$ 10$^{15}$ &32.49$_{-10.65}^{+9.80}$ &9.00$_{-4.14}^{+8.50}$ $\times$ 10$^{41}$ &4.47$_{-2.60}^{+6.41}$ $\times$ 10$^{44}$ &3.30\\ 
O5 & 2.46$_{-0.04}^{+0.02}$ & 3.88$_{-0.24}^{+0.45}$ & 4.25$_{-0.95}^{+1.41}$ $\times$ 10$^{5}$ & 7.90$_{-2.46}^{+4.14}$ $\times$ 10$^{-3}$ &4.75$_{-1.82}^{+2.31}$ $\times$ 10$^{16}$ &40.86$_{-5.60}^{+5.01}$ &8.13$_{-1.76}^{+2.26}$ $\times$ 10$^{41}$ &1.23$_{-0.44}^{+0.53}$ $\times$ 10$^{46}$ &5.87\\ 
O6 & 2.14$_{-0.03}^{+0.03}$ & 4.14$_{-0.11}^{+0.13}$ & 1.56$_{-0.11}^{+0.16}$ $\times$ 10$^{5}$ & 3.60$_{-0.35}^{+0.46}$ $\times$ 10$^{-2}$ &1.15$_{-0.11}^{+0.07}$ $\times$ 10$^{16}$ &48.86$_{-1.42}^{+0.77}$ &1.50$_{-0.21}^{+0.23}$ $\times$ 10$^{42}$ &9.31$_{-1.61}^{+1.64}$ $\times$ 10$^{44}$ &3.52\\ 
O7 & 2.21$_{-0.15}^{+0.12}$ & 3.71$_{-0.12}^{+0.23}$ & 4.00$_{-1.42}^{+1.40}$ $\times$ 10$^{5}$ & 3.76$_{-1.95}^{+5.13}$ $\times$ 10$^{-2}$ &1.82$_{-1.16}^{+1.55}$ $\times$ 10$^{16}$ &39.68$_{-9.64}^{+6.46}$ &2.00$_{-0.60}^{+0.75}$ $\times$ 10$^{42}$ &0.88$_{-0.63}^{+1.60}$ $\times$ 10$^{44}$ &3.81\\ 
O8 & 2.21$_{-0.09}^{+0.08}$ & 4.49$_{-0.42}^{+0.54}$ & 2.07$_{-0.48}^{+0.51}$ $\times$ 10$^{5}$ & 6.61$_{-2.75}^{+4.49}$ $\times$ 10$^{-2}$ &1.07$_{-0.38}^{+0.47}$ $\times$ 10$^{16}$ &42.68$_{-5.86}^{+4.50}$ &2.99$_{-1.02}^{+1.56}$ $\times$ 10$^{42}$ &6.98$_{-3.39}^{+6.53}$ $\times$ 10$^{44}$ &3.99\\ 
O10 & 2.33$_{-0.05}^{+0.03}$ & 4.15$_{-0.30}^{+0.46}$ & 5.27$_{-1.26}^{+1.24}$ $\times$ 10$^{5}$ & 1.35$_{-0.36}^{+0.56}$ $\times$ 10$^{-2}$ &2.91$_{-0.88}^{+0.81}$ $\times$ 10$^{16}$ &44.66$_{-4.38}^{+3.15}$ &1.03$_{-0.14}^{+0.20}$ $\times$ 10$^{42}$ &3.76$_{-1.26}^{+1.50}$ $\times$ 10$^{45}$ &4.85\\ 
O12 & 2.38$_{-0.06}^{+0.05}$ & 4.01$_{-0.41}^{+0.69}$ & 5.68$_{-1.90}^{+2.76}$ $\times$ 10$^{5}$ & 8.51$_{-3.56}^{+6.50}$ $\times$ 10$^{-3}$ &6.28$_{-3.14}^{+4.30}$ $\times$ 10$^{16}$ &36.11$_{-6.95}^{+7.58}$ &1.18$_{-0.34}^{+0.46}$ $\times$ 10$^{42}$ &6.34$_{-2.99}^{+4.88}$ $\times$ 10$^{45}$ &5.27\\ 
O13 & 2.33$_{-0.07}^{+0.07}$ & 3.71$_{-0.18}^{+0.41}$ & 3.14$_{-0.92}^{+1.51}$ $\times$ 10$^{5}$ & 2.21$_{-0.73}^{+1.08}$ $\times$ 10$^{-2}$ &2.85$_{-1.32}^{+1.85}$ $\times$ 10$^{16}$ &35.30$_{-7.19}^{+8.36}$ &1.63$_{-0.47}^{+0.64}$ $\times$ 10$^{42}$ &2.19$_{-0.83}^{+1.42}$ $\times$ 10$^{45}$ &4.85\\ 
O17 & 2.34$_{-0.13}^{+0.09}$ & 3.88$_{-0.23}^{+0.46}$ & 3.85$_{-1.32}^{+1.65}$ $\times$ 10$^{5}$ & 2.14$_{-0.98}^{+2.65}$ $\times$ 10$^{-2}$ &2.29$_{-1.39}^{+1.76}$ $\times$ 10$^{16}$ &41.07$_{-9.38}^{+5.62}$ &1.18$_{-0.46}^{+0.66}$ $\times$ 10$^{42}$ &2.41$_{-1.66}^{+2.94}$ $\times$ 10$^{45}$ &4.95\\ 
O18 & 2.25$_{-0.07}^{+0.06}$ & 3.88$_{-0.22}^{+0.45}$ & 5.09$_{-1.34}^{+1.93}$ $\times$ 10$^{5}$ & 1.36$_{-0.55}^{+1.00}$ $\times$ 10$^{-2}$ &3.42$_{-1.97}^{+3.48}$ $\times$ 10$^{16}$ &38.18$_{-9.74}^{+7.46}$ &9.51$_{-3.66}^{+5.65}$ $\times$ 10$^{41}$ &2.31$_{-1.15}^{+2.08}$ $\times$ 10$^{45}$ &4.10\\ 
O19 & 2.22$_{-0.09}^{+0.08}$ & 4.30$_{-0.41}^{+0.57}$ & 3.54$_{-1.23}^{+1.62}$ $\times$ 10$^{5}$ & 2.45$_{-1.15}^{+2.31}$ $\times$ 10$^{-2}$ &1.85$_{-1.09}^{+2.47}$ $\times$ 10$^{16}$ &35.68$_{-8.84}^{+8.13}$ &8.59$_{-2.72}^{+3.90}$ $\times$ 10$^{41}$ &1.45$_{-0.84}^{+1.67}$ $\times$ 10$^{45}$ &3.93\\ 
O20 & 2.35$_{-0.06}^{+0.06}$ & 4.07$_{-0.43}^{+0.63}$ & 6.09$_{-1.65}^{+2.30}$ $\times$ 10$^{5}$ & 6.76$_{-2.09}^{+2.90}$ $\times$ 10$^{-3}$ &8.45$_{-3.14}^{+4.15}$ $\times$ 10$^{16}$ &40.38$_{-6.16}^{+5.65}$ &1.75$_{-0.38}^{+0.55}$ $\times$ 10$^{42}$ &6.58$_{-2.49}^{+3.98}$ $\times$ 10$^{45}$ &4.99\\ 
O22 & 2.23$_{-0.11}^{+0.09}$ & 4.34$_{-0.43}^{+0.60}$ & 3.62$_{-1.13}^{+1.41}$ $\times$ 10$^{5}$ & 1.91$_{-0.92}^{+2.12}$ $\times$ 10$^{-2}$ &3.41$_{-1.69}^{+2.68}$ $\times$ 10$^{16}$ &39.65$_{-6.81}^{+5.95}$ &2.25$_{-0.82}^{+1.44}$ $\times$ 10$^{42}$ &1.87$_{-1.17}^{+2.53}$ $\times$ 10$^{45}$ &4.03\\ 
O23 & 2.18$_{-0.09}^{+0.09}$ & 4.10$_{-0.36}^{+0.54}$ & 2.15$_{-0.66}^{+0.90}$ $\times$ 10$^{5}$ & 5.69$_{-2.18}^{+3.77}$ $\times$ 10$^{-2}$ &7.66$_{-4.02}^{+7.83}$ $\times$ 10$^{15}$ &37.98$_{-8.51}^{+7.25}$ &9.10$_{-2.81}^{+3.99}$ $\times$ 10$^{41}$ &7.74$_{-3.89}^{+7.61}$ $\times$ 10$^{44}$ &3.73\\ 
\hline
steady & 2.33$_{-0.02}^{+0.02}$ & 3.84$_{-0.09}^{+0.09}$ & 2.13$_{-0.40}^{+0.52}$ $\times$ 10$^{5}$ & 2.29$_{-0.63}^{+0.90}$ $\times$ 10$^{-2}$ &7.98$_{-4.13}^{+7.91}$ $\times$ 10$^{16}$ &20.59$_{-4.29}^{+5.09}$ &5.21$_{-1.76}^{+2.20}$ $\times$ 10$^{42}$ &2.00$_{-0.46}^{+0.52}$ $\times$ 10$^{45}$ & 4.97\\ 
\hline
$\chi^2$/d.o.f & - & - & - & - & - & - & - & - & 696.9/254 \\
\hline
\hline 
\end{tabular}
}
\label{tab:one_zone model_params_list}
\end{table}

\section{Summary and conclusion}

In this paper, we present deep observations of the blazar Mrk~421 over three continuous years, from March 2021 to March 2024. The total significance of the source reaches 215 $\sigma$. These observations include 23 multi-timescale outbursts and the quasi-quiet states. For the first time, an unbiased sky survey has collected such an extensive sample of outbursts, providing an unprecedented opportunity to investigate the origin of flux variability and understand the emission mechanism. The main results of this work are summarized below.

\begin{itemize}

\item Over the three-year observation period of LHAASO, Mrk~421 exhibited increasing levels of activity in the TeV band. It has entered an active phase since 2022. The fractional variability rose from 52\% in 2021 to 91\% in 2024 according to LHAASO-WCDA observations. At lower energy observed by Fermi-LAT ($0.1-100$\,GeV) and higher energy observed by LHAASO-KM2A ($6-40$\,TeV), the fractional variability are notably weaker.

\item The temporal correlation of light curves between TeV and other wavelengths was analyzed. The variation of the X-ray flux is clearly correlated with the TeV gamma-ray flux, which is consistent with many previous observations in the past decades, supporting the idea that X-ray and VHE gamma-ray emissions originate from the same population of electrons. The GeV gamma-ray flux appears to be moderately correlated with the TeV gamma-ray flux. This result is consistent with the previous observation by ARGO-YBJ. 

\item The long-term VHE gamma-ray spectrum of Mrk~421 can be well described by an exponential cutoff power-law model, extending the maximum observed energy to 13 TeV. After the correcting for EBL absorption, the normalized flux at 3 TeV increases by approximately 30\%, and the spectrum hardens by $\Delta\Gamma\approx$ 0.12. Different EBL models give consistent absorption predictions within the observed energy range, having only a moderate influence on the derived spectral shape. The cutoff energies of the intrinsic and observed spectra remain consistent within uncertainties.

\item  Based on continuous observations in the TeV energy range by LHAASO-WCDA, we applied an iterative signal analysis combined with the Bayesian Block algorithm to identify outburst episodes of Mrk~421 systematically. We detected 23 significant outbursts with energy fluxes exceeding $3.66\times10^{-11}\ \mathrm{TeV cm^{-2} s^{-1}}$ in the 0.4–20 TeV lasting from 1 to 19 days. The spectrum of each outburst can be well described by a power-law model, with integral fluxes in $0.4–20$\, TeV varying from 1.2 to 2.6 Crab Units. The fitted normalization and spectral index exhibit a clear ‘harder-when-brighter’ trend, consistent with previous observations.

\item Thanks to the continuous and unbiased observations by WCDA in the TeV energy range, this study provides a more precise estimate of the outburst duty cycle than previous IACT-based measurements. The three-year-long monitoring enables us to calculate the gamma-ray duty cycle of Mrk~421 across different flux thresholds. Between March 2021 and March 2024, a total of 23 outbursts were identified with energy fluxes exceeding $3.66\times10^{-11}\ \mathrm{TeV/cm^{2}/s}$ in the 0.4–20 TeV range. Using this value as the flux threshold, the duty cycle is measured to be 14.6\%; increasing the threshold to $2.28\times10^{-10}\ \mathrm{TeV/cm^{2}/s}$ would reduce the duty cycle to 5\%.

\item Using the Fred function to characterize the rising ($T_{\rm r}$) and decay  timescales ($T_{\rm d}$), we categorize the 23 outbursts into four groups:
Group A (O1, O18, O22) has $T_{\rm r} > T_{\rm d}$; Group B (O2, O5, O8, O10, O17) has 
$T_{\rm r} < T_{\rm d}$; Group C (O3, O4, O7, O9, O21) has $T_{\rm r} \sim T_{\rm d}$, 
often on intraday timescales; Group-D, including the rest 10 outbursts, forms an ‘outburst forest’ with complex, multi-peaked light curves that cannot be decomposed into individual bursts given our measurement statistics. The variability behavior on different timescales can be used to constrain the spatial scale of the emission regions.

\item Modeling with a one-zone SSC framework successfully explains multi-epoch outbursts of Mrk~421, and suggests a conical jet with conserved magnetic power. Outbursts modeled with larger radiation zones generally require higher electron injection luminosities. It may imply that the energy has been accumulated in the jet before being dissipated, leading to  greater energy release at larger distances from the central engine.

\end{itemize}

\section{Acknowledgements}

The LHAASO Observatory, including its detector systems, was designed and constructed by the LHAASO project team and is operated and maintained by the LHAASO operations team. We sincerely thank all members of both teams, with special appreciation for those who work year-round at the LHAASO site at an altitude exceeding 4,400 meters. Their sustained dedication ensures the reliable operation of the detector systems and essential infrastructure, including the power supply. We sincerely acknowledge the Chengdu Management Committee of Tianfu New Area for its sustained financial support of research based on LHAASO data. We also thank the National High Energy Physics Data Center for providing the computing resources and data services that made the analysis in this work possible.

This work was supported by the National Key R\&D Program of China (Grants No. 2024YFA1611401--2024YFA1611404); the National Natural Science Foundation of China (Grants No. 12393851--12393854, 12173039, 12205314, 12105301, 12305120, 12261160362, 12105294, U1931201, and 12375107); the Department of Science and Technology of Sichuan Province (Grant No. 24NSFSC2319); the Project for Young Scientists in Basic Research of the Chinese Academy of Sciences (Grant No. YSBR-061); and, in Thailand, by the National Science and Technology Development Agency (NSTDA) and the National Research Council of Thailand (NRCT) under the High-Potential Research Team Grant Program (Grant No. N42A650868).

The optical data of this work is from ZTF. ZTF is supported by the National Science Foundation under grants Nos. AST-1440341 and AST-2034437 and a collaboration including current partners Caltech, IPAC, the Weizmann Institute for Science, The Oskar Klein Center at Stockholm University, the University of Maryland, Deutsches Elektronen-Synchrotron and Humboldt University, the TANGO Consortium of Taiwan, the University of Wisconsin at Milwaukee, Trinity College Dublin, Lawrence Livermore National Laboratories, IN2P3, University of Warwick, Ruhr University Bochum, Northwestern University and former partners the University of Washington, Los Alamos National Laboratories, and Lawrence Berkeley National Laboratories. Operations are conducted by COO, IPAC, and UW. This research has made use of the NASA/IPAC Extragalactic Database (NED), which is operated by Jet Propulsion Laboratory, California Institute of Technology, under contract with the National Aeronautics and Space Administration. Facilities: Fermi, Swift, MAXI. Software: astropy\citep{2013A&A...558A..33A, 2018AJ....156..123A, 2022ApJ...935..167A}, Numpy \citep{2020Natur.585..357H}.

\section{Author contributions}

M. Zha is the convener of this work. R. Wang, M. Zha, H.B. Tang and R.Y. Liu drafted the manuscript; R. Wang and M. Zha peformed WCDA data analysis, T. Wen performed KM2A data analysis, H.B. Tang and R.Y. Liu provided model interpretation; C.F. Feng and D.X. Xiao provided cross check analysis. And F. Aharonian provided suggestions on this work. The whole LHAASO collaboration contributed to the publication, with involvement at various stages ranging from the design, construction, and operation of the instrument to the development and maintenance of all software for data calibration, data reconstruction, and data analysis. All authors reviewed, discussed, and commented on the present results and on the manuscript.

\bibliographystyle{aasjournal}
\bibliography{mrk421_reference}{}

\newpage    
\section{Appendix-A}

\begin{figure}[htbp]
    \centering
    \begin{minipage}[b]{0.28\linewidth}
        \centering
        \includegraphics[width=\linewidth]{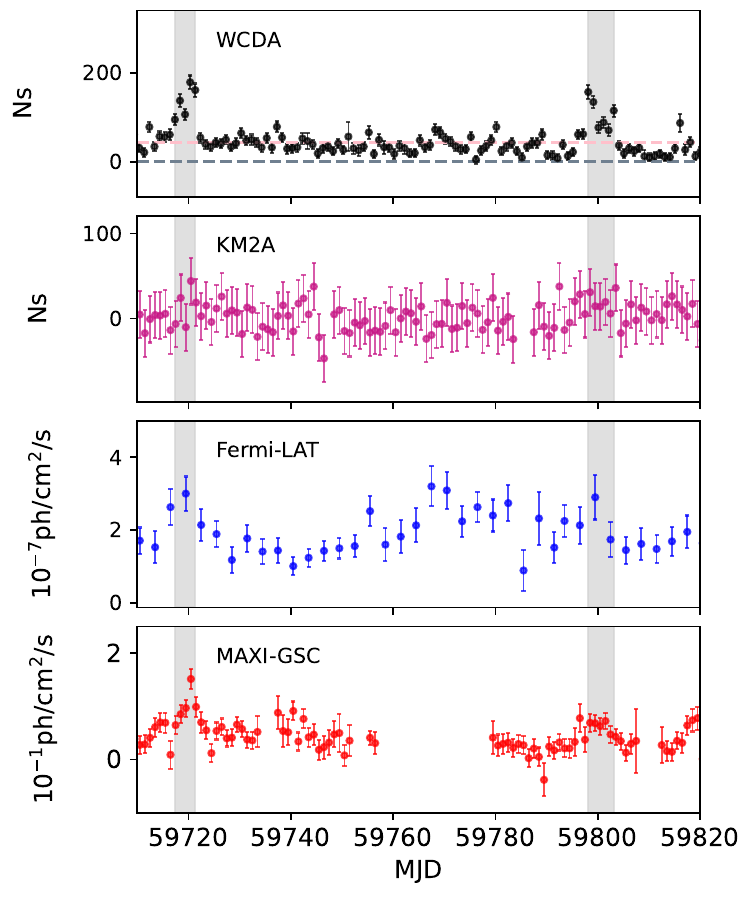} 
        \textbf{(a)} O1\_O2
    \end{minipage}
    \hfill       
    \begin{minipage}[b]{0.28\linewidth}
        \centering
        \includegraphics[width=\linewidth]{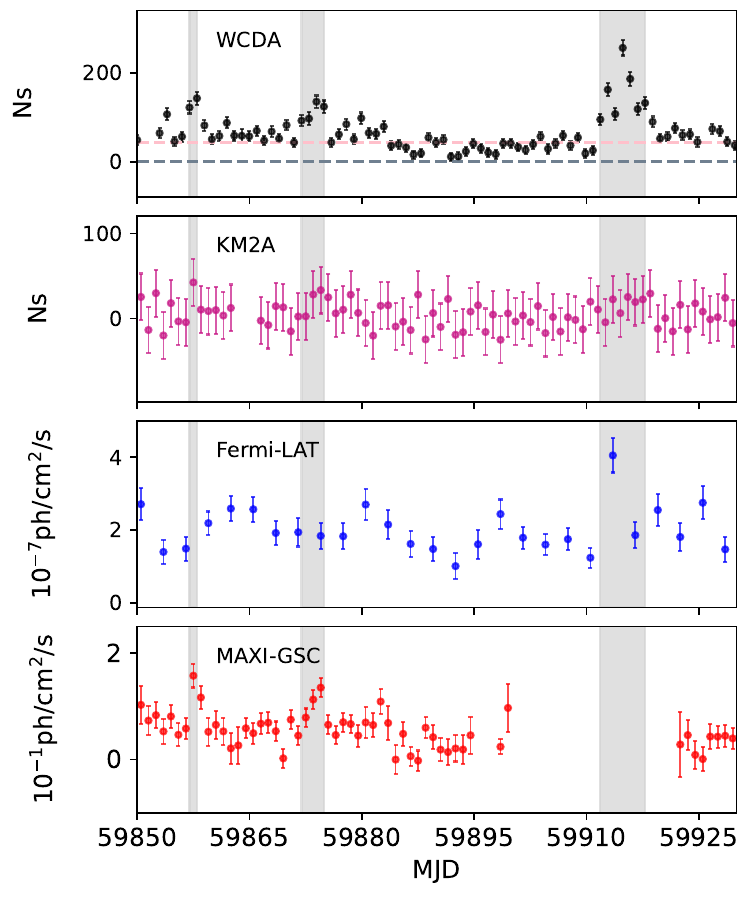} 
        \textbf{(b)} O3\_O5
    \end{minipage}
    \hfill  
    \begin{minipage}[b]{0.28\linewidth}
        \centering
        \includegraphics[width=\linewidth]{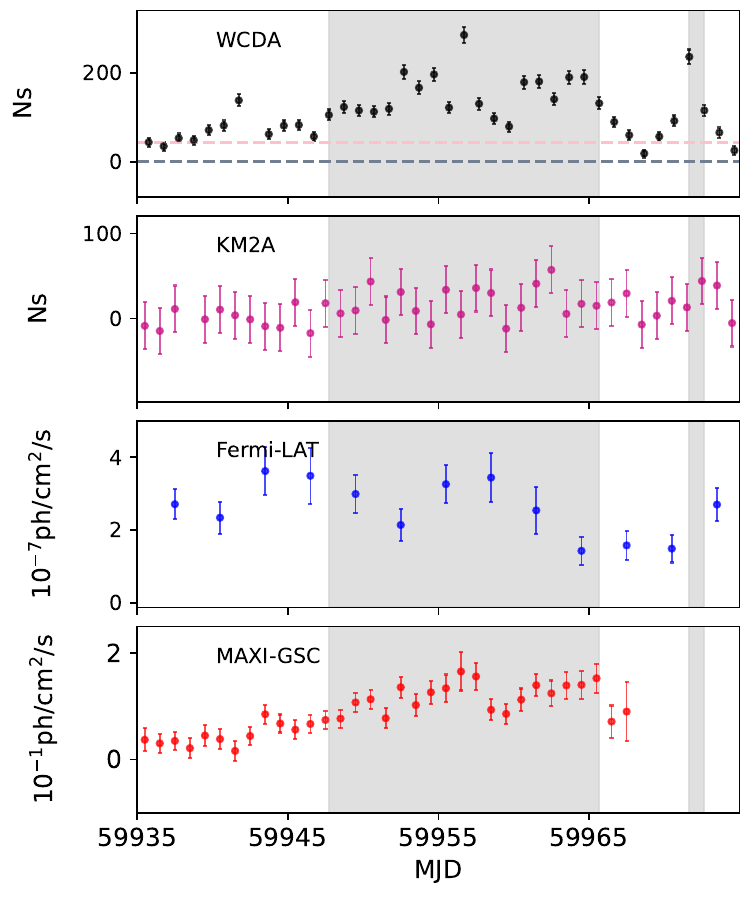} 
        \textbf{(c)} O6\_O7
    \end{minipage}

    \begin{minipage}[b]{0.28\linewidth}
        \centering
        \includegraphics[width=\linewidth]{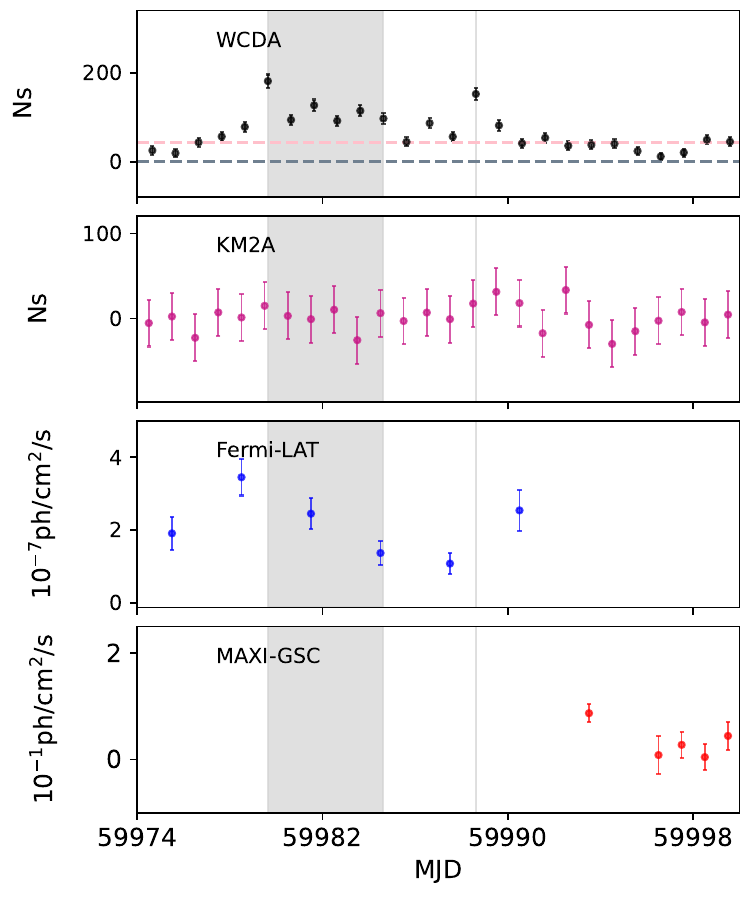} 
        \textbf{(d)} O8\_O9
    \end{minipage}
    \hfill 
    \begin{minipage}[b]{0.28\linewidth}
        \centering
        \includegraphics[width=\linewidth]{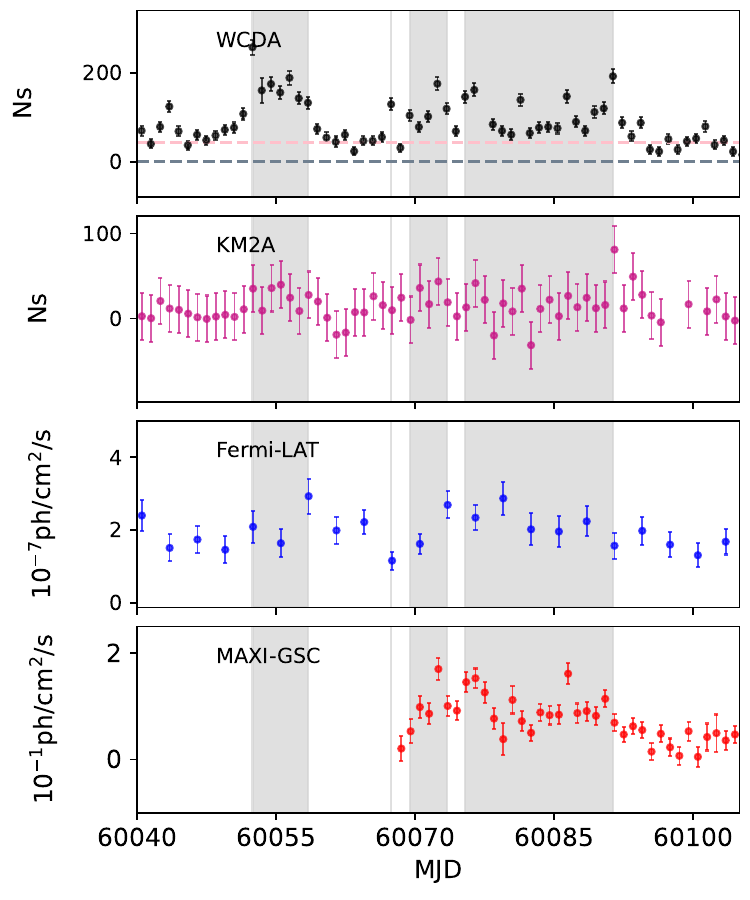} 
        \textbf{(e)} O10\_O13
    \end{minipage}
    \hfill 
    \begin{minipage}[b]{0.28\linewidth}
        \centering
        \includegraphics[width=\linewidth]{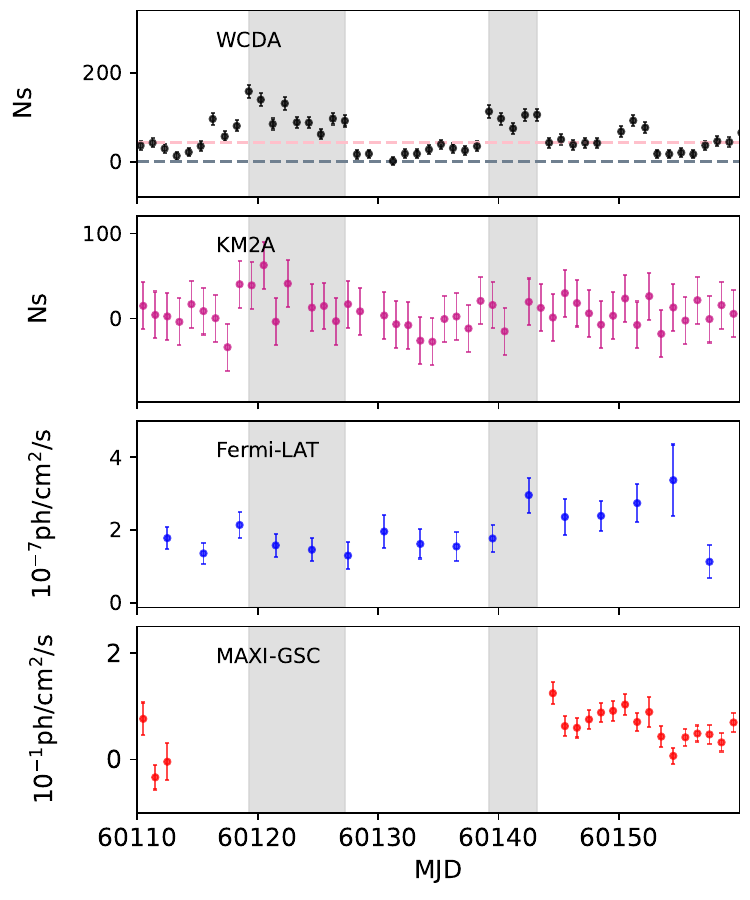} 
        \textbf{(f)} O14\_O15
    \end{minipage}

    \begin{minipage}[b]{0.28\linewidth}
        \centering
        \includegraphics[width=\linewidth]{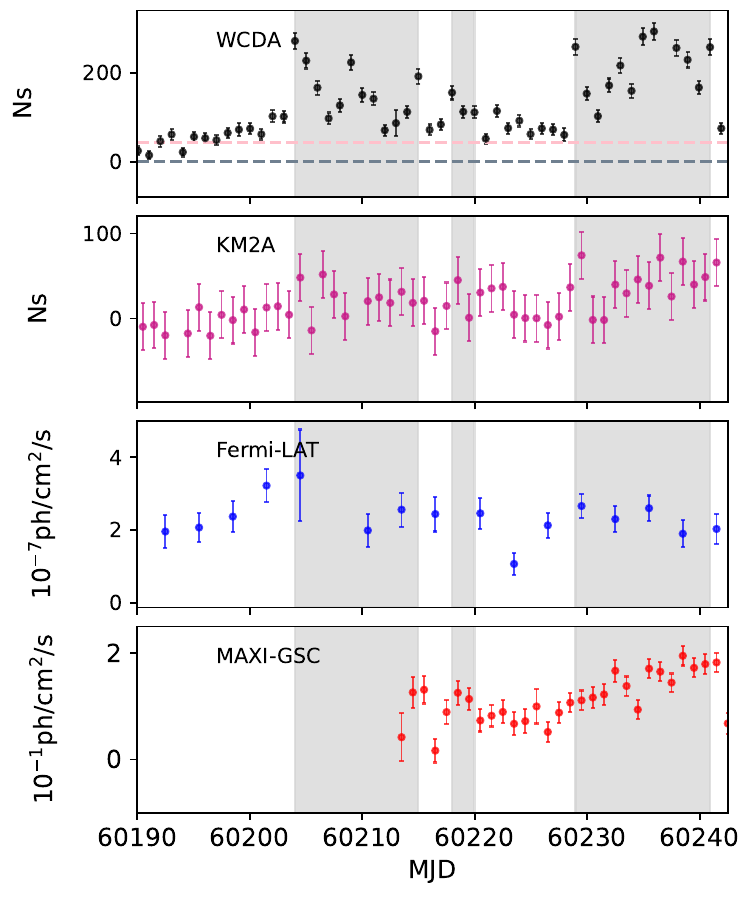} 
        \textbf{(g)} O16\_O18
    \end{minipage}
    \hfill  
    \begin{minipage}[b]{0.28\linewidth}
        \centering
        \includegraphics[width=\linewidth]{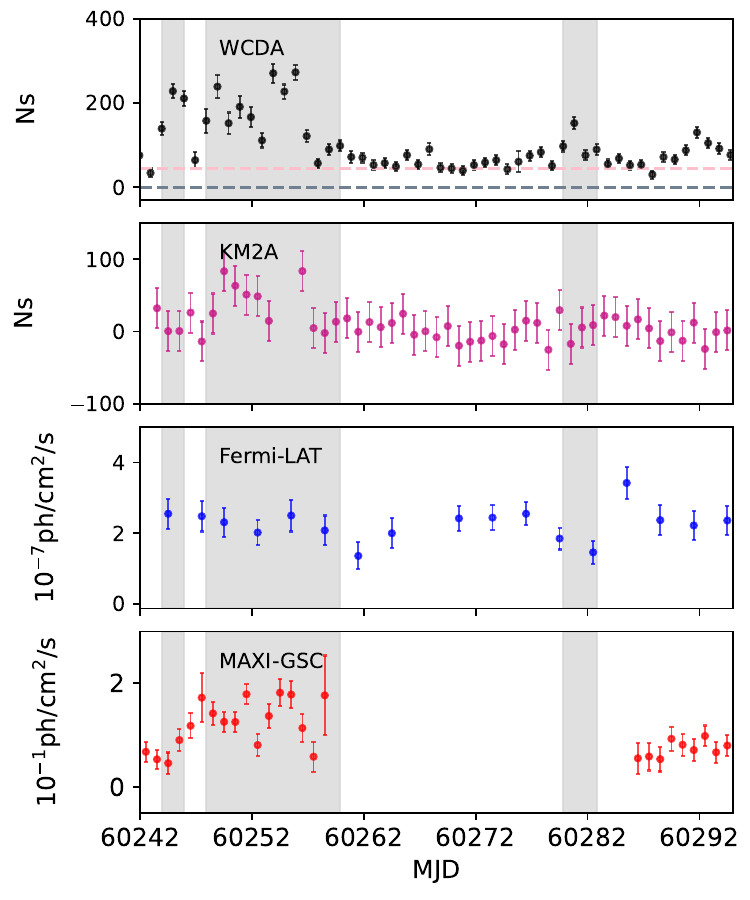} 
        \textbf{(h)} O19\_O21
    \end{minipage}
    \hfill  
     \begin{minipage}[b]{0.28\linewidth}
        \centering
        \includegraphics[width=\linewidth]{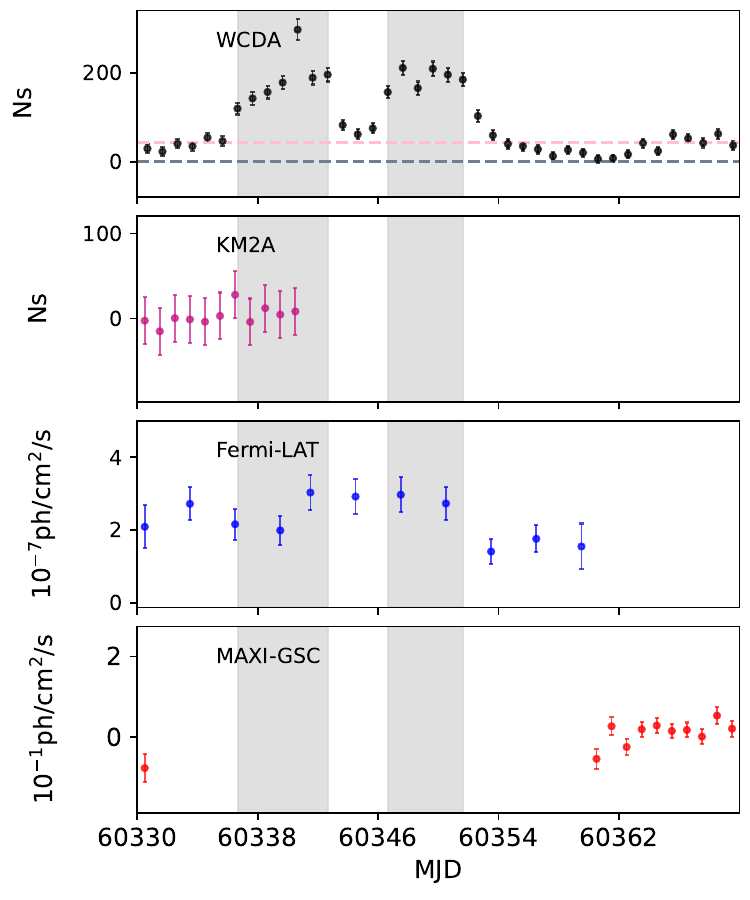} 
        \textbf{(i)} O22\_O23
    \end{minipage}
      
    \caption{The multi-wavelength light curves of 23 outburst episodes are constructed using data from LHAASO, Fermi-LAT, and MAXI-GSC. Shadowed regions demarcate the outburst durations, while the top subplot features gray and pink dashed lines corresponding to the zero-flux level and steady-state signal, respectively.}
    \label{fig:multi_lc_outburst}
\end{figure}

\newpage
\section{Appendix-B}

\begin{figure*}[!ht]
\centering
\includegraphics[width=0.5\textwidth]{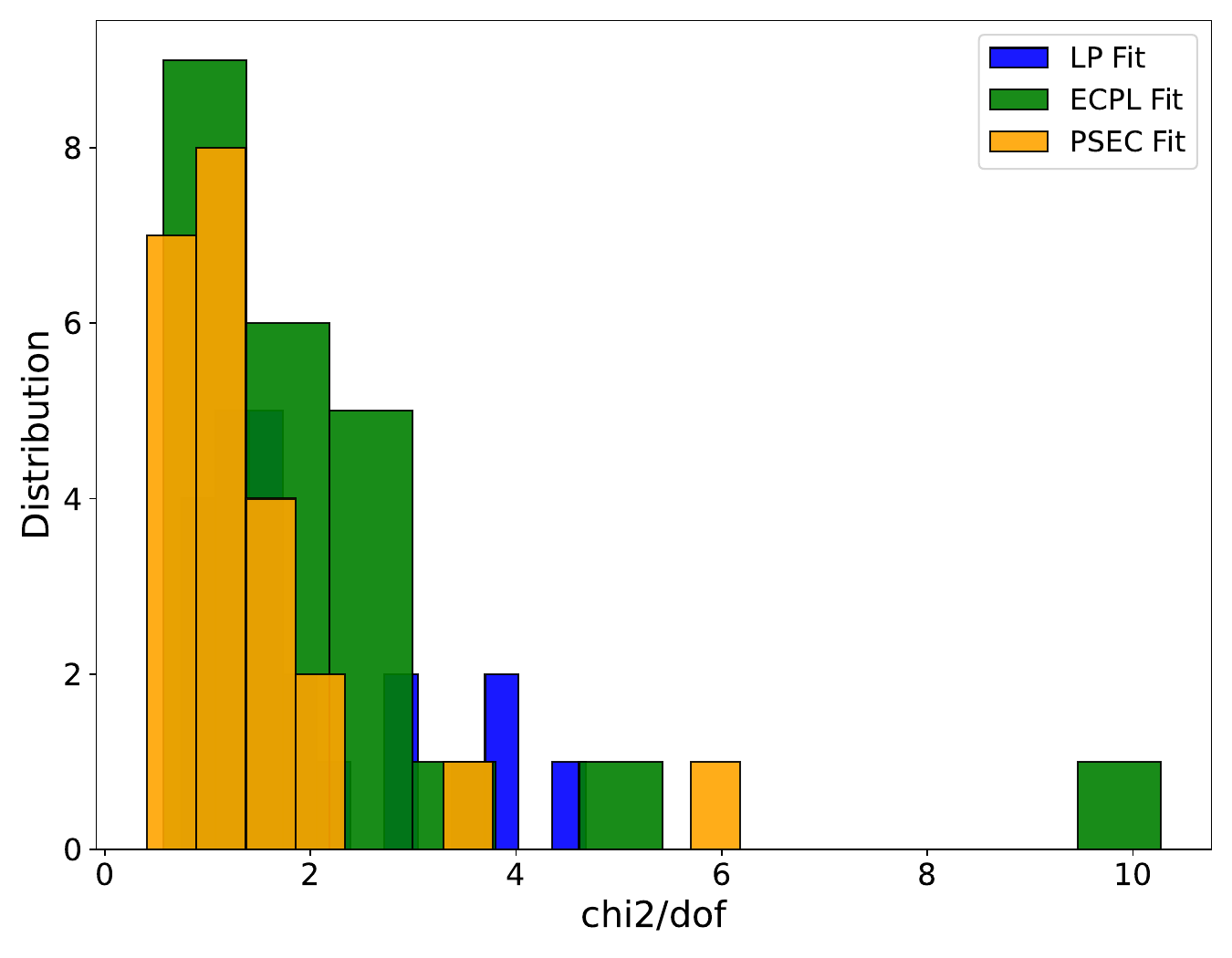}
\caption{The $\chi^{2}/dof$ distributions of gamma-ray energy spectrum fitted with three models during the outburst episodes are shown, where blue, green, and yellow represent the LP, ECPL, and PSEC models, respectively.}
\label{chi2_distribution}
\end{figure*}

\begin{table*}[!ht]
\centering
\caption{Parameters of the joint spectral fit to the gamma-ray SEDs obtained with LHAASO and Fermi-LAT data for Log-parabola function}
\begin{tabular}{lcccc}
\hline
\hline
\multicolumn{5}{c}{Log parabola : $dN/dE = N_{0} \times (\frac{E}{1 \mathrm{TeV}})^{-\alpha-\beta \times \log(\frac{E}{1 \mathrm{TeV}})}$} \\
\cline{1-5}
Epoch & $F_{0}$ & $\alpha$ & $\beta$ & $\chi^2/dof $ \\
& $(\times10^{-11}TeV^{-1}cm^{-2}s^{-1})$ &&&\\
\hline
O1 & $6.74\pm0.29$ & $2.33\pm 0.05$ & $0.06\pm 0.01$ & 0.81\\
O2 & $5.79\pm0.23$ & $2.67\pm 0.05$ & $0.12\pm 0.01$ & 2.23\\
O3 & $6.64\pm0.51$ & $2.08\pm 0.18$ & $0.38\pm 0.19$ & 3.75\\
O4 & $5.91\pm0.32$ & $2.45\pm 0.07$ & $0.09\pm 0.01$ & 2.85\\
O5 & $7.79\pm0.25$ & $2.32\pm 0.03$ & $0.06\pm 0.01$ & 1.31\\
O6 & $8.83\pm0.16$ & $2.49\pm 0.02$ & $0.10\pm 0.005$ & 3.03\\
O7 & $8.42\pm0.50$ & $2.20\pm 0.06$ & $0.05\pm 0.01$ & 0.75\\
O8 & $7.28\pm0.26$ & $2.52\pm 0.04$ & $0.09\pm 0.01$ & 1.83\\
O9 & $8.23\pm0.69$ & $2.33\pm 0.10$ & $0.06\pm 0.02$ & 1.17\\
O10 & $8.94\pm0.27$ & $2.36\pm 0.03$ & $0.09\pm 0.01$ & 3.90\\
O11 & $7.03\pm0.65$ & $2.35\pm 0.11$ & $0.07\pm 0.02$ & 1.21\\
O12 & $6.38\pm0.28$ & $2.39\pm 0.05$ & $0.08\pm 0.01$ & 1.07\\
O13 & $5.51\pm0.15$ & $2.41\pm 0.03$ & $0.08\pm 0.01$ & 1.32\\
O14 & $4.83\pm0.19$ & $2.36\pm 0.04$ & $0.07\pm 0.01$ & 1.48\\ 
O15 & $4.77\pm0.26$ & $2.46\pm 0.06$ & $0.07\pm 0.01$ & 1.78\\
O16 & $7.93\pm0.19$ & $2.44\pm 0.03$ & $0.09\pm 0.01$ & 1.17\\
O17 & $5.52\pm0.36$ & $2.34\pm 0.07$ & $0.08\pm 0.01$ & 1.52\\ 
O18 & $8.97\pm0.20$ & $2.27\pm 0.02$ & $0.07\pm0.004$ & 3.55\\
O19 & $9.12\pm0.40$ & $2.34\pm 0.05$ & $0.07\pm 0.01$ & 1.53\\
O20 & $7.33\pm0.21$ & $2.28\pm 0.02$ & $0.06\pm 0.01$ & 1.51\\
O21 & $5.11\pm0.32$ & $2.36\pm 0.07$ & $0.08\pm 0.01$ & 1.05\\
O22 & $9.43\pm0.27$ & $2.38\pm 0.04$ & $0.09\pm 0.01$ & 1.66\\
O23 & $10.21\pm0.28$ & $2.42\pm 0.03$ & $0.08\pm 0.01$ & 4.68\\
\hline
\hline
\end{tabular}
\label{tab:Fit_Logparabola}
\end{table*}

\begin{table*}[!ht]
\centering
\caption{Parameters of the joint spectral fit to the gamma-ray SEDs obtained with LHAASO and Fermi-LAT data for Power-law with exponential cutoff function}
\begin{tabular}{lcccc}
\hline
\hline
\multicolumn{5}{c}{Power-law with Exponential Cutoff : $dN/dE = N_{0} \times (\frac{E}{1 \mathrm{TeV}})^{-\alpha}\times exp(-\frac{E}{E_{c}})$} \\
\cline{1-5}
Epoch & $F_{0}$ & $\alpha$ & $E_{c}$ & $\chi^2/dof $ \\
& $(\times10^{-10}TeV^{-1}cm^{-2}s^{-1})$&&(TeV)&\\
\hline
O1 & $0.96\pm0.08$ & $1.83\pm0.02$ & $3.06\pm0.47$ & 1.00 \\
O2 & $1.23\pm0.13$ & $1.91\pm0.02$ & $1.53\pm0.20$ & 10.27 \\
O3 & $0.90\pm0.13$ & $1.84\pm0.07$ & $3.01\pm0.82$ & 1.82 \\
O4 & $1.04\pm0.12$ & $1.83\pm0.04$ & $2.01\pm0.32$ & 1.00 \\
O5 & $1.13\pm0.07$ & $1.87\pm0.01$ & $3.21\pm0.36$ & 2.98 \\
O6 & $1.41\pm0.05$ & $1.80\pm0.02$ & $2.34\pm0.14$ & 2.91 \\
O7 & $1.05\pm0.11$ & $1.85\pm0.03$ & $4.69\pm1.17$ & 0.57 \\
O8 & $1.29\pm0.10$ & $1.82\pm0.02$ & $2.01\pm0.21$ & 0.81 \\
O9 & $1.08\pm0.18$ & $1.86\pm0.07$ & $3.87\pm1.38$ & 1.42 \\
O10 & $1.31\pm0.09$ & $1.77\pm0.02$ & $2.76\pm0.32$ & 4.85 \\
O11 & $0.84\pm0.17$ & $1.86\pm0.06$ & $4.57\pm2.08$ & 1.89 \\
O12 & $0.93\pm0.08$ & $1.83\pm0.03$ & $2.81\pm0.44$ & 1.97 \\
O13 & $0.79\pm0.05$ & $1.85\pm0.02$ & $2.92\pm0.31$ & 2.53 \\
O14 & $0.71\pm0.06$ & $1.84\pm0.02$ & $2.87\pm0.39$ & 1.27 \\
O15 & $0.81\pm0.09$ & $1.89\pm0.03$ & $2.20\pm0.37$ & 0.79 \\
O16 & $1.13\pm0.06$ & $1.83\pm0.02$ & $2.88\pm0.26$ & 2.43 \\
O17 & $0.74\pm0.10$ & $1.83\pm0.05$ & $3.31\pm0.89$ & 2.04 \\
O18 & $1.14\pm0.04$ & $1.80\pm0.01$ & $4.29\pm0.30$ & 1.79 \\
O19 & $1.31\pm0.12$ & $1.83\pm0.03$ & $3.05\pm0.55$ & 0.81 \\
O20 & $0.89\pm0.04$ & $1.84\pm0.02$ & $4.77\pm0.46$ & 2.36 \\
O21 & $0.79\pm0.09$ & $1.83\pm0.02$ & $2.63\pm0.48$ & 0.60 \\
O22 & $1.48\pm0.09$ & $1.77\pm0.03$ & $2.39\pm0.24$ & 1.59 \\
O23 & $1.83\pm0.11$ & $1.78\pm0.02$ & $1.99\pm0.17$ & 3.25 \\
\hline
\hline 
\end{tabular}
\label{tab:Fit_ECPL}
\end{table*}

\begin{table*}[!ht]
\centering
\caption{Parameters of the joint spectral fit to the gamma-ray SEDs obtained with LHAASO and Fermi-LAT data for Power-law with subexp cutoff function}
\begin{tabular}{lcccc}
\hline
\hline
\multicolumn{5}{c}{Power-law: $dN/dE = N_{0} \times(\frac{E}{1 \mathrm{TeV}})^{-\alpha}\times exp(-\sqrt{\frac{E}{E_{c}}})$} \\
\cline{1-5}
Epoch & $F_{0}$ & $\alpha$ & $E_{c}$ & $\chi^2/dof $ \\
& $(\times10^{-10}TeV^{-1}cm^{-2}s^{-1})$&&(TeV)&\\
\hline
O1 & $2.04\pm0.35$ & $1.73\pm0.03$ & $0.86\pm0.23$ & 0.72 \\
O2 & $4.48\pm0.90$ & $1.70\pm0.04$ & $0.26\pm0.05$ & 6.18 \\
O3 & $1.81\pm0.60$ & $1.73\pm0.09$ & $0.93\pm0.51$ & 2.07 \\
O4 & $2.71\pm0.68$ & $1.68\pm0.06$ & $0.46\pm0.14$ & 1.40 \\
O5 & $2.46\pm0.30$ & $1.75\pm0.02$ & $0.84\pm0.16$ & 1.58 \\
O6 & $3.88\pm0.30$ & $1.65\pm0.02$ & $0.48\pm0.05$ & 1.09 \\
O7 & $1.79\pm0.39$ & $1.77\pm0.04$ & $1.84\pm0.85$ & 0.56 \\
O8 & $3.78\pm0.57$ & $1.66\pm0.03$ & $0.40\pm0.07$ & 0.54 \\
O9 & $2.18\pm0.75$ & $1.76\pm0.09$ & $1.12\pm0.68$ & 1.13 \\
O10 & $3.08\pm0.40$ & $1.63\pm0.03$ & $0.67\pm0.12$ & 3.55 \\
O11 & $1.77\pm0.69$ & $1.74\pm0.09$ & $1.19\pm0.84$ & 1.57 \\
O12 & $2.20\pm0.39$ & $1.71\pm0.04$ & $0.69\pm0.18$ & 1.18 \\
O13 & $1.83\pm0.20$ & $1.74\pm0.02$ & $0.71\pm0.11$ & 1.02 \\
O14 & $1.64\pm0.27$ & $1.71\pm0.04$ & $0.72\pm0.17$ & 0.79 \\
O15 & $2.03\pm0.46$ & $1.77\pm0.04$ & $0.52\pm0.15$ & 0.59 \\
O16 & $2.84\pm0.30$ & $1.69\pm0.03$ & $0.63\pm0.09$ & 0.95 \\
O17 & $1.60\pm0.43$ & $1.71\pm0.07$ & $0.90\pm0.39$ & 1.62 \\
O18 & $2.38\pm0.17$ & $1.69\pm0.02$ & $1.14\pm0.13$ & 0.98 \\
O19 & $2.80\pm0.50$ & $1.72\pm0.04$ & $0.86\pm0.25$ & 0.67 \\
O20 & $1.82\pm0.16$ & $1.73\pm0.02$ & $1.26\pm0.20$ & 1.19 \\
O21 & $1.87\pm0.44$ & $1.69\pm0.04$ & $0.64\pm0.20$ & 0.41 \\
O22 & $3.75\pm0.51$ & $1.63\pm0.04$ & $0.55\pm0.10$ & 0.95 \\
O23 & $5.08\pm0.62$ & $1.64\pm0.02$ & $0.42\pm0.06$ & 2.07 \\
\hline
\hline 
\end{tabular}
\label{tab:Fit_PSEC}
\end{table*}


\end{document}